%% file: 00_main.tex
\documentclass[fleqn,usenatbib,useAMS]{mnras}
\usepackage[utf8]{inputenc}
\usepackage{graphicx}
\usepackage{newtxtext,newtxmath}
\usepackage[T1]{fontenc}
\usepackage{hyperref}
\usepackage{amsmath}
\usepackage{amsfonts}
\usepackage{bm}
\usepackage{threeparttable}
\usepackage{natbib}
\usepackage[normalem]{ulem}
\defcitealias{Pierel_2021}{P21}
\allowdisplaybreaks

\title[Time-Delay Estimation with GausSN]{GausSN: Bayesian Time-Delay Estimation for Strongly Lensed Supernovae}

\author[E. E. Hayes et al.]{Erin E. Hayes,$^{1}$\thanks{Contact e-mail: \href{mailto:eeh55@cam.ac.uk}{eeh55@cam.ac.uk}}
Stephen Thorp,$^{2}$
Kaisey S. Mandel,$^{1,3}$
Nikki Arendse,$^{2}$
Matthew Grayling,$^{1}$
\newauthor
and Suhail Dhawan$^{1}$
\\
$^{1}$Institute of Astronomy and Kavli Institute for Cosmology, Madingley Road, Cambridge CB3 0HA, UK\\
$^{2}$The Oskar Klein Centre, Department of Physics, Stockholm University, AlbaNova University Centre, SE 106 91 Stockholm, Sweden\\
$^{3}$Statistical Laboratory, DPMMS, University of Cambridge, Wilberforce Road, Cambridge, CB3 0WB, UK\\
}

\date{Accepted XXX. Received YYY; in original form ZZZ}

\pubyear{2023}

\begin{document}
\label{firstpage}
\pagerange{\pageref{firstpage}--\pageref{lastpage}}
\maketitle

\newcommand{\Cov}[1]{\operatorname{Cov}\left(#1\right)}

\begin{abstract}
    We present \textsc{GausSN}, a Bayesian semi-parametric Gaussian Process (GP) model for time-delay estimation with resolved systems of gravitationally lensed supernovae (glSNe). \textsc{GausSN} models the underlying light curve non-parametrically using a GP. Without assuming a template light curve for each SN type, \textsc{GausSN} fits for the time delays of all images using data in any number of wavelength filters simultaneously. We also introduce a novel time-varying magnification model to capture the effects of microlensing alongside time-delay estimation. In this analysis, we model the time-varying relative magnification as a sigmoid function, as well as a constant for comparison to existing time-delay estimation approaches. We demonstrate that \textsc{GausSN} provides robust time-delay estimates for simulations of glSNe from the Nancy Grace Roman Space Telescope and the Vera C. Rubin Observatory's Legacy Survey of Space and Time (Rubin-LSST). We find that up to 43.6\% of time-delay estimates from Roman and 52.9\% from Rubin-LSST have fractional errors of less than 5\%. We then apply \textsc{GausSN} to SN Refsdal and find the time delay for the fifth image is consistent with the original analysis, regardless of microlensing treatment. Therefore, \textsc{GausSN} maintains the level of precision and accuracy achieved by existing time-delay extraction methods with fewer assumptions about the underlying shape of the light curve than template-based approaches, while incorporating microlensing into the statistical error budget. \textsc{GausSN} is scalable for time-delay cosmography analyses given current projections of glSNe discovery rates from Rubin-LSST and Roman.
\end{abstract}


\section{Introduction}
\input{01_introduction}

\section{Modelling glSNe with Gaussian Processes in Flux-Space}
\label{sec:gp-math}
\input{02_modeling}

\section{Microlensing}
\label{sec:microlensing}
\input{03_microlensing}

\section{Computation}
\label{sec:sampling}
\input{04_sampling}

\section{Tests on Simulated Data}
\label{sec:example}
\input{05_0_tests}

\section{Application to SN Refsdal}
\label{sec:refsdal}
\input{06_refsdal}

\section{Discussion}
\label{sec:discussion}
\input{07_discussion}

\section{Conclusion}
\label{sec:conclusion}
\input{08_conclusions}

\section*{Acknowledgements}

We thank Vidhi Lalchand, Sam Ward, and Ben Boyd for useful discussions. We also thank the reviewer for their insightful feedback on this work.

Supernova and astrostatistics research at Cambridge University is supported by the European Union’s Horizon 2020 research and innovation programme under European Research Council Grant Agreement No 101002652 and Marie Skłodowska-Curie Grant Agreement No 873089.
ST was supported by the European Research Council (ERC) under the European Union's Horizon 2020 research and innovation programme (grant agreement no.\ 101018897 CosmicExplorer).
EEH is supported by a Gates Cambridge Scholarship (\#OPP1144).
NA is supported by the research project grant ‘Understanding the Dynamic Universe’ funded by the Knut and Alice Wallenberg Foundation under Dnr KAW 2018.0067.

\section*{Data Availability}

The simulated Roman data from \cite{Pierel_2021} are publicly available in the MAST Archive at \url{https://dx.doi.org/10.17909/t9-k8w7-zk32}. The SN Refsdal data from \cite{Kelly_2023_ApJ} are publicly available from The Astrophysical Journal at \url{https://dx.doi.org/10.3847/1538-4357/ac4ccb}. The simulated Rubin-LSST data from \cite{Arendse_2023} are publicly available from Github at \url{https://github.com/Nikki1510/lensed_supernova_simulator_tool/tree/main}. The simulated constant magnification Rubin-LSST data are available upon request to the corresponding author.

\bibliographystyle{mnras}
\bibliography{bib}

\appendix
\section{The Effect of Microlensing on Time-Delay Estimation}
\label{sec:appendixA}
\input{09_appendixA}

\section{Flexible Sinusoidal Microlensing Parameterization}
\label{sec:appendixB}
\input{10_appendixB}

\bsp
\label{lastpage}
\end{document}

%% file: 01_introduction.tex
Strong lensing of a background variable source, such as a supernova (SN), by a foreground galaxy or galaxy cluster, can result in the appearance of multiple images of the source. \cite{Refsdal_1964} was the first to demonstrate that the time delay between the multiple images of a gravitationally lensed SN (glSN) images could be related to $H_{0}$ when combined with a model of the lens mass distribution. A independent $H_{0}$ estimate from strong lensing is motivated by the persistent $5\sigma$ tension between the present-day expansion rate of the universe, $H_{0}$, measured using light from early-times (i.e. the Cosmic Microwave Background; \citealp{Planck_2020}) and from late-times (i.e. the distance ladder; \citealp{Riess_2022}), known as the Hubble tension. If determined to be physical, this tension may be suggestive of new physics beyond our present model of the universe, $\Lambda$CDM \citep{Mortsell_2018, DiValentino_2021}. As time-delay distances inferred from strong lensing are completely independent of the luminosity distances used in the distance ladder, time-delay cosmography represents a promising technique for cross-checking local measurements of $H_{0}$ (see e.g. \citealp{Treu_2022, Suyu_2023} for recent reviews). In this paper, we present \href{https://github.com/erinhay/gaussn}{\textsc{GausSN}}: a Bayesian semi-parametric approach for the extraction of time delays from glSN systems using Gaussian Processes (GPs). \textsc{GausSN} is a publicly available package that can be found at: \url{https://github.com/erinhay/gaussn}.

Because of the rarity of glSNe, the first cosmological constraints from strong lensing came from quasars \citep[e.g.][]{Keeton_1997, Wong_2019, Birrer_2020}. However, SNe have several advantages over quasars for strong lensing analyses which have the potential to push the field further in precision and accuracy. Firstly, that SNe disappear after a few months allows for the study of the lens and host galaxy in greater detail. Improved knowledge of the lens helps to break the mass-sheet degeneracy \citep{Falco_1985}, which was cited as the main source of uncertainty in the H0LiCOW estimate of $H_{0}$ \citep{Wong_2019, Birrer_2020}. If the lensed object is a Type Ia SN (SN~Ia), further leverage can be gained from the fact that SNe~Ia are standardizable candles \citep{FoxleyMarrable_2018, Birrer_2022}. Secondly, the relatively simple light curves of SNe, in contrast to the stochastic variability of quasars, makes it easier to extract time delays from the systems. In particular, the achromatic expansion phase of SNe~Ia, during which microlensing has been shown to cancel in color curves \citep{Goldstein_2018, Huber_2021}, may contribute to improved constraints on the microlensing affecting the system \citep{Bonvin_2019_microlens}. Microlensing, which can distort the shape of an image's light curve and lead to artificial shifts in the apparent peak of the light curve, is an important source of uncertainty that has proven difficult to disentangle from the underlying variability of the source \citep{FoxleyMarrable_2018, Pierel_2022, Weisenbach_2024}. Lastly, time-delay estimation with glSNe can also be done with a shorter time frame of observations because their variability is constrained to a period of a few weeks to months, compared to the years required to measure reliable time delays from lensed quasars \citep{Bonvin_2017, Bonvin_2018, Bonvin_2019_lensedqso}. Of course, the rarity of SNe and short window in which they are active make glSNe more difficult to discover.

Therefore, it was not until November 2014 that the first resolved multiply-imaged SN, SN Refsdal, was discovered \citep{Kelly_2015}. SN Refsdal, a peculiar Type II SN at $z=1.49$ \citep{Kelly_2016_spec}, was discovered with four images in an Einstein cross due to lensing by an elliptical galaxy in the MACS J1149.6+2223 galaxy cluster at $z=0.54$. Unfortunately, the time delays from galaxy-scale lensing, originally estimated in \citep{Rodney_2016}, were determined to be too short and imprecise to obtain a measurement of $H_{0}$. However, the host galaxy of SN Refsdal was also imaged multiple times. It was predicted that a $5^{\text{th}}$ image of SN Refsdal would later appear in another image of the host galaxy with a time delay of just under a year \citep{Kelly_2015, Oguri_2015, Sharon_2015, Grillo_2016, Diego_2016, Jauzac_2016, Kawamata_2016, Treu_2016}. As predicted, SN Refsdal reappeared less than a year later in October 2015 \citep{Kelly_2016_reappearance}. Using the time delay between SN Refsdal's fifth image and the other four images \citep{Kelly_2023_ApJ}, the first measurement of $H_{0}$ was published in May 2023 \citep{Kelly_2023_Science}. Finding $H_{0} = 64.8^{+4.4}_{-4.3} \, \text{km} \, \text{s}^{-1} \, \text{Mpc}^{-1}$ -- a 6.0\% precision estimate -- SN Refsdal alone has demonstrated the importance of glSNe as precise local cosmological probes.

In addition to SN Refsdal, seven glSNe have been found: SN PS1-10afx \citep{Chornock_2013, Quimby_2013}, SN 2016geu \citep{Goobar_2017}, SN Requiem \citep{Rodney_2021}, SN C22 \citep{Chen_2022}, SN Zwicky \citep{Goobar_2022}, SN 2022riv \citep{Kelly_2022_riv}, SN H0pe \citep{Frye_2023, Frye_2023_ApJ, Polletta_2023}, and SN Encore \citep{Pierel_2023_Encore}. However, six of these glSNe have suffered from either having extremely short time delays (SN 2016geu and SN Zwicky), having extremely long time delays (SN Requiem), being found only in an archival search and therefore having insufficient data (SN PS1-10afx, SN Requiem, and SN C22), or having only one observed image (SN 2022riv), making them unsuitable for cosmological analyses. Recently, SN H0pe, discovered in March 2023, became the first glSN~Ia to be used for a measurement of $H_{0}$ \citep{Pascale_2024, Chen_2024, Pierel_2024}. \cite{Pascale_2024} estimate $H_{0} = 75.4^{+8.1}_{-5.5} \, \text{km} \, \text{s}^{-1} \, \text{Mpc}^{-1}$ when leveraging information about the absolute magnitude of SNe~Ia, and more conservatively $H_{0} = 71.8^{+9.8}_{-7.6} \, \text{km} \, \text{s}^{-1} \, \text{Mpc}^{-1}$ without using absolute magnification information. In addition, SN Encore, discovered in December 2023, is a strong candidate for a measurement of $H_{0}$. The analysis of SN Encore is currently ongoing and more information about this glSN is expected in the near future.

In the next decade, the Vera C. Rubin Observatory's Legacy Survey of Space and Time (Rubin-LSST) and Roman Space Telescope (Roman) are expected to discover tens to hundreds of glSNe, if we adopt the right search strategies \citep{Huber_2019, Wojtak_2019, Pierel_2021, Craig_2021}. With this data, it will be possible to reach percent level constraints on the cosmological parameters competitive with SH0ES and Planck measurements \citep{Huber_2019, Arendse_2023}. As samples of glSNe grow, so does their potential to resolve the Hubble tension with precise local measurements of $H_{0}$.

The relatively simple and well-understood variability of SNe lends itself to time-delay estimation techniques based around SN light curve templates -- empirical models for the light curves or spectral energy distributions (SEDs) of SNe. Supernova Time Delays (\texttt{sntd}; \citealp{Pierel_2019}) -- which uses light curve templates or parametric models -- has emerged as the standard for glSNe time-delay estimation due to its flexibility and accessibility. The package is built upon \texttt{sncosmo} \citep{Barbary_2022}, which contains a diverse library of SN light curve templates. A variety of methods have been used to construct SN light curve templates, so now there are a selection of templates for each of the many types of SNe, e.g. Type Ia, Type Ib, or Type IIP. Of course, the use of templates therefore requires knowledge of the SN type, which is most reliably determined via classification of an observed SN spectrum \citep{Filippenko_1997, GalYam_2017}. Also, the redshift of the SN is always required to use templates to shift the rest-frame template to the observer-frame in wavelength and time. Alternatively, flexible parametric models, such as the Bazin function \citep{Bazin_2009}, have been created to fit SN light curves if spectroscopic classification is unavailable or ambiguous. Given a specified SN light curve template and redshift, \texttt{sntd} uses the nested sampling functionality available through \texttt{sncosmo} to yield posterior distributions for the light curve template parameters, the time delay, and the magnification. 

However, uncertainties relating to template choice and the unknown redshift evolution of SN light curves, especially for types of SNe other than SNe~Ia, may introduce systematic errors into the time-delay analysis that are difficult to quantify. While template-based approaches are motivated by the well-described nature of some SN light curves, a complementary method which is independent of these potential systematics is needed. \textsc{GausSN} is a template-independent approach for time-delay estimation that models the underlying light curve non-parametrically using a Gaussian Process (GP). This model is motivated by previous work on quasar time-delay estimation with GPs, in particular that of \cite{Tak_2017}, as well as that of \cite{Hojjati_2013, Hojjati_2014, Meyer_2023}. 

By modelling the underlying light curve non-parametrically with a GP, \textsc{GausSN} is able to fit the light curves of any type of SN without prior knowledge of the type or the redshift of the object. In addition, the model naturally fits for the time delays of all images, observed in any number of bands, simultaneously. By fitting all images in all filters together, we take advantage of the full information available from the data by leveraging the knowledge that for each band, the multiple images' time-series are time-shifted and magnified realization of the same underlying light curve. \textsc{GausSN} also implements a novel microlensing treatment which occurs alongside time-delay estimation for a one-step time-delay measurement. A time-varying magnification term is used to account for both macrolensing and microlensing simultaneously with the time-delay estimate. Because the time delay and microlensing are jointly inferred, \textsc{GausSN} does not require post-processing to account for a systematic error budget due to microlensing, but instead incorporates uncertainty from microlensing in the statistical uncertainty. Therefore, \textsc{GausSN} provides a coherent Bayesian time-delay estimate in one-step, without the need for post-processing to account for systematics from template choice or microlensing.

The peculiar nature of SN Refsdal's light curve exemplifies the need for a template-independent approach for time-delay estimation for glSNe, such as \textsc{GausSN}. Although most similar to SN~1987A \citep{Woosley_1988, Kelly_2016_spec}, SN Refsdal's light curves are not well described by any existing templates \citep[although some theoretical models have been developed, e.g.][]{Baklanov_2021}. While the light curves of SNe~Ia are highly uniform, the shapes of the light curves of other types of SNe are less well studied and more highly varied from object to object. GPs are well suited for this application, as they are flexible and data driven, with minimal assumptions about the properties of the true underlying light curve. 

Indeed, in the analysis of the time delays of SN Refsdal's five images, the Bayesian Gaussian Process method, upon which \textsc{GausSN} is based, was highly successful \citep{Kelly_2023_ApJ}. Four methods, including \texttt{sntd}, used with a Bazin function as the parametric model because an SED template does not exist for SN Refsdal, and a custom piece-wise polynomial template approach, were tested on simulations of SN Refsdal-like light curves to determine the quality of each method's fits. In simulations, the Bayesian Gaussian Process method outperformed the parametric model-based approaches for extracting the time delays of the 2 best-sampled images S2 and S3 relative to the first image. On the other hand, \texttt{sntd} and the custom piece-wise polynomial approach were more successful in estimating the time delay of the more poorly sampled images S4 and SX relative to the first image. Ultimately, the custom piece-wise polynomial approach was the most generally successful in fitting for time delays and magnifications for all the images in simulations.

\textsc{GausSN} provides coherent Bayesian time-delay estimates with complementary strengths to existing template-based time-delay estimation methods. As a template-independent approach, \textsc{GausSN} does not require prior knowledge of SN type or redshift, making the model quick and easy to implement. Furthermore, \textsc{GausSN} is not subject to potential sources of systematic uncertainties and biases from template choice; fine tuning of generic SN models, such as the Bazin function; or the need to construct custom models, which may not be feasible for larger samples of glSNe and makes validation of an analysis difficult. Finally, the model provides Bayesian time-delay estimates in one step, without the need for post-processing to account for systematic uncertainty from microlensing, by incorporating a time-varying magnification treatment.

In \S\ref{sec:gp-math}, we describe how we model glSN in flux-space with GPs. We describe the microlensing model in \S\ref{sec:microlensing} and the sampling algorithms implemented in \textsc{GausSN} in \S\ref{sec:sampling}. In \S\ref{sec:example}, we show the fits to simulated doubly-imaged glSN light curves as expected to be seen from Rubin-LSST and Roman. We then apply \textsc{GausSN} to the five images of SN Refsdal in \S\ref{sec:refsdal}. In \S\ref{sec:discussion}, we discuss the performance of \textsc{GausSN} compared to other leading time-delay extraction techniques, as well as comment on future work possible with \textsc{GausSN}. In \S\ref{sec:conclusion}, we conclude.

%% file: 02_modeling.tex
Strong lensing of a SN by a foreground galaxy or galaxy cluster results in the appearance of multiple images, which are copies of the same underlying light curve. These multiple images are observed with some shift in time and magnification or de-magnification relative to each another. We model the true underlying light curve in flux space, $f(t)$, as a draw from a Gaussian Process:
\begin{equation}
    f(t) \sim \mathcal{GP}(c(t), k(t, t'))
\end{equation}
where $c(t)$ is the mean function, which can depend on time, t, in the case of time series data, and $k(t, t')$ is the covariance (or kernel) function, which gives the covariance between $f(t)$ and $f(t')$.

Throughout this work, we use a zero mean function, such that $c(t) = 0$. The zero mean function encourages the inferred function to go to zero flux outside of where there is data, which reflects the physical expectation that before the SN explosion and after the SN has faded, we expect an average of zero background-subtracted flux. For the covariance function, we choose the squared exponential kernel:
\begin{equation}
    k(t, t') = A^{2} e^{-(t-t')^2/ 2\tau^2}
    \label{eq:kernel}
\end{equation}
where $A$, the amplitude, and $\tau$, the length scale, are the two hyperparameters controlling the kernel. This kernel enforces strong correlation between points nearby to one another, with weaker correlation as points further away are considered. It is also stationary, or invariant to overall time shifts, and produces smooth functions \citep{Rasmussen_2006}. The squared exponential kernel has proven to be well-suited for fitting SN light curve data, as it has been used to do so in previous work with success \citep{Kim_2013, Boone_2019, Vincenzi_2019, Qu_2021}.

The GP is defined such that any vector $\bm{f}$ consisting of the evaluations of this continuous function, $f(t)$, at the finite set of points, $\bm{t}$, has a joint multivariate Gaussian prior distribution:
\begin{equation}
    \bm{f}(\bm{t}) \sim \mathcal{N}(\bm{0}, \bm{K}(\bm{t}, \bm{t}))
\end{equation}
where the elements of the covariance matrix, $\bm{K}(\bm{t}, \bm{t})$, are given by $K_{ij} = k(t_{i}, t_{j})$ for $t_{i}, t_{j} \in \bm{t}$. Note that we use $\bm{X} \sim \mathcal{N}(\bm{\mu}, \bm{\Sigma})$ to denote a multivariate normal vector with mean $\bm{\mu}$ and covariance matrix $\bm{\Sigma}$. Then, $P(\bm{X}) = \mathcal{N}(\bm{X}| \ \bm{\mu}, \bm{\Sigma})$ gives the PDF of the vector.

\subsection{Two Images in One Band}
We first consider the simplest case of two images observed in one band, i.e. a filter covering a defined range of wavelengths. Taking image 1 to be the arbitrarily chosen reference image, the light curves of the two images are described by:
\begin{equation}
    \begin{split}
        f_{1}(t) &= f(t) \\
        f_{2}(t) &= \beta f(t - \Delta)
    \end{split}
\end{equation}
where $\Delta$ is the relative time delay of image 2 compared to image 1 and $\beta$ is the relative magnification of image 2 compared to image 1.

Say we observe image 1 at a set of $N_{1}$ times, $\bm{\hat{t}}_{1}$, and image 2 at a set of $N_{2}$ times, $\bm{\hat{t}}_{2}$. We define $\bm{\hat{f}}_{1}(\bm{\hat{t}}_{1})$ and $\bm{\hat{f}}_{2}(\bm{\hat{t}}_{2})$, the observations of image 1 and 2 respectively, such that for the $i^{\text{th}}$ observation of image 1:
\begin{align}
    \label{eq:image-fluxes-1}
    \hat{f}_{1, i} &= \hat{f}_{1}(t_{i}) = f(t_{i}) + \epsilon_{1, i} \\
    \epsilon_{1, i} &\sim \mathcal{N}(0, \sigma_{1, i}^{2})
\end{align}
and for the $j^{\text{th}}$ observation of image 2:
\begin{align}
    \hat{f}_{2, j} &= \hat{f}_{2}(t_{j}) = \beta f(t_{j} - \Delta) + \epsilon_{2, j} \\
    \label{eq:image-fluxes-2}
    \epsilon_{2, j} &\sim \mathcal{N}(0, \sigma_{2, j}^{2})
\end{align}
where $\sigma_{1, i}$ and $\sigma_{2, j}$ are the standard deviations of the measurement errors of image 1 and image 2, respectively. We assume that the measurement errors are independent. We re-scale the flux data and measurement standard deviations for each band and image by a single, constant factor determined by the range of fluxes over all images and bands.

We concatenate these two vectors to give the flux data vector, $\bm{\hat{F}}$, as:
\begin{equation}
    \bm{\hat{F}} = ( \bm{\hat{f}}_{1}, \bm{\hat{f}}_{2} )^{\top}
\end{equation}
and the time data vector, $\bm{\hat{T}}$, as:
\begin{equation}
    \bm{\hat{T}} = ( \bm{\hat{t}}_{1}, \bm{\hat{t}}_{2} )^{\top}.
    \label{eq:time-vector}
\end{equation}
In addition, we define the de-shifted time vector, $\bm{\hat{T}}_{\Delta}$, which depends on the time delay, $\Delta$:
\begin{equation}
    \bm{\hat{T}}_{\Delta} = ( \bm{\hat{t}}_{1}, \bm{\hat{t}}_{2}-\Delta)^{\top}.
    \label{eq:deshifted-time-vector}
\end{equation}
To infer $f(t)$ from the data $\bm{\hat{F}}$ and $\bm{\hat{T}}$, we fit for a time delay, $\Delta$, and magnification, $\beta$, in addition to the two kernel hyperparameters. Therefore, $\bm{\theta} = (A, \tau, \Delta, \beta)$.

The marginal likelihood over $\bm{\theta}$ is given as:
\begin{equation}
    P(\bm{\hat{F}} | \ \bm{\hat{T}}, \bm{\theta}) = \mathcal{N}(\bm{\hat{F}} | \  \bm{0}, \bm{\Sigma_{\theta}})
    \label{eq:marginal-likelihood}
\end{equation}
where we can think of $\bm{\Sigma_{\theta}}$ in four quadrants:
\begin{equation}
    \bm{\Sigma_{\theta}} = 
    \begin{pmatrix}
        \Cov{\bm{\hat{f}}_{1}, \bm{\hat{f}}_{1}} & \Cov{\bm{\hat{f}}_{1}, \bm{\hat{f}}_{2}}\\
        \Cov{\bm{\hat{f}}_{2}, \bm{\hat{f}}_{1}} & \Cov{\bm{\hat{f}}_{2}, \bm{\hat{f}}_{2}}
    \end{pmatrix}.
    \label{eq:sigma-quadrants}
\end{equation}
Recall that for a GP, the covariance is given as $\Cov{f(t), f(t')} = k(t, t')$. With Equations \ref{eq:image-fluxes-1}-\ref{eq:image-fluxes-2}, the covariances in the first quadrant are:
\begin{equation}
    \begin{split}
        \Cov{\hat{f}_{1, i}, \hat{f}_{1, j}} &= \Cov{f_{1, i} + \epsilon_{1, i}, f_{1, j} + \epsilon_{1, j}} \\
        &= k(\hat{t}_{1, i}, \hat{t}_{1, j}) + \delta_{ij} \sigma_{1, i}^2
    \end{split}
\end{equation}
assuming there is no covariance between $f_{1, i}$ and $\epsilon_{1, j}$, as the physical process generating the light curve is independent of the measurement process. In the second quadrant, the covariances are:
\begin{equation}
\begin{split}
    \Cov{\hat{f}_{1, i}, \hat{f}_{2, j}} &= \Cov{f_{1, i} + \epsilon_{1, i}, f_{2, j} + \epsilon_{2, j}} \\
    &= \beta k(\hat{t}_{1, i}, \hat{t}_{2, j} - \Delta).
\end{split}
\end{equation}
It follows that:
\begin{equation}
    \Cov{\hat{f}_{2, i}, \hat{f}_{1, j}} = \beta k(\hat{t}_{2, i} - \Delta, \hat{t}_{1, j})
\end{equation}
and
\begin{equation}
    \begin{split}
    \Cov{\hat{f}_{2, i}, \hat{f}_{2, j}} &= \beta^{2} k(\hat{t}_{2, i} - \Delta, \hat{t}_{2, j}  - \Delta) + \delta_{ij} \sigma_{2, i}^2 \\
    &= \beta^{2} k(\hat{t}_{2, i}, \hat{t}_{2, j}) + \delta_{ij} \sigma_{2, i}^2
    \end{split}
\end{equation}
for the remaining two quadrants. Because our choice of kernel is stationary, $k(\hat{t}_{2, i} - \Delta, \hat{t}_{2, j} - \Delta) = k(\hat{t}_{2, i}, \hat{t}_{2, j})$.

Based on this formulation, we can decompose $\bm{\Sigma_{\theta}}$ into three factors:
\begin{equation}
    \bm{\Sigma_{\theta}} = (\bm{M} \odot \bm{K}) + \bm{W}
\end{equation}
where $\odot$ represents a Hadamard product (i.e. the elementwise product). The first factor, $\bm{M}$, depends only on $\beta$. Defining $\bm{1}$ as a matrix of ones, $\bm{M}$ is given as:
\begin{equation}
    \bm{M} = 
    \begin{pmatrix}
        \bm{1}_{N_{1} \times N_{1}} & \beta \bm{1}_{N_{1} \times N_{2}} \\
        \beta \bm{1}_{N_{2} \times N_{1}} & \beta^{2} \bm{1}_{N_{2} \times N_{2}}
    \end{pmatrix}
    \label{eq:M}
\end{equation}
where, again $N_{1}$ is the number of observations of image 1 and $N_{2}$ is the number of observations of image 2. The second factor, $\bm{K}$, depends on $\Delta$ and the kernel parameters and is defined as:
\begin{equation}
    \bm{K} = \bm{k}(\bm{\hat{T}}_{\Delta}, \bm{\hat{T}}_{\Delta}) = 
    \begin{pmatrix}
        \bm{k}(\bm{\hat{t}}_{1}, \bm{\hat{t}}_{1}) & \bm{k}(\bm{\hat{t}}_{1}, \bm{\hat{t}}_{2} - \Delta) \\
        \bm{k}(\bm{\hat{t}}_{2} - \Delta, \bm{\hat{t}}_{1}) & \bm{k}(\bm{\hat{t}}_{2}, \bm{\hat{t}}_{2})
    \end{pmatrix}
    \label{eq:K-oneband}
\end{equation}
where $\bm{k}(\bm{\hat{t}}, \bm{\hat{t}}')$ is the matrix whose $(i,j)^{\text{th}}$ element is $k(\hat{t}_{i}, \hat{t}_{j}')$. Finally, the diagonal matrix $\bm{W}$ contains the variances of the measurement errors, $\bm{\hat{\sigma}}^{2}$, along the diagonal.

With a specified prior on $\theta$ and Equation \ref{eq:marginal-likelihood}, we can construct the posterior:
\begin{equation}
    P(\bm{\theta} | \ \bm{\hat{F}}, \bm{\hat{T}}) \propto P(\bm{\hat{F}} | \  \bm{\hat{T}}, \bm{\theta}) P(\bm{\theta}| \ \bm{\hat{T}})
    \label{eq:posterior}
\end{equation}
which we can sample from and marginalize over to determine the probability distributions over the parameters. The dependence of $P(\bm{\theta}| \ \bm{\hat{T}})$ on $\bm{\hat{T}}$ arises because we will restrict the time delay, $\Delta$, to be within the range of observations.

\subsection{Two Images in Two Bands}
We will now consider the case with two images in two bands -- band A and band B. We now have two functions we wish to model as a draw from a GP:
\begin{equation}
    f_{A}(t) \sim \mathcal{GP}(0, k(t,t'))
\end{equation}
and
\begin{equation}
    f_{B}(t) \sim \mathcal{GP}(0, k(t,t'))
\end{equation}
where the draws from the GP for each band are independent. While we choose to neglect correlations between the underlying light curves in different bands, these correlations can in principle be included, for example as in \cite{Hu_2020}. Alternatively, the model could be used to fit color curves, which would encode correlations between bands, instead of the light curves in individual bands. Indeed, \cite{Pierel_2021} found that fitting color curves with \texttt{sntd} gives better performance for time-delay estimation in simulations.

The light curves of the images are therefore defined by:
\begin{equation}
    \begin{split}
        f_{1, \text{A}}(t) &= f_{\text{A}}(t)\\
        f_{1, \text{B}}(t) &= f_{\text{B}}(t)\\
        f_{2, \text{A}}(t) &= \beta f_{\text{A}}(t - \Delta)\\
        f_{2, \text{B}}(t) &= \beta f_{\text{B}}(t - \Delta)
    \end{split}
\end{equation}
We note that the magnification and time delay of the second image relative to the first is the same in band A as it is in band B. Therefore, there is still only one $\beta$ and one $\Delta$ for which to fit, so it remains that $\bm{\theta} = (A, \tau, \Delta, \beta)$. It is possible to adopt unique GP kernel hyperparameters for each band, but we find that these extra parameters are unnecessary for the examples that follow in this paper. 

We construct the flux data vector, $\bm{\hat{F}}$, as:
\begin{equation}
    \bm{\hat{F}} = ( \bm{\hat{f}}_{1, \text{A}}, \bm{\hat{f}}_{2, \text{A}}, \bm{\hat{f}}_{1, \text{B}}, \bm{\hat{f}}_{2, \text{B}} )^{\top}
\end{equation}
and the time data vector, $\bm{\hat{T}}$, as:
\begin{equation}
    \bm{\hat{T}} = ( \bm{\hat{t}}_{1, \text{A}}, \bm{\hat{t}}_{2, \text{A}}, \bm{\hat{t}}_{1, \text{B}}, \bm{\hat{t}}_{2, \text{B}} )^{\top}
\end{equation}
where $\bm{\hat{f}}_{m, b}$ are the observations of image $m$ in band $b$ at times $\bm{\hat{t}}_{m, b}$. We also define the de-shifted time vector, $\bm{\hat{T}_{\Delta}}$, as:
\begin{equation}
    \bm{\hat{T}}_{\Delta} = ( \bm{\hat{t}}_{1, \text{A}}, \bm{\hat{t}}_{2, \text{A}} - \Delta, \bm{\hat{t}}_{1, \text{B}}, \bm{\hat{t}}_{2, \text{B}} - \Delta )^{\top}.
\end{equation}

The marginal likelihood is:
\begin{equation}
    P(\bm{\hat{F}} | \ \bm{\hat{T}}, \bm{\theta}) = \mathcal{N}(\bm{\hat{F}} | \  \bm{0}, \bm{\Sigma_{\theta}})
\end{equation}
where, again, $\bm{\Sigma_{\theta}} = (\bm{M} \odot \bm{K}) + \bm{W}$.

We make the simplifying assumption that there is no shared information between bands beyond the time delay and magnification. In other words, there is no covariance of either image in band A with either image in band B. With this in mind, we write $\bm{K}$ as:
\begin{equation}
    \bm{K} = 
    \begin{pmatrix}
        \bm{K}_{\text{A}} & \bm{0} \\
        \bm{0} & \bm{K}_{\text{B}} \\
    \end{pmatrix}
\end{equation}
where $\bm{0}$ denotes a matrix of all zeros and $\bm{K}_{\text{A}}$ and $\bm{K}_{\text{B}}$ are defined as in Equation \ref{eq:K-oneband} for each band separately. We point out that there does not need to be the same number of observations of the images in band A as there are in band B. Therefore, the shape of $\bm{0}$ need not necessarily be square.

Finally, the posterior is the same in this case as in Equation \ref{eq:posterior} with the new marginal likelihood terms defined for this data case. The case of two images in two bands can be generalized to fit multiple ($>2$) images in multiple ($>2$) bands. We fit simulated multi-band data as expected from Rubin-LSST and Roman in \S\ref{sec:example} and multi-image data from SN Refsdal in \S\ref{sec:refsdal}.

%% file: 03_microlensing.tex
The effect of microlensing from stars and substructure in the lensing galaxy or galaxy cluster can additionally introduce additional time-varying magnification to individual images. Microlensing in systems of glSNe is extremely difficult to model because of complex layouts of stars and dark substructure in the foreground of lensing systems. The discrepancy between the observed and the model-predicted brightnesses of the four images from SN Zwicky without considerations of microlensing and the observed brightnesses of the images -- up to 1.5 mag for one image -- demonstrates the difficulty of modeling this effect and the significant impact which microlensing/substructure can have on a system \citep{Pierel_2022, Goobar_2023}.

Following \cite{Tak_2017}, we model microlensing as a time-dependent extension of $\beta$, so the previously constant $\beta$ becomes $\beta(t)$. Because we can only learn relative magnification effects from the light curve data available, we take all magnification to be relative to an arbitrarily chosen image 1. Therefore, Equations \ref{eq:image-fluxes-1}-\ref{eq:image-fluxes-2} become:
\begin{align}
    \label{eq:image-fluxes-ml-1}
    \hat{f}_{1, i} &= \hat{f}_{1}(t_{i}) = f(t_{i}) + \epsilon_{1, i} \\
    \epsilon_{1, i} &\sim \mathcal{N}(0, \sigma_{1, i}^{2}) \\
    \hat{f}_{2, j} &= \hat{f}_{2}(t_{j}) = \beta(t_{j}) f(t_{j} - \Delta) + \epsilon_{2, j} \\
    \label{eq:image-fluxes-ml-2}
    \epsilon_{2, j} &\sim \mathcal{N}(0, \sigma_{2, j}^{2})
\end{align}
where again $\sigma_{1, i}$ and $\sigma_{2, j}$ are the standard deviations of the measurement errors of image 1 and image 2, respectively.

Figure 3 in \cite{FoxleyMarrable_2018} and Figure 3 in \cite{Pierel_2021} illustrate from microlensing maps examples of microlensing as a function of time. In these example microlensing curves, a period of constant magnification is typically interrupted by one change in brightness which occurs over a range of timescales, from less than a day to weeks. Physically, this shape corresponds to the SN crossing a microcaustic as the photosphere expands. After this change, the observed magnification appears to remain constant, as the small size of the SN and microlensing sub-structures means the photosphere most often crosses only one significant microcaustic. Therefore, we have chosen to model microlensing as a sigmoid function, given by:
\begin{equation}
    \beta(t) = \beta_{0} + \frac{\beta_{1}}{1+e^{-r (t-t_{0})}}
\end{equation}
where $\beta_{0}$ is the macrolensing effect, $\beta_{1}$ is the scale of the microlensing effect, $r$ is the rate of change in the microlensing effect, and $t_{0}$ is the location of a change in microlensing.

We make the simplifying assumption of achromatic microlensing, or microlensing which is identical across bands for a given image. It is demonstrated by \cite{Goldstein_2018, Huber_2021} that this assumption is valid for the first three rest-frame weeks of lensed SN~Ia, in which time microlensing is predicted to be largely achromatic. This model for the time delay and magnification therefore takes 5 parameters per image. The parameters we fit for in a doubly-imaged system are now $\bm{\theta} = (A, \tau, \Delta, \beta_{0}, \beta_{1}, r, t_{0})$.

Consider again the case of two images observed in one band, where image 1 is observed at a set of $N_{1}$ times and image 2 is observed at a set of $N_{2}$ times. For this system, the matrix $\bm{M}$ from Equation \ref{eq:M} is now:
\begin{equation}
    \bm{M} =
    \begin{pmatrix}
        \bm{1} \, \bm{1}^{\top} & \bm{1} \, \bm{\beta}_{2}^{\top} \\
        \bm{\beta}_{2} \, \bm{1}^{\top} & \bm{\beta}_{2} \, \bm{\beta}_{2}^{\top} \\
    \end{pmatrix}
\end{equation}
where $\bm{1}$ is a vector of ones of length $N_{1}$ and $\bm{\beta}_{2}$ is a vector of length $N_{2}$ whose $j^{\text{th}}$ element is given by $\beta_{2, j} = \beta(\hat{t}_{2,j} - \Delta)$. This formulation generalizes to an arbitrary number of images observed in an arbitrary number of bands. We will refer to the model with a constant $\beta$ as the ``constant magnification'' model and the model with a time-varying magnification term as the ``sigmoid magnification'' model.

\begin{figure}
    \centering
    \includegraphics[width=\linewidth]{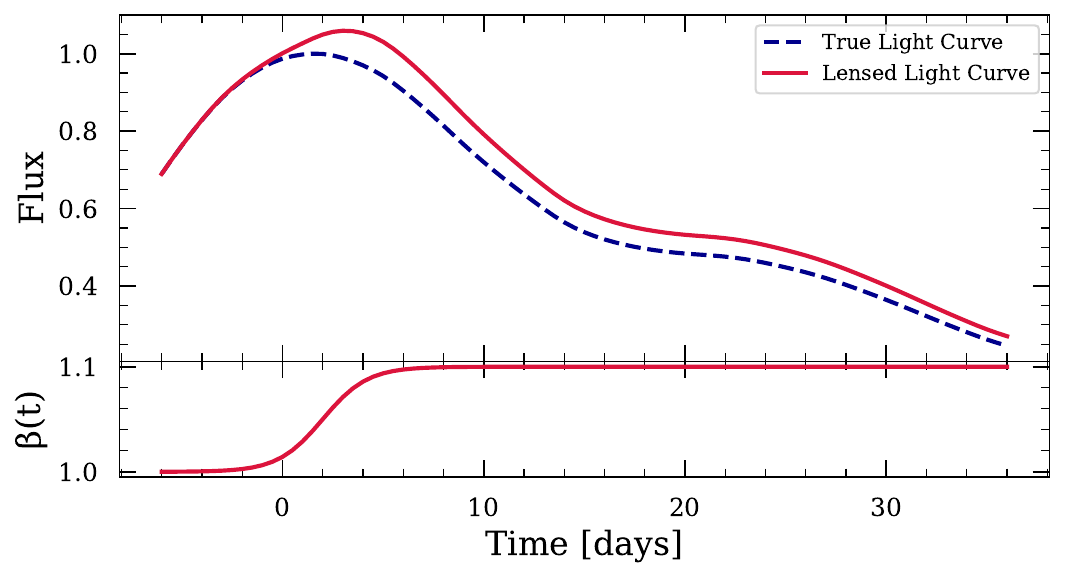}
    \caption{The effect that a 10\% increase in brightness due to a time-varying magnification term over 5 days can have on the shape of a SN light curve. The upper panel shows the $r$-band flux of an unlensed SN~Ia light curve from the \citet{Hsiao_2007} template (blue, dashed line) and a microlensed version of the same underlying light curve (red, solid line). We emphasize that the true time delay is $\Delta = 0$, although the peak in the microlensed light curve appears to be later because of the time-varying magnification term. The bottom panel shows the time-varying relative magnification affecting the second image, $\beta(t)$.}
    \label{fig:microlensing-example}
\end{figure}

Figure \ref{fig:microlensing-example} demonstrates how a time-varying magnification term can have a significant impact on the shape of a SN light curve, to the point that the inferred time delay could be significantly skewed if the glSN system is fit with a constant magnification model. A combined treatment of macrolensing and microlensing mitigates the potential source of bias that arises if these magnification effects are considered separately. We explore this effect in greater detail in Appendix \ref{sec:appendixA}.

We emphasize that the choice of a single parameterization to describe the microlensing may introduce a source of systematic uncertainty, which should be carefully considered in a cosmological analysis. In \S\ref{sec:discussion}, we discuss future work which will mitigate systematics that may arise from a choice of parameterization of microlensing. In addition, we test a second parameterization of microlensing -- a flexible sinusoidal function -- on a subset of the data used in the main analysis in Appendix \ref{sec:appendixB}.

%% file: 04_sampling.tex
\subsection{Sampling Algorithms}
Within the publicly available \textsc{GausSN} framework, we implement three methods for sampling from the posterior distributions:
\begin{itemize}
    \item \texttt{emcee}: affine invariant MCMC ensemble sampler \citep{ForemanMackey_2013, Goodman_2010}

    \item \texttt{dynesty}: nested sampling \citep{Speagle_2020, Skilling_2004, Skilling_2006}

    \item \texttt{zeus}: MCMC ensemble slice sampler \citep{karamanis2020ensemble, karamanis2021zeus}
\end{itemize}
We choose to sample parameters with \texttt{dynesty} for the analyses in this paper. The nested sampling approach is best able to handle the multi-modal nature of the posteriors on the time delay and other magnification parameters. Therefore, for the most accurate time-delay estimates and associated uncertainties, we opt for \texttt{dynesty}.

Within \texttt{dynesty}, we use both uniform sampling with multi-ellipsoidal bounding \citep{Feroz_2009}, for systems with fewer than 100 data points and fewer than 5 parameters, and random slice sampling, for systems with more than 100 data points or more than 5 parameters, with 500 live points. The slice sampling technique was developed by \cite{Neal_2003} and first implemented within the nested sampling framework by \cite{Handley_2015a, Handley_2015b}. We retain the default stopping criteria in \texttt{dynesty}, which is met when the remaining, unaccounted for evidence is less than a threshold which depends on the number of live points used \citep{Speagle_2020}. We note that the specifications of \texttt{dynesty} sampling, and all other sampling methods, can be easily adjusted within the \textsc{GausSN} parameter optimization function.

Given these specifications, \textsc{GausSN} performs as follows. For an object with 320 total data points, i.e. 40 observations of 2 images in 4 bands, a single evaluation of the marginal likelihood takes $\sim2$ ms using standard CPU resources on a desktop computer. The stopping criteria is typically met after 200,000-800,000 evaluations of the likelihood function, which roughly corresponds to 5,000-15,000 nested sampling iterations. Therefore, the nested sampling algorithm takes between 5 and 30 minutes to run for such an object. Depending on the quantity and quality of the data, sampling may take anywhere from 30 s to an hour for the simulated Rubin-LSST and Roman data. Given the rarity of glSNe, even in the best case scenarios for future observatories, such run times will not be a barrier for future applications of our model to real data.

\subsection{Plotting the Fits}
\label{sec:plotting}
When considering example glSNe data in \S\ref{sec:example} and \S\ref{sec:refsdal}, we will plot draws of the fitted light curves to demonstrate the quality of the \textsc{GausSN} fits. Here, we describe the procedure for creating the fitted curves for a system with $M$ images. First, we take a random draw of the time delay, $\Delta$, and relative magnification, $\beta$, from the joint posterior and de-shift and un-magnify the data from the $(M-1)$ trailing images according to the drawn parameters. Setting the mean and covariance of a multivariate normal distribution to the posterior predictive mean and covariance conditioned on the now overlapping data from all images, we take a realization of the fit. We plot a copy of this realization for each of the $M$ images, where each copy of the realization is time-shifted and magnified back to the frame of the observed data, so that it overlaps with the raw light curve data of each image. We repeat this process for 200 samples of the joint posterior to get a representative visualization of the fitted light curve and uncertainty for each image. Therefore, the final plot shows the raw data for all $M$ images with 200 realizations of the fitted light curve per image.

The procedure follows the same steps when considering a time-varying magnification term. In the first step, the data from each image will be un-magnified according to the drawn time-varying magnification function. When the realization of the light curve is being plotted, it is then put back into the observed data frame by magnifying according to the time-varying magnification term to match the original data. In this case, we will additionally plot realization of the magnification function for each posterior sample in a separate panel of the figure. The realizations of the magnification function alongside the data and fitted light curves allows for physical interpretations of the magnification term.

%% file: 05_0_tests.tex
\subsection{Roman Simulations}
\input{05_1_roman}

\subsection{Rubin-LSST Simulations}
\input{05_2_rubin}

%% file: 05_1_roman.tex
\subsubsection{Data}
\label{sec:roman-data}
Recently, \cite{Pierel_2021} (hereafter \citetalias{Pierel_2021}) simulated 2.4 million glSN light curves -- 600,000 each of Type Ia, Ib/c, IIn, and IIP -- as expected from the Roman Space Telescope.\footnote{Publicly available at: \url{https://dx.doi.org/10.17909/t9-k8w7-zk32}.} The cadence, depth, and detection thresholds for the simulations are based on the Roman SN survey "All-z" strategy described in \cite{Hounsell_2018}, with modifications made for more recent survey updates. Based on the current plans for the instrument, \citetalias{Pierel_2021} predicts Roman will discover glSNe up to $z=4$.

\begin{figure*}
    \centering
    \includegraphics[width=\linewidth]{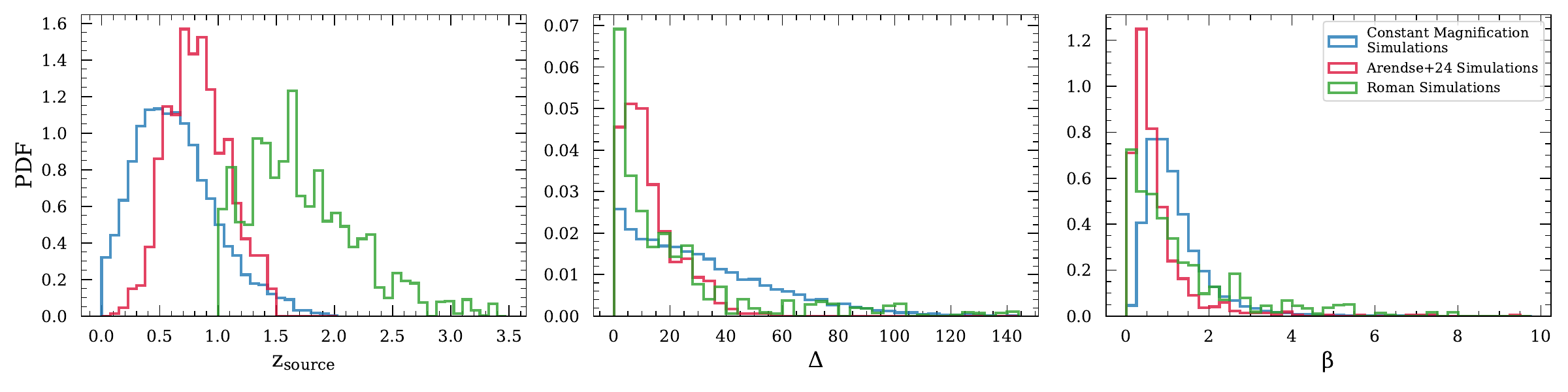}
    \vspace{-10pt}
    \caption{The distributions of $z_{\text{source}}$, $\Delta$, and $\beta$ for the the three simulated glSNe samples.}
    \label{fig:sim-dists}
\end{figure*}

The simulation pipeline works as follows: for each SN type, a sample of 50 simulated galaxy-scale lenses are used to simulate 10 distinct glSN light curves with 2-4 images based on the structure of the lens. Each of these 500 systems are then subject to 100 iterations of microlensing for each of 12 microlensing maps, yielding 1200 variations of each glSN light curve. \citetalias{Pierel_2021} assumes achromatic microlensing, so any individual image experiences the same microlensing effects across wavelength space. The 12 microlensing maps are based on different choices of stellar mass model, which vary the effective radius, initial mass function, and Sérsic index of the galaxy profile. Objects are required to pass a series of data cuts, which are described in \citetalias{Pierel_2021}. For simplicity, we consider only the doubly-imaged glSNe from the \citetalias{Pierel_2021} Roman simulations. The distributions of parameters for the \citetalias{Pierel_2021} Roman simulations are shown in Figure \ref{fig:sim-dists}.

We fit each object with both the constant magnification model and the sigmoid magnification model. For the constant magnification model, we use the following priors on the two kernel parameters, the time delay, and the magnification\footnote{We use $\mathcal{U}(a, b)$ to denote a uniform distribution over ($a$, $b$) and $\mathcal{TN}(\mu, \sigma^{2}, c, d)$ to denote a truncated normal distribution where $\mu$ and $\sigma$ are the mean and variance of an untruncated normal distribution, which is then truncated on the left at location $c$ and on the right at location $d$.}:
\begin{align}
    A &\sim \mathcal{U}(0, 5) \\
    \tau &\sim \mathcal{U}(10, 40), \\
    \Delta | \, \bm{\hat{T}} &\sim \mathcal{TN}(\mu_{\Delta}, \, 50^{2}, \, \min(\bm{\hat{t}}_{1}) - \max(\bm{\hat{t}}_{2}), \notag \\
    &\hspace{4cm}\max(\bm{\hat{t}}_{1}) - \min(\bm{\hat{t}}_{2})) \\
    \beta_{0} &\sim \mathcal{TN}(\mu_{\beta}, 10^{2}, 0, \infty)
    \label{eq:roman-priors-1}
\end{align}
where $\mu_{\Delta}$ is the difference between the times of the brightest observations of image 2 and 1 and $\mu_{\beta}$ is the ratio of the brightest observation of image 2 relative to the brightest observation of image 1. We define $\bm{\hat{t}}_{1}$ and $\bm{\hat{t}}_{2}$ to be all the times of observation of image 1 and 2, respectively. The prior on $\Delta$ therefore requires that there always be overlap between data from image 1 and data from image 2. Although the scaled flux data has a maximum value of 1, we allow the prior on $A$ to range from 0 to 5 so as not to overly restrict the amplitude of the underlying light curve.

We adopt wide priors on all parameters to ensure the diversity of behavior present in the simulations is represented in our parameter space. With real glSN events, which will be fit on an individual basis because of their rarity, these priors can be adjusted to account for additional contextual information for each specific event.

For the sigmoid model, we set the following priors on the three additional magnification parameters:
\begin{gather}
    \beta_{1} \sim \mathcal{N}(0, 0.5^{2}) \\
    r \sim \mathcal{N}(0, 0.5^{2}) \\
    t_{0}|\, \bm{\hat{T}}, \Delta \sim \mathcal{U}(\min(\bm{\hat{T}}_{\Delta}), \max(\bm{\hat{T}}_{\Delta}))
\end{gather}
with $\bm{\hat{T}}$ as defined in Equation \ref{eq:time-vector} and $\bm{\hat{T}}_{\Delta}$ as defined in Equation \ref{eq:deshifted-time-vector}.

We compare the results from the two \textsc{GausSN} models to those from \texttt{sntd} presented in \citetalias{Pierel_2021}. While the results from \citetalias{Pierel_2021} provide a helpful benchmark, these fits assume the correct redshift and the correct SN template, which may not always be well-known for real glSNe. However, the Roman objects were fit by \citetalias{Pierel_2021} using \texttt{sntd}'s ``parallel'' method, which fits each band separately, because it is less computationally expensive. The ``series'' method, which fits all bands simultaneously, and the ``color'' method, which fits color curves simultaneously with the lensing parameters, have been shown to be more robust. For this reason, these two methods are the only two that have been used in the analyses of real glSNe events with \texttt{sntd} \citep{Kelly_2023_ApJ, Pierel_2022}. Therefore, the \texttt{sntd} results may be improved by using a different fitting method.

\subsubsection{Analysis}

\begin{figure}
    \centering
    \includegraphics[width=\linewidth]{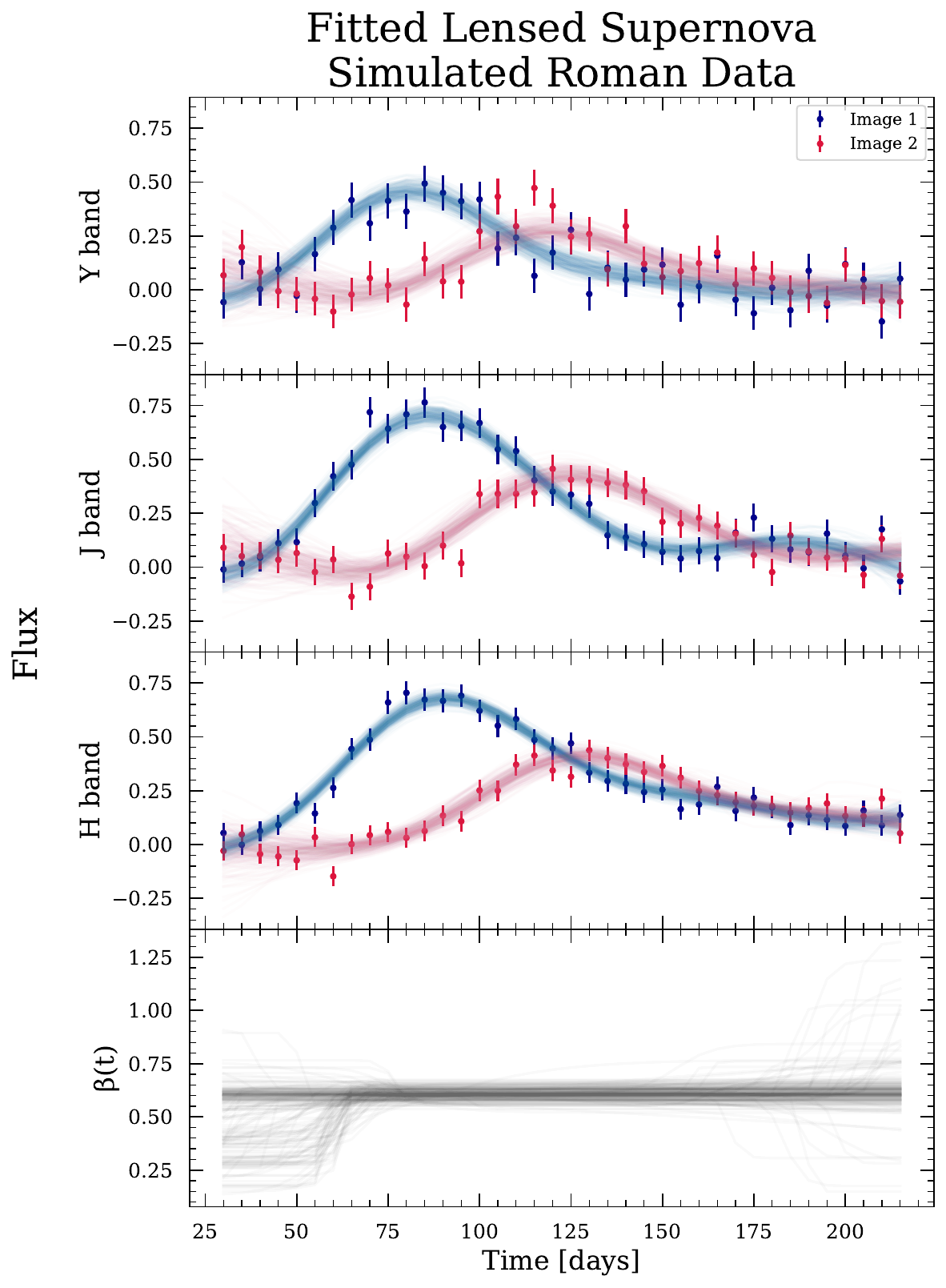}
    \caption{An example Roman-like simulated glSN system fitted with the sigmoid magnification model. The first image is shown in blue and the second in red. This object is at $z = 1.93$ and has time delay, $\Delta = 38.53$ days and relative magnification, $\beta = 0.24$. At this redshift, the Roman $Y$-, $J$-, and $H$-bands roughly cover $3000 - 6000$ \AA\, in the rest frame, or the $u$-, $g$-, and $r$-band wavelength regimes. The top three panels show the fitted flux data and GP fit and the bottom panel of the figure shows the fitted sigmoid magnification, $\beta(t)$, plotted according to the procedure described in \S\ref{sec:plotting}. The red fitted curves are time-shifted and magnified copies of the blue fitted curves, which have been conditioned on the data from both images.}
    \label{fig:Roman-after}
\end{figure}

On an individual object basis, the GP fits to the data and posterior distributions show that \textsc{GausSN} is effectively and accurately inferring the time delays from the glSNe systems. In Figure \ref{fig:Roman-after}, we demonstrate the quality of the fit on the light curve level from the sigmoid magnification model. We plot the observed data with draws from the posterior predictive distribution. The bottom panel of the figure shows realizations of the magnification function for each posterior sample. Notably, because we are only able to constrain the relative magnification, the magnification function shows greater uncertainty around the beginning and end of the time series because these regions are where the light curve lacks overlapping data from the two images.

\begin{figure}
    \centering
    \includegraphics[width=\linewidth]{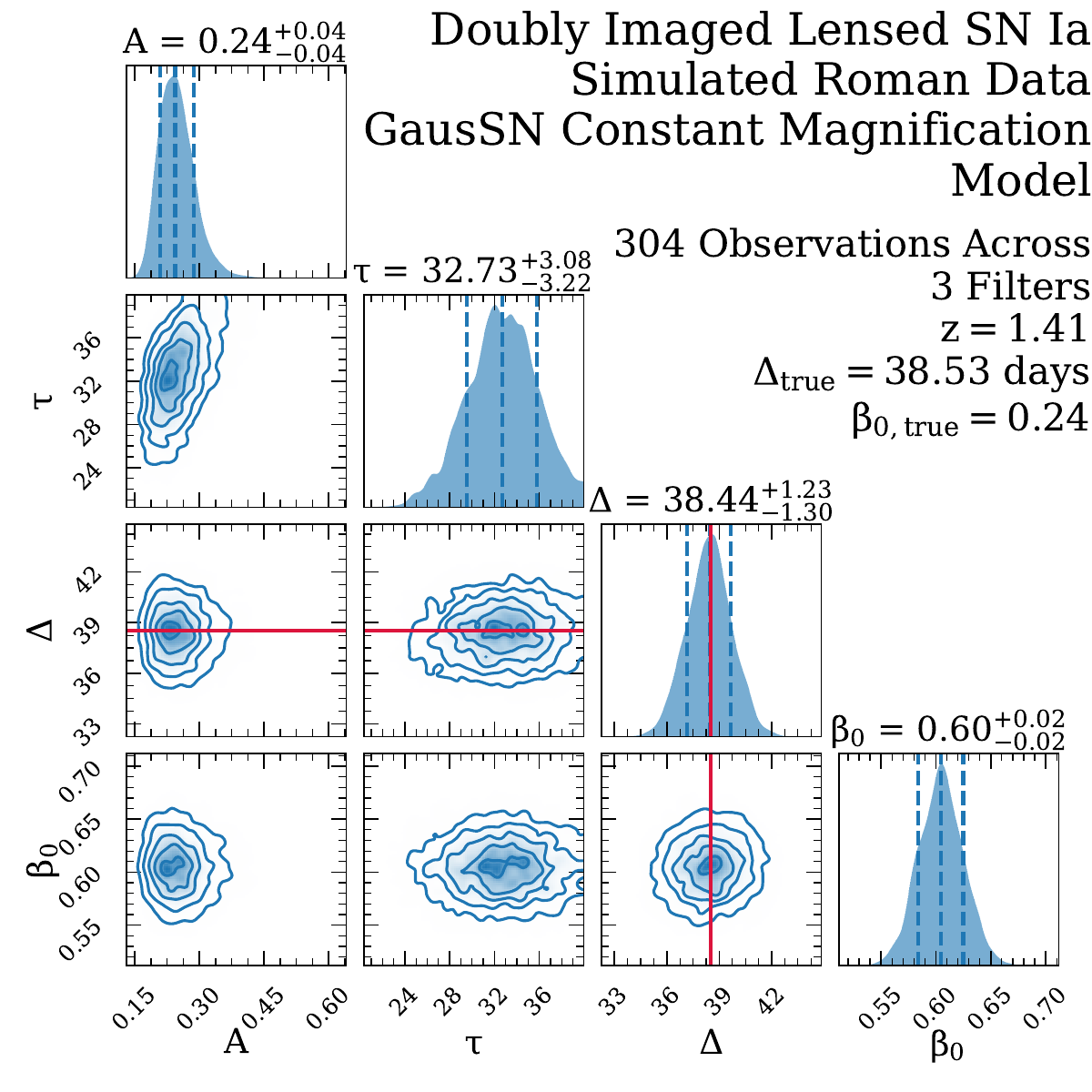}
    \caption{The corner plot for the constant magnification fit to the glSN data shown in Figure \ref{fig:Roman-after}. The constant magnification model has four parameters: the kernel amplitude, $A$, the kernel length scale, $\tau$, the time delay, $\Delta$, and the relative magnification, $\beta$. The three dashed lines in each of the 1D posterior distribution plots shows the 16th, 50th, and 84th percentiles from left to right, respectively, and the red line shows the true time delay and relative magnification. The time delay is well recovered and to 3.38\% precision. The magnification is not as well recovered, potentially indicating additional magnification effects from microlensing that are not captured by the constant magnification model.}
    \label{fig:Roman-corner}
\end{figure}

In Figure \ref{fig:Roman-corner}, we show the posteriors of the constant magnification fit. \textsc{GausSN} recovers $\Delta = 38.44^{+1.23}_{-1.30}$ days, whcih captures the true time delay of 38.53 days well within the 68\% credible interval (CI). As we don't necessarily expect the posteriors to be Gaussian, we calculate the 68\% CI from the $16^{\text{th}}$ and $84^{\text{th}}$ percentiles of the posterior samples. The magnification is not well recovered, as \textsc{GausSN} finds $\beta = 0.60 \pm 0.02$, when the true $\beta = 0.24$. We attribute this discrepancy to the presence of microlensing from lens substructure. Indeed, there is more uncertainty in the recovered magnification when fitting with the sigmoid magnification model. With this model, \textsc{GausSN} finds $\beta_{0} = 0.59^{+0.05}_{-0.30}$, which is more consistent with the truth. The uncertainty on the time delay is however increased when fitting with the sigmoid magnification model, as expected, with $\Delta = 37.72^{+1.76}_{-1.97}$. This represents a 5.22\% precision time-delay estimate with the sigmoid magnification model, up from 3.38\% precision with the constant magnification model. Of course, the constant magnification model gives only a statistical uncertainty, so it does not take into consideration systematic uncertainty from microlensing. A separate systematic microlensing error would have to be accounted for in post-processing. Therefore, the time delay estimate with the sigmoid magnification model may, in the end, be more precise.

\begin{figure}
    \centering
    \includegraphics[width=\linewidth]{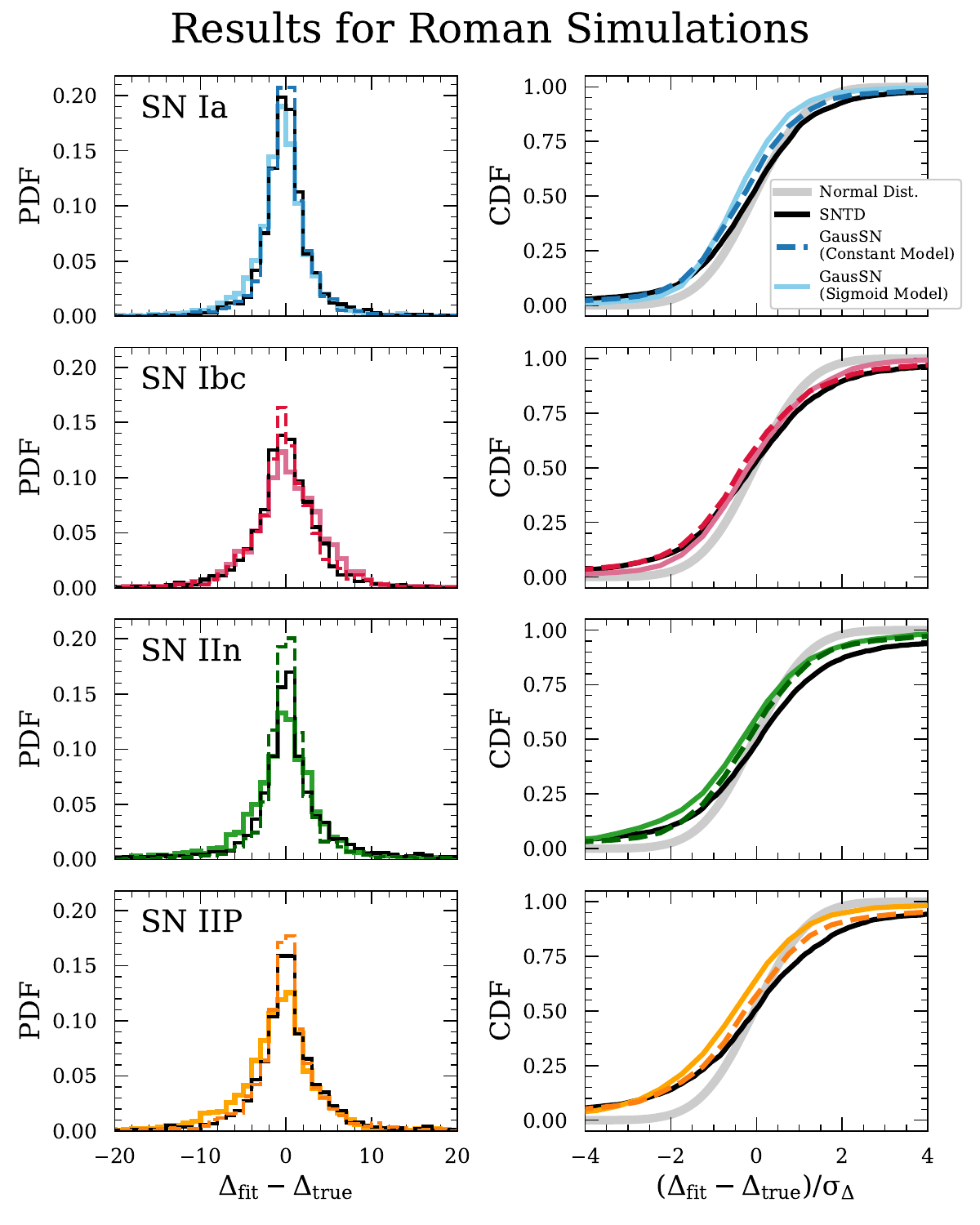}
    \caption{\textbf{Left:} The difference between the measured time delay and the true time delay for the Roman simulations by type of SN. \textbf{Right:} The CDF of the standardised error, $(\Delta_{\text{fit}} - \Delta_{\text{true}}) / \sigma_{\Delta}$, distribution by type of SN. If the error bars are well calibrated, we expect this distribution to resemble that of a Gaussian CDF, shown by the gray line. The solid colored line shows the results from the \textsc{GausSN} sigmoid magnification model, the dashed colored line shows that of the \textsc{GausSN} constant magnification model, and the solid black line shows the results from \texttt{sntd}.}
    \label{fig:Roman-combined-pdf-cdf}
\end{figure}

On a population level, \textsc{GausSN} performs well across all types of SNe and mass models. We note that the true parameters for the SN~IIP SC2 and SC4 mass models were not available, so these results are excluded from the reported statistics. The left column of Figure \ref{fig:Roman-combined-pdf-cdf} shows the distribution of $\Delta_{\text{fit}} - \Delta_{\text{true}}$ for the sigmoid and constant magnification models. We also compare to the results from \texttt{sntd}, which are included in the \citetalias{Pierel_2021} simulated Roman data release. With the constant magnification model we find 35.32\% of time delays within $\pm 1$ day, 70.33\% within $\pm 3$ days, and 83.73\% within $\pm 5$ days. With the sigmoid magnification model, we find that 27.13\% of time delays are recovered within $\pm 1$ day, 62.12\% are recovered within $\pm 3$ days and 79.07\% are recovered within $\pm 5$ days. For comparison, \texttt{sntd} recovers 31.95\% within $\pm 1$ day, 66.25\% within $\pm 3$ days, and 81.28\% within $\pm 5$ days.

We also find that the distribution of $\Delta_{\text{fit}} - \Delta_{\text{true}}$ is generally unbiased. The median of the distribution is -0.168 days for the sigmoid magnification model and -0.249 days for the constant magnification model. The diminished bias of the sigmoid magnification model relative to the constant model reiterates the need to account for microlensing alongside time-delay estimation. Future work on improved methods for marginalizing over a choice of microlensing parameterization to further reduce biases are discussed in \S\ref{sec:discussion}.

We also consider the fractional error, $|\Delta_{\text{fit}} - \Delta_{\text{true}}| / \Delta_{\text{true}}$, on the time-delay estimates. This metric gives a better sense of closeness to the truth scaled by the size of the time delay. For the constant magnification model, we find that 32.55\% of systems have fractional errors of less than 5\% and for the sigmoid magnification model, 26.48\% of systems have fractional errors of less than 5\%. These results are comparable to the \texttt{sntd} results, which find 30.84\% of systems to have a fractional error of less than 5\%.

\textsc{GausSN}, therefore, is effective at estimating time delays close in absolute and fractional value to the true time delay. That the sigmoid magnification model is in general further from the truth according to this metric than the constant magnification model or \texttt{sntd} is not unexpected. As this model has more flexibility, often leading to more complex posteriors, the point estimate for the time delay may be further from the truth. The full posteriors give a more accurate sense of the time-delay estimate by a given model.

In the right column of Figure \ref{fig:Roman-combined-pdf-cdf}, we show the CDF of the standardised error, $(\Delta_{\text{fit}} - \Delta_{\text{true}}) / \sigma_{\Delta}$ distribution for the sigmoid and constant \textsc{GausSN} magnification models, \texttt{sntd}, and a unit normal distribution for reference. If the uncertainties are well-calibrated, we expect that across the sample of SNe, 68\% of $\Delta_{\text{true}}$ will fall within the 68\% CI and 95\% of $\Delta_{\text{true}}$ will fall within the 95\% CI. Although the full posterior distributions from \texttt{sntd} are not available, we can compare these statistics to the fraction of SNe which have time delays recovered within $1\sigma$ and $2\sigma$ of the truth from the \texttt{sntd} point estimate and uncertainty.

For the constant magnification model, we find 54.46\% in 68\% CI and 81.31\% in 95\% CI. For the sigmoid magnification model, we find 54.65\% of SNe with $\Delta_{\text{true}}$ in 68\% CI and 82.65\% in 95\% CI. For \texttt{sntd}, 50.92\% of SNe have $\Delta_{\text{true}}$ which fall in the $1\sigma$ CI and 78.33\% fall in the $2\sigma$ CI. While these statistics do not meet the benchmark set by the normal distribution, they are comparable to, if not a slight improvement on, the results from \texttt{sntd}. Therefore, \textsc{GausSN} provides time-delay estimates that are as close, if not closer, in absolute value to the truth with relatively well-calibrated uncertainties compared to \texttt{sntd}, the leading time-delay estimation technique for glSNe. As there are some variations for different types of SNe, we report the above statistics broken down by type of SN for all three models in Table \ref{tab:Roman-results}.

\begin{table*}
    \centering
    \caption{The results of \textsc{GausSN} fits to the Roman simulations by type of SN and model used. We compare the \textsc{GausSN} results to those of \texttt{sntd}, a template-based approach. While \textsc{GausSN} is agnostic to SN type and redshift, \texttt{sntd} has assumed the correct template of the correct SN type and redshift. Note that the results from \texttt{sntd} were not available for two mass models (SC4 and SC7) of SNe~IIP. However, there is little variation in the results when further broken down by mass model, so we would not anticipate significant changes in the SNe~IIP results if these mass models were included.}
    \begin{threeparttable}
        \begin{tabular}{c|c|c|c|c|c|c|c|c|c}
            & & & $|\Delta_{\text{fit}} - \Delta_{\text{true}}|$ & $|\Delta_{\text{fit}} - \Delta_{\text{true}}|$ & $|\Delta_{\text{fit}} - \Delta_{\text{true}}|$ & Median & $|\Delta_{\text{fit}} - \Delta_{\text{true}}|/\Delta_{\text{true}}$ & $\Delta_{\text{true}}$ & $\Delta_{\text{true}}$ \\
            SN Type & N\tnote{a} & Model & < 1 day & < 3 days & < 5 days & $\Delta_{\text{fit}} - \Delta_{\text{true}}$ & < 5\% & in 68\% CI\tnote{b} & in 95\% CI\tnote{b} \\
            \hline
            \hline
            Ia &  &  &  &  &  &  &  & \\
             & 2340 & Sigmoid & 0.358 & 0.741 & 0.873 & -0.331 & 0.368 & 0.636 & 0.912 \\
             & 2340 & Constant & 0.414 & 0.788 & 0.901 & -0.075 & 0.436 & 0.608 & 0.880 \\
             & 2340 & SNTD & 0.381 & 0.762 & 0.884 & -0.113 & 0.437 & 0.588 & 0.853 \\
             & 599901 & SNTD & 0.386 & 0.764 & 0.883 & -0.001 & 0.421 & 0.568 & 0.842 \\
            Ibc &  &  &  &  &  &  &  & \\
             & 2243 & Sigmoid & 0.226 & 0.572 & 0.774 & 0.067 & 0.229 & 0.558 & 0.849 \\
             & 2243 & Constant & 0.278 & 0.636 & 0.802 & -0.144 & 0.263 & 0.529 & 0.786 \\
             & 2243 & SNTD & 0.274 & 0.638 & 0.820 & -0.158 & 0.261 & 0.504 & 0.789 \\
             & 599820 & SNTD & 0.284 & 0.637 & 0.811 & -0.021 & 0.269 & 0.498 & 0.779 \\
            IIn &  &  &  &  &  &  &  & \\
             & 2517 & Sigmoid & 0.257 & 0.589 & 0.763 & -0.201 & 0.250 & 0.502 & 0.767 \\
             & 2517 & Constant & 0.391 & 0.729 & 0.832 & 0.075 & 0.339 & 0.543 & 0.829 \\
             & 2517 & SNTD & 0.321 & 0.629 & 0.764 & 0.165 & 0.287 & 0.501 & 0.766 \\
             & 600000 & SNTD & 0.325 & 0.632 & 0.762 & 0.052 & 0.279 & 0.492 & 0.751 \\
            IIP &  &  &  &  &  &  &  & \\
             & 1609 & Sigmoid & 0.244 & 0.582 & 0.753 & -0.529 & 0.212 & 0.49 & 0.778 \\
             & 1609 & Constant & 0.329 & 0.660 & 0.814 & -0.050 & 0.264 & 0.498 & 0.757 \\
             & 1609 & SNTD & 0.302 & 0.620 & 0.783 & 0.019 & 0.249 & 0.443 & 0.724 \\
             & 455873 & SNTD & 0.332 & 0.643 & 0.783 & 0.051 & 0.258 & 0.477 & 0.729 \\
            \hline
        \end{tabular}
        \begin{tablenotes}\footnotesize
            \item[a] The number of SNe included in total. The SNe fit by \textsc{GausSN} were selected at random from the full sample. While there are not equal numbers of each mass model fit by each model, we weight by the number from each mass model, so there is equal contribution from all mass models.
        
            \item[b] The fraction of SNe for which $\Delta_{\text{true}}$ falls within the 68\% and 95\% credible intervals (CI) computed from the $16^{\text{th}}$ and $84^{\text{th}}$ percentiles of the posterior samples. Note that because the full posteriors are not available for the SNTD fits, we use the 1 and 2 $\sigma$ intervals as a proxy for this value.
        \end{tablenotes}
    \end{threeparttable}
    \label{tab:Roman-results}
\end{table*}

\textsc{GausSN} seems to have an improved calibration of uncertainties compared to \texttt{sntd}, which does not just arise from large uncertainties from \textsc{GausSN}. The mean uncertainty on the time delay is 2.90 days from the constant magnification model -- comparable to the mean uncertainty of 3.02 days from \texttt{sntd} -- and 3.81 days from the sigmoid magnification model. Furthermore, we can compare the fraction of glSNe with time delays measured to a certain level of precision, or which have $|\sigma_{\Delta} / \Delta_{\text{fit}}|$ less than a threshold. We find that with the constant magnification model, 5.10\% of fitted Roman objects are measured to 1\% precision -- or have $|\sigma_{\Delta} / \Delta_{\text{fit}}| < 0.01$ -- and 25.50\% are measured to 5\% precision. For the sigmoid magnification model, 2.88\% and 17.87\% of objects are measured to 1\% and 5\% precision, respectively.

Finally, \texttt{sntd} measures 5.94\% of glSNe to 1\% precision and 25.85\% to 5\% precision. The median precision is 14.55\% for the constant magnification model, 22.44\% for the sigmoid magnification model, and 14.57\% for \texttt{sntd}. The uncertainties from \textsc{GausSN} are still underestimated in general, though, possibly due to the single microlensing parameterization considered. Future work to reduce systematic uncertainties from microlensing is described in \S\ref{sec:discussion}.

We note that 2.30\% and 4.81\% of glSNe have $(|\Delta_{\text{fit}} - \Delta_{\text{true}}|)/\sigma_{\Delta}$ > 5 for the sigmoid and constant magnification model, respectively. It is unsurprising that the more conservative sigmoid magnification treatment has fewer catastrophic outliers compared to the constant magnification model. However, we still exceed the rate of 5$\sigma$ outliers consistent with a normal distribution. Concerningly, many of the outliers, particularly from the constant magnification fit, have fitted light curves that appear convincing based on visual inspection. It is likely that significant effects from microlensing, which may cause a time-varying magnification that is not well described by a sigmoid function, are at play in these simulations. Testing additional parameterizations of microlensing and using model comparison metrics will be necessary in analyses of real glSNe with significant microlensing effects. That said, these outliers will remain an issue for cosmological analyses of glSNe and uncertainties due to the presence of such outliers should be propagated through further analyses.

There are some additional outliers which arise from multi-modal posteriors on the time delay which are not well-described by the mean and standard deviation of the samples. These are of less concern for two reasons. Firstly, for many objects, a more informative prior on the time delay can resolve this issue. More careful tailoring of priors based on visual inspection of the light curves and other contextual information will be possible with real glSNe, due to their rarity, though it is difficult to do on the large scales needed for this analysis. Secondly, the full posterior, rather than a point estimate, should be used in the cosmological analysis of any glSN. There are also some outliers with very low SNR for one or both images, making any constraint on the time delay difficult. These two types of outliers are also expected at the low rates we see and are not of concern because of the ease with which they can be identified as needing more careful treatment. 

Furthermore, such rates of outliers are not inconsistent with existing methods for time-delay estimation. For reference, 5.96\% of \texttt{sntd} fits to the Roman simulations have $(|\Delta_{\text{fit}} - \Delta_{\text{true}}|)/\sigma_{\Delta}$ > 5. That \textsc{GausSN} has fewer outliers is consistent with the statistics reported in Table \ref{tab:Roman-results}, which show the \textsc{GausSN} uncertainties are better calibrated. In addition, the methods presented in \cite{Kelly_2023_ApJ} show similar rates of outliers. Therefore, \textsc{GausSN} again meets, if not exceeds, the benchmarks set by existing time-delay estimation methods.

Together, these results demonstrate that \textsc{GausSN} provides competitive time-delay estimates. When comparing to the results from \texttt{sntd}, the \textsc{GausSN} constant magnification model, which has the more similar magnification treatment to \texttt{sntd}, consistently retrieves time delays that are as close, if not closer depending on the type of SN, to the true time delay. The estimates also tend to have better calibrated uncertainties, with minimal differences in the percent of time delays estimated to 1\% and 5\% precision. Furthermore, \textsc{GausSN} is agnostic to the SN type and redshift, while \texttt{sntd} has assumed the correct template of the correct SN type and redshift to achieve the reported results. Because the Roman simulations were not generated from the \textsc{GausSN} forward model, our results demonstrate the ability of \textsc{GausSN} to generalize.

Including a time-varying magnification model with \textsc{GausSN}, as in the sigmoid magnification model, leads to an increase in the uncertainties on the time delay, as expected given our uncertainty in the effect of microlensing. Unsurprisingly, the fraction of SNe with time delays recovered to 1\% and 5\% precision is lower for this model with minimal decreases also seen in the time-delay recovery within 1, 3, and 5 days. However, the uncertainties are the best calibrated out of all three models, which makes sense given this model is the most conservative approach to microlensing considered in this work. The performance of this model suggests promise for the ability to constrain a time-varying magnification function with \textsc{GausSN}. We further comment on future work possible with time-varying magnification models in \S\ref{sec:discussion}.

%% file: 05_2_rubin.tex
\subsubsection{Constant Magnification Simulations}
We apply this method to LSST-like simulations of doubly-imaged SNe. The objects are simulated using templates from \texttt{sncosmo} \citep{Barbary_2022}. Included in the sample of simulated objects are SNe~Ia, SNe~Ib, SNe~Ic, SNe~IIb, SNe~IIn, and SNe~IIP. We use the \cite{Hsiao_2007} template to simulate SNe~Ia, the \cite{Pierel_2018} templates to simulate SNe~IIP, and the \cite{Vincenzi_2019} templates to simulate SNe~Ib, SNe~Ic, SNe~IIb, and SNe~IIn. Note that some templates are excluded due to the presence of unphysical artifacts in the light curves or poor NIR fits.

To achieve a realistic survey cadence and depth, we use the simulated survey strategy from the Rubin Operations Simulator (OpSim), which simulates the field selection and image acquisition process of Rubin-LSST over the 10-year duration of the planned survey \citep{Delgado_2014, Naghib_2019}. We use \texttt{OpSimSummary} \citep{Biswas_2019, Biswas_2020} to retrieve the times of observations of a specific point on the sky with the baseline v3.0 strategy from Rubin-LSST's Survey Cadence Optimization Committee \citep{PSTN-055}. While baseline v3.3 was recently released, we do not expect the updates to have a significant impact on the results of this analysis, as the biggest change has resulted in a decrease in the $u$-band sensitivity and increase in the sensitivity of the other bands. While 10 years of data can be simulated, we choose to clip the data in time to just the time frame covered by the template to avoid artifacts in the light curves that might make fitting for the time delay easier. The objects are observed in as many of Rubin's $ugrizy$ as possible at each object's simulated redshift given the template limitations in wavelength and the survey cadence.

The distribution for the absolute luminosities of CC~SNe are taken from \cite{Grayling_2023}. The redshift distributions and absolute magnifications of glSNe as expected to be discovered by Rubin-LSST are approximated from \cite{Goldstein_2019}. For each simulated object, a redshift, magnification for the first image, and magnification for the second image are randomly assigned from these distributions. The time delays are randomly drawn from a normal distribution with a mean of 0 and standard deviation of 50 days. The distributions of simulated parameters are shown in Figure \ref{fig:sim-dists}. Simulating realistic microlensing and dust extinction effects is beyond the scope of this work, so we choose not to include these effects in the simulations. We also do not consider correlations between the redshift, magnification, and time delay distributions. These simulations are intended to test a wide range of parameter space to probe the limits of the method.

The simulated objects must pass a series of data cuts to ensure the objects are high enough quality to firstly, produce an alert in the LSST data stream and secondly, fit for a time delay which may be reasonably informative. We require at least two observations of each image and the total number of observations to be greater than two times the number of bands in which there is data. The light curves must also have two data points with signal to noise ratio greater than 5 and at least two data points in any band within $\pm 5$ days of the time of peak magnitude for each image. We furthermore impose a requirement that $\beta$ be less than 15. These quality cuts are conservative, as we expect to need higher quality data than the minimum required in our simulations for a competitive time-delay constraint.

For the Rubin-LSST simulations, we fit the objects using the priors given in Table \ref{tab:rubin-priors}. Again, the prior on the time delay requires that there always be overlap between data from image 1 and data from image 2. Although the simplified Rubin-LSST simulations created for this analysis do not include microlensing, we fit with both the constant magnification model, which is the model the objects were simulated from, and the sigmoid magnification model to test if any biases arise from using an incorrect magnification model to fit the data.

\begin{table}
    \centering
    \caption{Priors on the hyperparameters used when fitting the Rubin-LSST simulations. If a simulation model is specified, with ``C'' denoting the constant magnification simulations and ``A'' denoting the \citet{Arendse_2023} simulations, then the prior applies to the objects simulated from that model. If a simulation model is not specified, then the prior on that hyperparameter is the same for both sets of simulations. Note that $\mu_{\beta}$, $\bm{\hat{t}}_{1}$, and $\bm{\hat{t}}_{2}$ are defined as in Equations \ref{eq:roman-priors-1}, and $\bm{\hat{T}}$ and $\bm{\hat{T}}_{\Delta}$ are defined as in Equations \ref{eq:time-vector} and \ref{eq:deshifted-time-vector}, respectively.}
    \begin{tabular}{c|c|c}
        Hyperparameter & Sim. Model & Prior \\
        \hline
        $A$ & - & $\mathcal{U}(0, 5)$ \\
        $\tau$ & - & $\mathcal{U}(10, 40)$ \\
        $\Delta$ & C & $\mathcal{U}(\min(\bm{\hat{t}}_{1}) - \max(\bm{\hat{t}}_{2}), $ \\
         &  & $\max(\bm{\hat{t}}_{1}) - \min(\bm{\hat{t}}_{2}))$ \\
         & A & $\mathcal{U}(0, \max(\bm{\hat{t}}_{1}) - \min(\bm{\hat{t}}_{2}))$ \\
        $\beta_{0}$ & - & $\mathcal{TN}(\mu_{\beta}, 10^{2}, 0, \infty)$ \\
        $\beta_{1}$ & - & $\mathcal{N}(0, 0.5^{2})$ \\
        $r$ & - & $\mathcal{N}(0, 0.5^{2})$ \\
        $t_{0}$ & - & $\mathcal{U}(\min(\bm{\hat{T}}_{\Delta}), \max(\bm{\hat{T}}_{\Delta}))$
    \end{tabular}
    \label{tab:rubin-priors}
\end{table}

\subsubsection{Realistic glSN~Ia Simulations from \citet{Arendse_2023}}

Recently, \citet{Arendse_2023} released the publicly available ``lensed Supernova Simulation Tool'' (\texttt{lensedSST}\footnote{\url{https://github.com/Nikki1510/lensed_supernova_simulator_tool}}), which simulates glSN~Ia with a realistic microlensing treatment. The \texttt{lensedSST} pipeline simulates glSNe~Ia from the \texttt{SALT3} model \citep{Guy_2007, Kenworthy_2021} using \texttt{sncosmo}, where the cadence and observing conditions are drawn from \texttt{OpSimSummary} using the baseline v3.0 observing strategy. The redshift of each SN is drawn jointly with the lens redshift and the Einstein radius of the system to produce systems with realistic time delays and magnifications using the \texttt{Lenstronomy} lens modeling package \citep{Birrer_2018, Birrer_2021}. In Figure \ref{fig:sim-dists}, the distributions of simulated parameters are shown.

Microlensing, which is caused by the expanding photosphere of the SN, is accounted for using realistic SN~Ia explosion models from \texttt{ARTIS} \citep{Kromer_2009} and microlensing maps from \texttt{GERLUMPH} \citep{Vernardos_Fluke_2014, Vernardos_2014, Vernardos_2015}. Given the microlensing parameters at the position of the images on the microlensing maps from \texttt{Lenstronomy}, a realization of microlensing is drawn for each wavelength filter and applied to the light curves. Finally, the simulated glSNe~Ia are subject to selection effects, described in detail in \citet{Arendse_2023}, to determine whether an object would be detected via several detection pipelines.

The explosion models for other types of SNe are less well-studied, making the simulation of realistic chromatic microlensing for non-SNe~Ia difficult. \cite{Pierel_2021} assume a simple achromatic model for the expansion of the SN photosphere to produce realistic achromatic microlensing. The simulation of chromatic microlensing, however, is only possible for glSNe~Ia. Therefore, in addition to the constant magnification simulations, which probe the sparse data regime for all types of glSNe, we use the \texttt{lensedSST} simulations to test \textsc{GausSN}'s performance in the sparse data regime with microlensing for glSNe~Ia. In this work, we continue to only focus on simulations of doubly imaged glSNe.

We use the same priors on the hyperparameters as the constant magnification Rubin-LSST simulations, as given in Table \ref{tab:rubin-priors}, with the exception of the time delay. In the \texttt{lensedSST} simulations, the leading image is assigned as image 1, so we restrict the time delay to be positive. In addition, we remove objects for which the two images do not overlap in phase space. This choice is motivated firstly by the need to set some general prior to fit the simulated objects in batches. The chosen prior would exclude the true value of the time delay for these objects, so we remove them. Secondly, GPs are data-driven, so their ability to extrapolate outside of where there is data is limited. If the multiple images of a glSN do not overlap in phase space, the \textsc{GausSN} fit may be unreliable because of this limitation of the model. We comment further on future work with \textsc{GausSN} on very sparsely sampled light curves in \S\ref{sec:discussion}. As with the constant magnification simulations, we fit with both the constant magnification model and the sigmoid magnification model, both of which assume achromatic microlensing, to test the bias introduced by incorrect assumptions about the underlying microlensing.

\subsubsection{Analysis}
\begin{figure}
    \centering
    \includegraphics[width=\linewidth]{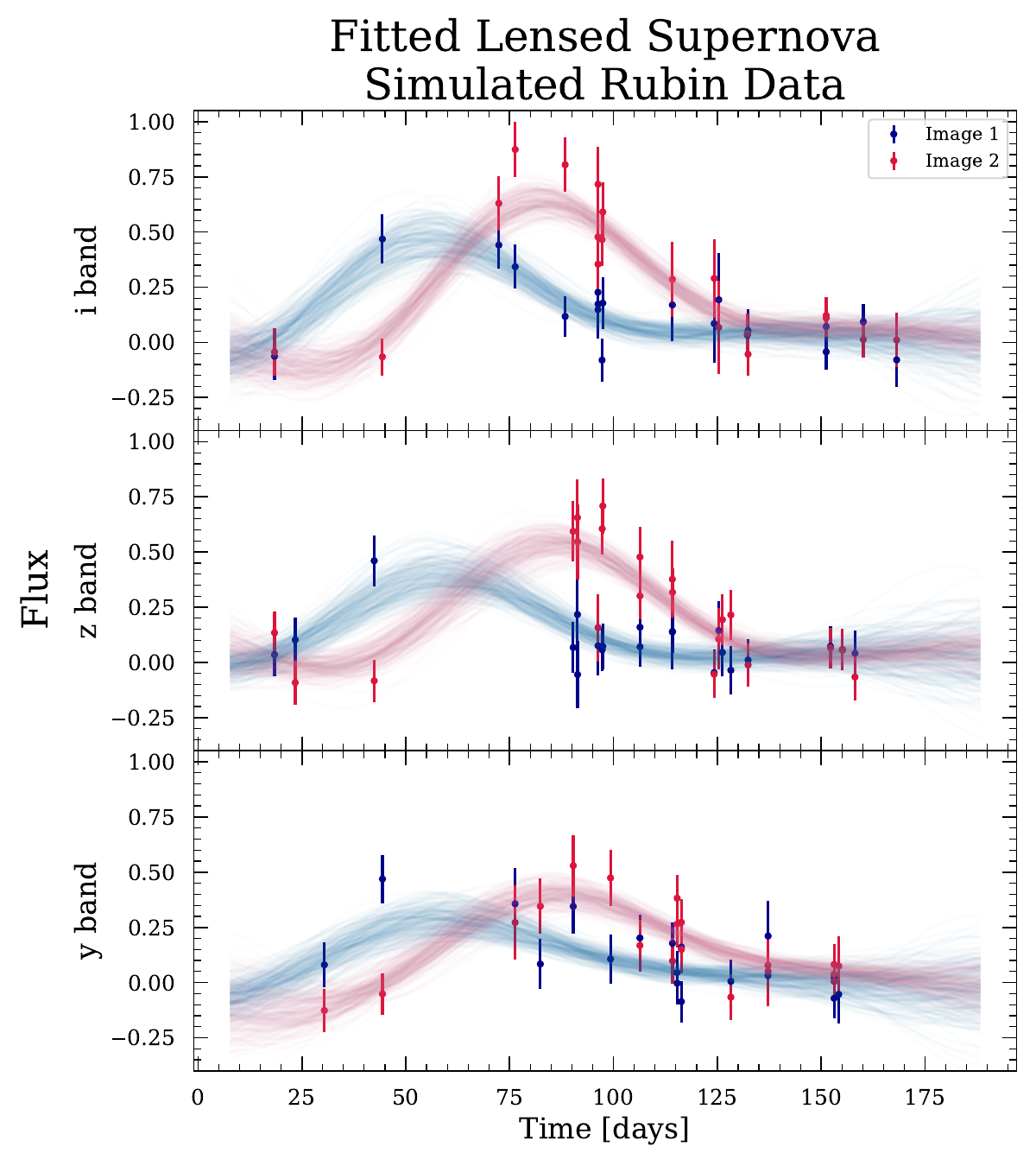}
    \caption{The flux data and fitted light curve for a simulated Rubin-LSST doubly imaged glSN using the constant magnification \textsc{GausSN} model. This object is a glSN~Ia at $z = 1.39$ with $\Delta = 25.62$ and $\beta = 1.30$. As described in \S\ref{sec:plotting}, the red fitted curves are time-shifted and magnified copies of the blue fitted curves, which have been conditioned on the data from both images.}
    \label{fig:Rubin-after}
\end{figure}

As with the Roman simulations, we first demonstrate the quality of \textsc{GausSN}'s performance on an individual object -- a glSN~Ia from the constant magnification simulations at $z = 1.39$ and with $\Delta = 25.62$ and $\beta = 1.30$. We demonstrate the quality of the \textsc{GausSN} fit at the data level in Figure \ref{fig:Rubin-after}. As in Figure \ref{fig:Roman-after}, we plot draws of the posterior predictive distribution over the light curve data. Even with the sparse Rubin-LSST data, \textsc{GausSN} recovers a realistic fitted light curve.

\begin{figure}
    \centering
    \includegraphics[width=\linewidth]{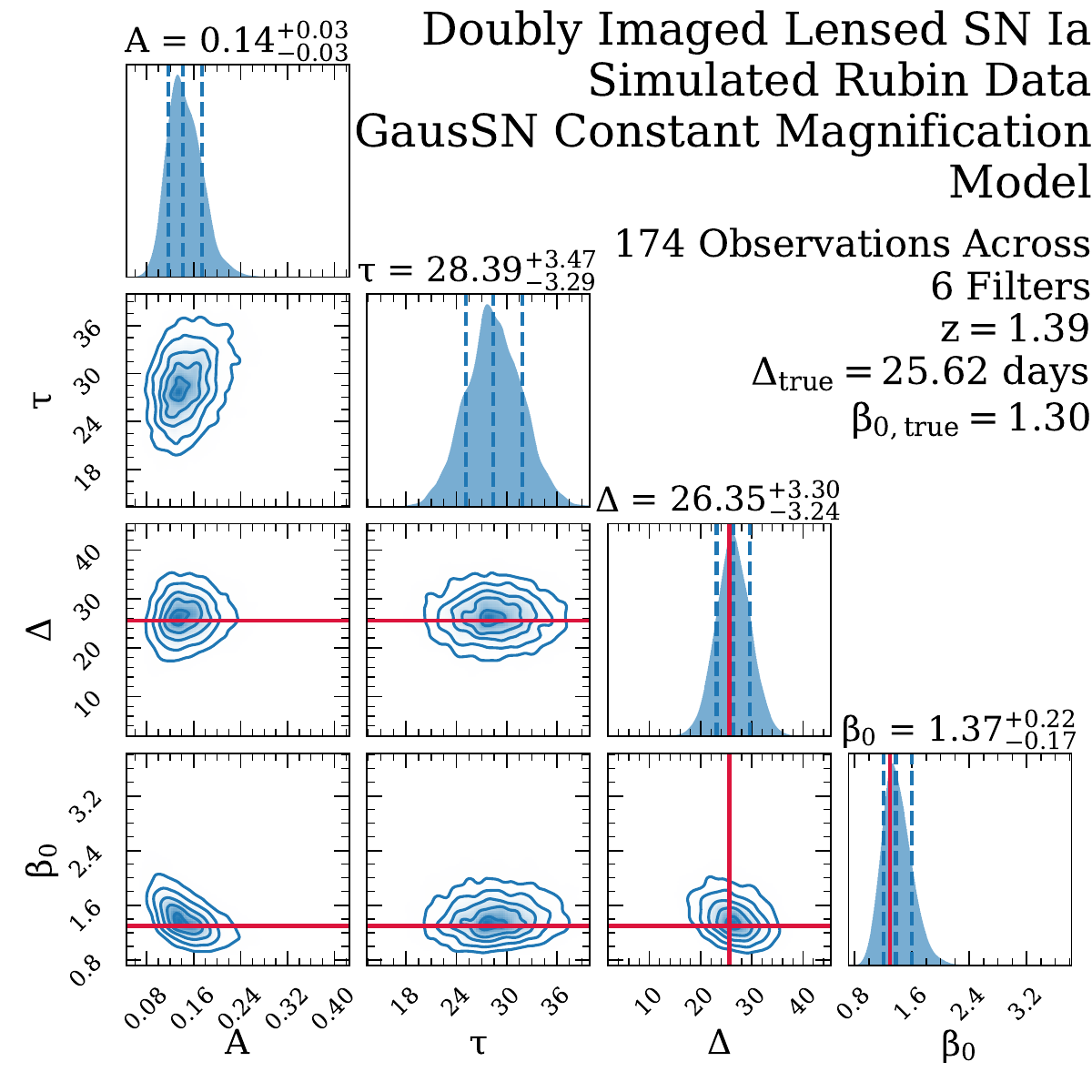}
    \caption{The corner plot for the \texttt{dynesty} fit to an example Rubin-LSST-like light curve at $z = 1.39$. From left to right, the parameters fit are $A$, the kernel amplitude, $\tau$, the kernel length scale, $\Delta$, the time delay, and $\beta$, the magnification. $\Delta$ and $\beta$ are retrieved within $1\sigma$ of the truth.}
    \label{fig:Rubin-corner}
\end{figure}

In Figure \ref{fig:Rubin-corner}, we show the joint posteriors for the four fitted parameters from the constant magnification fit in a corner plot. \textsc{GausSN} estimates $\Delta = 26.41^{+3.33}_{-3.29}$ days, a 12.61\% precision estimate, and $\beta = 1.37^{+0.21}_{-0.17}$. Although modeling this object is simpler than modeling the Roman objects, for example, because it is not subject to microlensing effects, the sparsity of the data clearly provides a challenge to the model. Even so, the time delay and magnification are still accurately recovered by \textsc{GausSN}.

\begin{figure*}
    \centering
    \includegraphics[width=\linewidth]{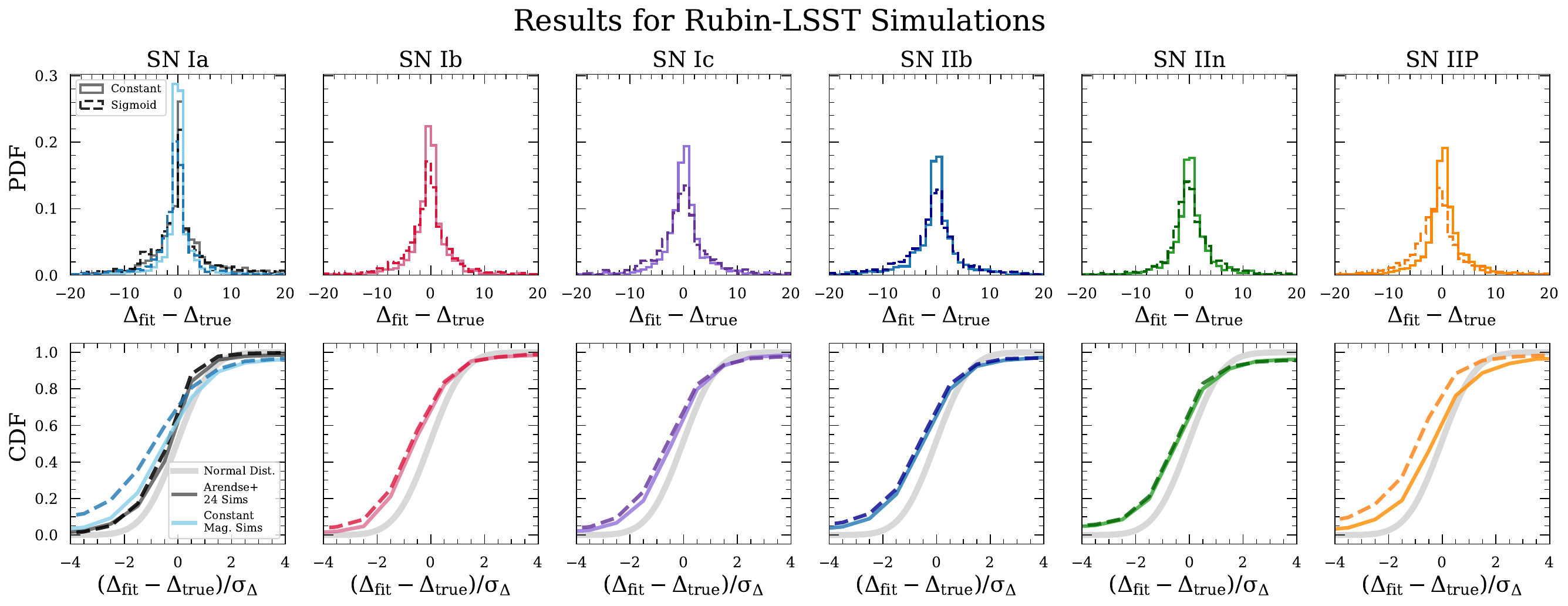}
    \caption{\textbf{Top:} The distribution of $\Delta_{\text{fit}} - \Delta_{\text{true}}$ by type of SN for the Rubin-LSST simulations. The solid lines show the results from the constant magnification model and the dashed lines show the results from the sigmoid magnification model. The results from fitting the constant magnification simulations are shown in color and the results from fitting the \citet{Arendse_2023} glSN~Ia simulations are shown in black. We take the mean of the posterior distribution for $\Delta$ to be $\Delta_{\text{fit}}$. \textbf{Bottom:} The CDF of the standardized error, $(\Delta_{\text{fit}} - \Delta_{\text{true}})/\sigma_{\Delta}$, distribution for the Rubin-LSST simulations by type of SN. The CDF of a normal distribution is shown in gray for reference.}
    \label{fig:Rubin-recovery}
\end{figure*}

On a population level for the constant magnification simulations, \textsc{GausSN} shows effectivenesss in recovering time delays that are close in absolute value to the true time delays. In the top row of Figure \ref{fig:Rubin-recovery}, we show the distribution of $\Delta_{\text{fit}} - \Delta_{\text{true}}$ by type of SN for both the constant magnification fits and sigmoid magnification fits. With the constant magnification model, \textsc{GausSN} retrieves 40.78\% of time delays within $\pm 1$ day of the truth, 69.20\% within $\pm 3$ days, and 79.92\% within $\pm 5$ days. We also find that 40.67\% of systems have fractional errors of less than 5\%. With the sigmoid magnification model, the performance is slightly reduced, but follow the same trends seen in the results on the Roman simulations. Notably, there does appear to be a slight bias in the median of the distribution of $\Delta_{\text{fit}} - \Delta_{\text{true}}$ of -0.417 days that is introduced when fitting with the sigmoid magnification model. As discussed further in \S\ref{sec:discussion}, it will be important to marginalize over the choice of microlensing parameterization to mitigate any potential bias introduced by a single microlensing function, as may be present in these results. On the whole, though, the \textsc{GausSN} time-delay estimate is consistently close to the truth, with only a slight and anticipated performance decrease when assuming an incorrect magnification model.

As with the Roman simulations, we check that the uncertainties on the time delays are well-calibrated by evaluating the distribution of $(\Delta_{\text{fit}} - \Delta_{\text{true}})/\sigma_{\Delta}$. We plot the CDF of this distribution in the bottom row of Figure \ref{fig:Rubin-recovery}. The fraction of SNe for which the true time delay is captured within the 68\% and 95\% CI, calculated from the $16^{\text{th}}$ and $84^{\text{th}}$ percentiles of the posterior samples, is 51.88\% and 80.57\%, respectively, for the constant magnification fits. With the sigmoid magnification model, 50.54\% and 78.62\% of true time delays are recovered within the 68\% and 95\% CI. The uncertainties are generally underestimated, as was seen in the results from the Roman simulations. We report these results broken down by type of SN in Table \ref{tab:Rubin-results}.

When fitting the \texttt{lensedSST} simulations, \textsc{GausSN} maintains a high quality performance, despite the greater complexity of these simulations due to their inclusion of realistic microlensing. Slight reductions are seen in the fraction of objects with time delays measured within 1 day -- 26.70\% for the constant magnification fit and 23.23\% for the sigmoid magnification fit. However, the fraction of objects with time delays measured within three and five days remain consistent at 58.67\% and 72.70\% for the constant magnification fit and 54.00\% and 70.59\% for the sigmoid magnification fit. There is a reduction in the number of objects with fractional errors of less than 5\% on the time delay, but such behavior is expected given that microlensing increases uncertainty in time delay estimation. The distribution of $(\Delta_{\text{fit}} - \Delta_{\text{true}})$ shows a slight bias, with a median of 0.106 days for the constant magnification fits and of -0.223 days for the sigmoid magnification fits. Interestingly, this offset in the median is reduced relative to the results on the constant magnification simulations.

Although the model tends to be further from the truth for the \texttt{lensedSST}, the uncertainties on the time delay accurately reflect the model's ability to recover the true time delay; with the constant magnification model, 61.54\% and 88.54\% are recovered within the 68\% and 95\% CI and with the sigmoid magnification model, 65.01\% and 91.10\% are recovered within the 68\% and 95\% CI. Further improvements on the calibration of uncertainties can be made through future work discussed in \S\ref{sec:discussion}. Despite the challenge of sparsely sampled light curves and realistic microlensing effects, \textsc{GausSN} continues to provide precise and accurate time-delay estimates with reliable uncertainties when fitting the \texttt{lensedSST} simulations.

\begin{table*}
    \centering
    \caption{The results of \textsc{GausSN} fits to the Rubin-LSST simulations by type of SN.}
    \begin{threeparttable}
        \begin{tabular}{c|c|c|c|c|c|c|c|c|c|c}
            & & Sim. & Fitting & $|\Delta_{\text{fit}} - \Delta_{\text{true}}|$ & $|\Delta_{\text{fit}} - \Delta_{\text{true}}|$ & $|\Delta_{\text{fit}} - \Delta_{\text{true}}|$ & Median & $|\Delta_{\text{fit}} - \Delta_{\text{true}}|/\Delta_{\text{true}}$ & $\Delta_{\text{true}}$ & $\Delta_{\text{true}}$ \\
            SN Type & N & Model\tnote{a} & Model\tnote{b} & < 1 day & < 3 days & < 5 days & $\Delta_{\text{fit}} - \Delta_{\text{true}}$ & < 5\% & in 68\% CI\tnote{c} & in 95\% CI\tnote{c} \\
            \hline
            \hline
            Ia & & & & & & & & & & \\
             & 1471 & C & C & 0.56 & 0.747 & 0.812 & 0.009 & 0.53 & 0.476 & 0.763 \\
             & 1471 & C & S & 0.369 & 0.608 & 0.708 & -0.387 & 0.354 & 0.394 & 0.668 \\
             & 663 & A & C & 0.267 & 0.587 & 0.727 & 0.106 & 0.166 & 0.615 & 0.885 \\
             & 663 & A & S & 0.232 & 0.54 & 0.706 & -0.223 & 0.134 & 0.650 & 0.911 \\
            Ib & & & & & & & & & & \\
             & 1498 & C & C & 0.421 & 0.729 & 0.84 & -0.134 & 0.432 & 0.569 & 0.876 \\
             & 1498 & C & S & 0.31 & 0.617 & 0.758 & -0.385 & 0.325 & 0.535 & 0.828 \\
            Ic & & & & & & & & & & \\
             & 1440 & C & C & 0.373 & 0.694 & 0.814 & 0.045 & 0.385 & 0.55 & 0.837 \\
             & 1440 & C & S & 0.263 & 0.564 & 0.717 & -0.27 & 0.278 & 0.528 & 0.806 \\
            IIb & & & & & & & & & & \\
             & 1528 & C & C & 0.354 & 0.624 & 0.74 & -0.006 & 0.355 & 0.494 & 0.8 \\
             & 1528 & C & S & 0.26 & 0.531 & 0.675 & -0.254 & 0.262 & 0.495 & 0.778 \\
            IIn & & & & & & & & & & \\
             & 1730 & C & C & 0.369 & 0.669 & 0.778 & 0.004 & 0.356 & 0.522 & 0.795 \\
             & 1730 & C & S & 0.282 & 0.616 & 0.743 & -0.205 & 0.292 & 0.528 & 0.787 \\
            IIP & & & & & & & & & & \\
             & 1748 & C & C & 0.37 & 0.688 & 0.81 & 0.169 & 0.382 & 0.501 & 0.764 \\
             & 1748 & C & S & 0.253 & 0.55 & 0.703 & -0.998 & 0.262 & 0.473 & 0.735 \\
            \hline
        \end{tabular}
        \begin{tablenotes}\footnotesize
        \item[a] The model from which the data was simulated, where ``C'' denotes the constant magnification model simulations from this work and ``A'' denotes fully realistic glSN~Ia simulations from \citet{Arendse_2023}.

        \item[b] The \textsc{GausSN} model that was used to fit the data, where again ``C'' denotes the constant magnification model and ``S'' denotes the sigmoid magnification model.
        
        \item[c] The fraction of SNe for which $\Delta_{\text{true}}$ falls within the 68\% and 95\% credible intervals (CI) computed from the $16^{\text{th}}$ and $84^{\text{th}}$ percentiles of the posterior samples.
        \end{tablenotes}
    \end{threeparttable}
    \label{tab:Rubin-results}
\end{table*}

Additionally, with the current observing strategy, \citet{Arendse_2023} predict that an appreciable fraction of glSNe discovered with Rubin-LSST will have few to no observations before light-curve peak for one or more images. In Table \ref{tab:Rubin-missedpeak}, we summarize the performance on \textsc{GausSN} on the subset of objects for which the peak in the light curve of one or both images is missed. Unsurprisingly, the recovered time delays for this subset of glSNe are on average further from the truth, as indicated by the reduced number of objects with fractional errors in the measured time delay of less than 5\%. Similar drops in performance are also seen in the rate of time delays within 1, 3, and 5 days of the true time delay. Importantly, though, the fraction of objects with a true time delay that falls within the 68\% and 95\% CI remains equally well-calibrated, indicating an appropriate increase in uncertainty when estimating the time delays of these objects.

\begin{table}
    \centering
    \caption{The results from \textsc{GausSN} on the subset of Rubin-LSST data for which there are no observations before peak for one or both images.}
    \begin{tabular}{c|c|c|c|c|c|c}
        & Sim. & Fit. & $|\Delta_{\text{fit}} - \Delta_{\text{true}}|/\Delta_{\text{true}}$ & $\Delta_{\text{true}}$ in & $\Delta_{\text{true}}$ in \\
        N & Model & Model & < 5\% & 68\% CI & 95\% CI \\
        \hline
        \hline
        256 & C & C & 0.309 & 0.52 & 0.805 \\
        256 & C & S & 0.266 & 0.566 & 0.855 \\
        93 & A & C & 0.054 & 0.591 & 0.903 \\
        93 & A & S & 0.097 & 0.634 & 0.914 \\
        \hline
    \end{tabular}
    \label{tab:Rubin-missedpeak}
\end{table}

Only 5.73\% and 7.87\% of the constant magnification simulations have fitted time delays which are more than 5$\sigma$ away from the true value for the constant magnification and sigmoid magnification models, respectively. Based on visual inspection of the fits, it is very easy to tell when there has been a catastrophic failure with \textsc{GausSN} for the Rubin-LSST simulations. For the majority of these objects, the posteriors show that the nested sampling chains have run up against the prior bounds, either for the GP length scale or the time delay, suggesting that the prior was not well-chosen for the data. Real glSNe will not be fit in bulk, as the simulated objects are for this analysis, so again more informative priors can be selected for objects on an individual basis to mitigate many of these extreme outliers. Indeed, fitting a few of these objects with more informative priors enabled more accurate time-delay recovery. Therefore, these outliers are not of great concern, as they could be easily flagged and refit with more informative priors to get reliable time-delay estimates. 

As with the Roman simulations, there are also several outliers with multi-modal posteriors for which the mean and standard deviation of the samples are simply a poor summary of the posterior. Again, fitting with more informative priors easily solves this problem. Furthermore, we re-emphasize that the full posterior should be used in the cosmological analysis of any glSN. These types of outliers are, therefore, expected and not as concerning.

A similarly small fraction of \texttt{lensedSST} objects -- 2.11\% for the constant magnification fit and 0.90\% for the sigmoid magnification fit -- have fitted time delays which are more than 5$\sigma$ away from the true value. Outliers appear to be dominated by objects that experience extreme microlensing effects, which appear to alter the location of the peak in the light curve. Because the sigmoid microlensing model is not flexible enough to capture all possible effects of microlensing, these outliers are unsurprising. Future work to account for uncertainty in the microlensing model choice is discussed in \S\ref{sec:discussion} and would reduce the frequency of these fits with very poorly calibrated uncertainties. Unfortunately, these outliers are difficult to identify based on visual inspection of the fitted light curves.

The results in Table \ref{tab:Rubin-results} are roughly consistent with those reported for the Roman simulations. However, the mean uncertainty on the time-delay estimate from \textsc{GausSN} for the constant magnification simulations is 7.53 days for the constant magnification model and 10.22 days for the sigmoid magnification model. For the \texttt{lensedSST} simulations, the mean uncertainty on the time-delay estimate is 5.02 days for the constant magnification model and 5.84 days for the sigmoid magnification model. Relative to the Roman simulations, the time-delay estimates for the Rubin-LSST simulations are more uncertain in general. Given that \textsc{GausSN} is a data driven approach, and the Rubin-LSST data is more sparse than the Roman data, this increase in uncertainty is expected. Most importantly, the uncertainties remain equally well-calibrated across two very different sets of data.

Finally, we note that there are differences in the constant magnification and \texttt{lensedSST} simulations beyond the microlensing treatment, which may also have an effect on the results. The \texttt{lensedSST} pipeline retrieves time delays and magnifications from physical modeling of the lens given the source redshift and lens geometry. Therefore, these parameters are highly correlated in these simulations. On the other hand, the constant magnification simulations from this work independently sample from the distributions of redshift, time delay, and magnification. Therefore, the constant magnification simulations may contain parameter combinations that would be unlikely to occur in reality. The regions of parameter space which the \texttt{lensedSST} simulations fall into appear to lend themselves better to fitting in the \textsc{GausSN} model, as seen in the improved calibration of the time delay uncertainties. The distributions of source redshifts, time delays, and relative magnifications for these two sets of simulations are shown in Figure \ref{fig:sim-dists}. However, the number of objects with fractional errors on the time delay estimate of less than 5\% is reduced for the \texttt{lensedSST} simulations, which we attribute to microlensing uncertainty. On the whole, \textsc{GausSN} shows strength in fitting Rubin-LSST simulations, despite the sparse nature of the light curves.

%% file: 06_refsdal.tex
We apply \textsc{GausSN} to the five images of SN Refsdal to demonstrate the model's performance on real data, as well as on an object with more than two images. Following the publication of \cite{Kelly_2023_Science} and \cite{Kelly_2023_ApJ}, the data was made public\footnote{Available at \url{https://dx.doi.org/10.3847/1538-4357/ac4ccb}.}. The five images of SN Refsdal, images S2-S4 and SX, were observed by the Hubble Space Telescope (HST) in three filters -- F105W, F125W, and F160W. As the data in the F105W filter is limited and of worse quality relative to the other two bands, we follow \cite{Kelly_2023_ApJ} and exclude this data from the time-delay fit. 

\begin{figure*}
    \centering
    \includegraphics[width=\linewidth]{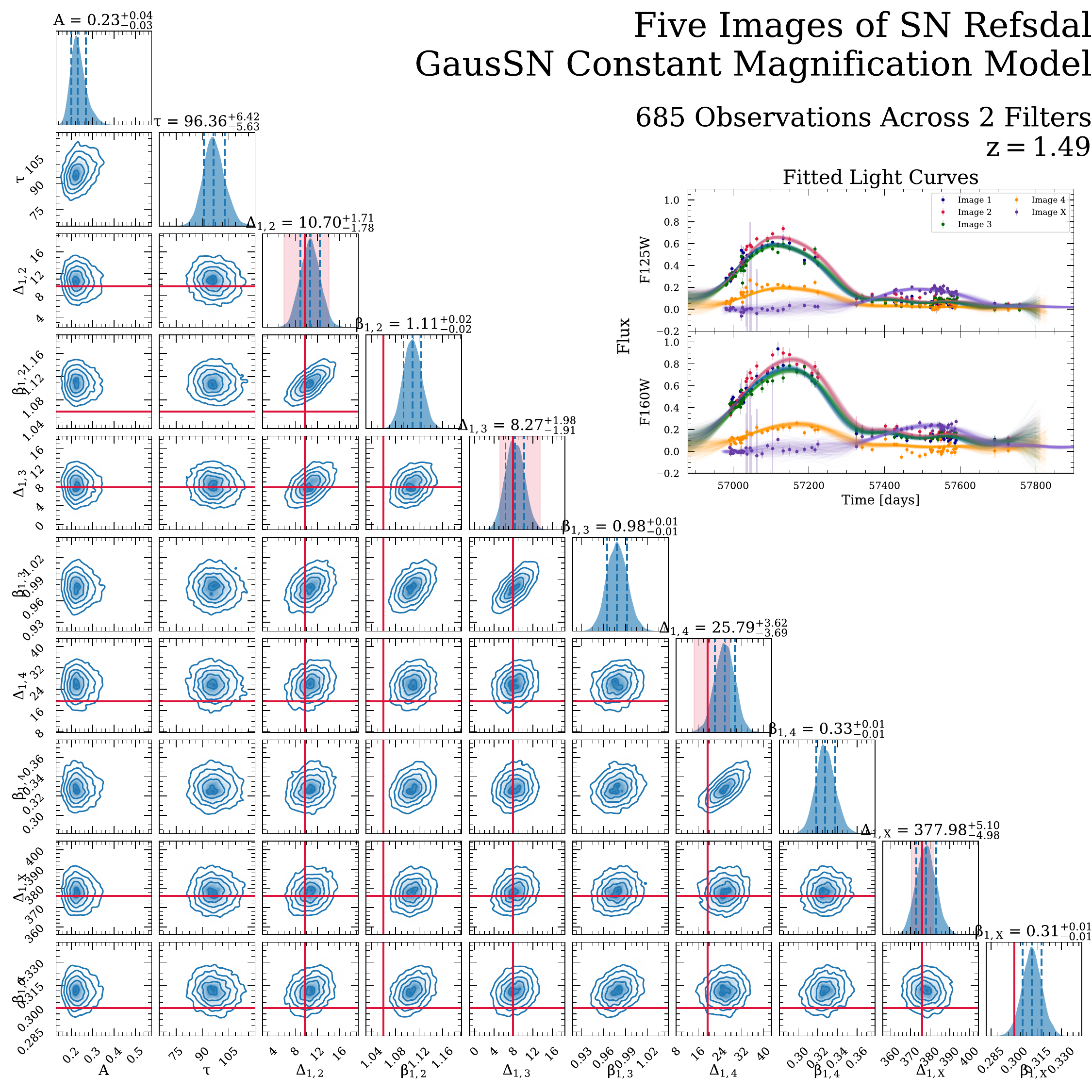}
    \caption{The posterior distribution from \textsc{GausSN} over parameters from the constant magnification model fit to SN Refsdal's five images. For comparison, we show the 68\% CI for the time delays from the $16^{\text{th}}$ and $84^{\text{th}}$ percentiles reported in \citet{Kelly_2023_ApJ} as the shaded red region, with the reported point estimate shown by the red line. We do not show the shaded region for the magnifications because they cover all of the parameter space shown in each 1D posterior plot. We use the 68\% CI as a proxy for the full posterior distributions from \citet{Kelly_2023_ApJ}, which is not publicly available. The time delays from \textsc{GausSN} are all in good agreement with \citet{Kelly_2023_ApJ}. \textbf{Inset:} The flux data and fitted light curves of SN Refsdal's five images from the constant magnification model. As described in detail in \S\ref{sec:plotting}, the red fitted curves are time-shifted and magnified copies of the blue fitted curves, which have been conditioned on the data from both images.}
    \label{fig:Refsdal-corner-fit}
\end{figure*}

We first fit SN Refsdal with the constant magnification model. The priors on the hyperparameters of this fit are given in Table \ref{tab:refsdal-priors}. We note that this analysis is not blinded in the way that the analysis in \cite{Kelly_2023_ApJ} was. The corner plot for the fit is shown in Figure \ref{fig:Refsdal-corner-fit}. We measure the time delay of the fifth image, SX, which is the most cosmologically interesting due to its long time delay, to be $\Delta_{1, X} = 377.98^{+5.10}_{-4.98}$ days -- a 1.35\% precision measurement. In addition, our results are in very good agreement with those reported by \cite{Kelly_2023_ApJ} for all five images. As shown in Figure \ref{fig:Refsdal-corner-fit}, the time-delay estimates from \textsc{GausSN} fall within the $16^{\text{th}}$ and $84^{\text{th}}$ percentiles of the posterior distributions on these parameters given in \cite{Kelly_2023_ApJ}. The fitted light curves are shown as an inset in Figure \ref{fig:Refsdal-corner-fit}.

\begin{table}
    \centering
    \caption{Priors on the hyperparameters describing the SN Refsdal system. If an image is specified for a prior, the hyperparameter is taken as relative to image 1. If an image is not specified, then the prior on that hyperparameter is the same for every image.}
    \begin{tabular}{c|c|c}
        Hyperparameter & Image & Prior \\
        \hline
        $A$ & - & $\mathcal{U}(0, 5)$ \\
        $\tau$ & - & $\mathcal{U}(10, 120)$ \\
        $\Delta$ & Image 2 & $\mathcal{U}(-30, 50)$ \\
         & Image 3 & $\mathcal{U}(-30, 50)$ \\
         & Image 4 & $\mathcal{U}(-50, 150)$ \\
         & Image X & $\mathcal{U}(100, 500)$ \\
        $\beta_{0}$ & - & $\mathcal{U}(0.01, 1.50)$ \\
        $\beta_{1}$ & - & $\mathcal{N}(0, 0.5^{2})$ \\
        $r$ & - & $\mathcal{N}(0, 0.5^{2})$ \\
        $t_{0}$ & - & $\mathcal{U}( \text{min}(\bm{\hat{T}}), \text{max}(\bm{\hat{T}}))$
    \end{tabular}
    \label{tab:refsdal-priors}
\end{table}

As noted in \cite{Kelly_2023_ApJ}, there is evidence for microlensing affecting the light curves, particularly those of images S2, S4, and SX. Therefore, we additionally fit SN Refsdal with the sigmoid magnification function, with the priors on the additional hyperparameters given in Table \ref{tab:refsdal-priors}. With the sigmoid magnification function, the time delay for SX, $\Delta_{1,X} = 376.46^{+5.37}_{-5.29}$ days, is still highly consistent with the previous results. There is only a slight reduction in the precision, with the time delay now measured to 1.43\% precision. The time delay of image S3 is also still in very good agreement with \cite{Kelly_2023_ApJ}.

For images S2 and S4, the results are more discrepant from those in the constant magnification fit to the data. With the sigmoid magnification model, \textsc{GausSN} estimates $\Delta_{1,2} = -1.95^{+2.53}_{-2.54}$ days and $\Delta_{1,4} = -0.40^{+5.45}_{-6.02}$ days. On the other hand, \cite{Kelly_2023_ApJ} reports $\Delta_{1,2} = 9.7^{+4.4}_{-3.7}$ days and $\Delta_{1,4} = 19.44^{+8.0}_{-4.9}$ days. We emphasize, though, that the results from the individual methods used in the original analysis show additional dispersion in their time-delay estimates that is not well-represented in the values reported above. Therefore, the discrepancy seen in the \cite{Kelly_2023_ApJ} time-delay estimates and the time-delay estimates reported from the sigmoid magnification model fit in this analysis is not unexpected or of great concern. Most importantly, the results for image SX remain consistent.

\begin{figure*}
    \centering
    \includegraphics[width=\linewidth]{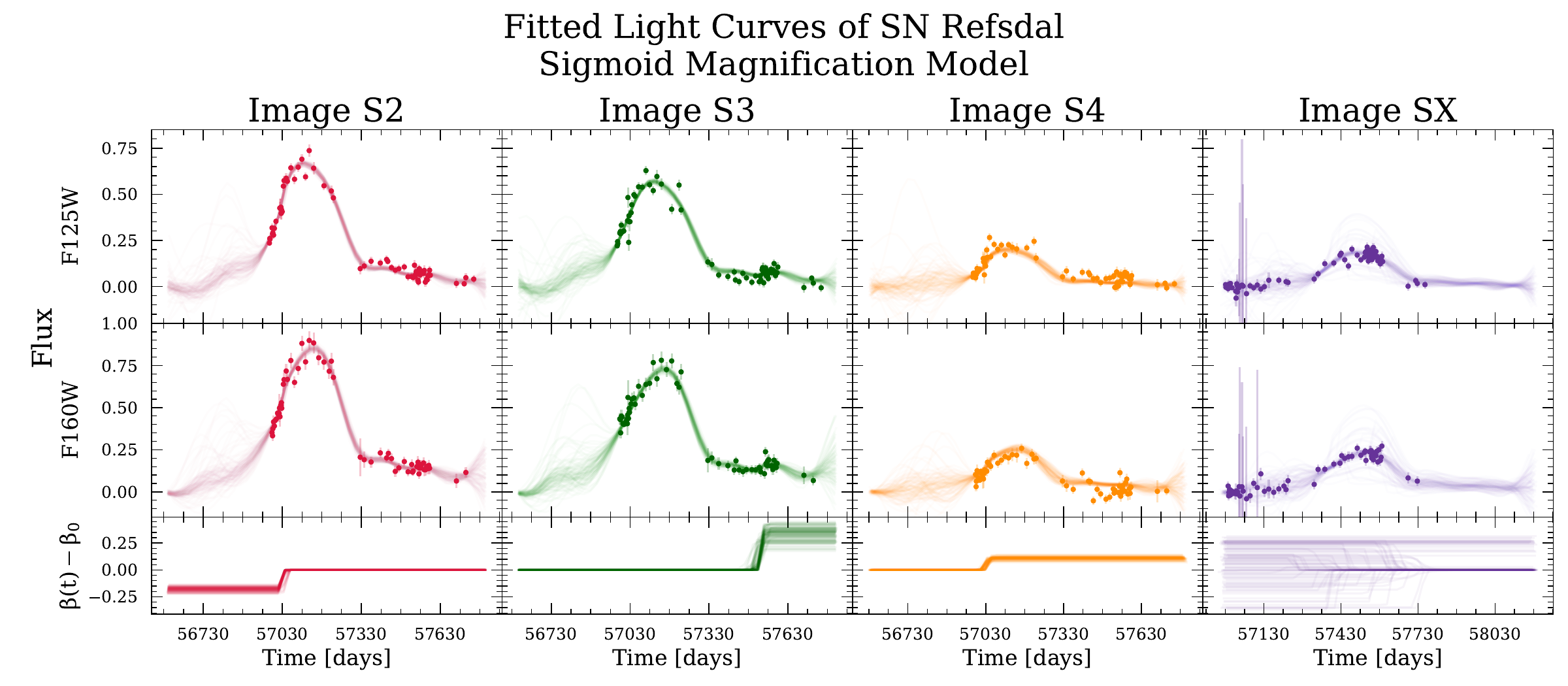}
    \caption{The sigmoid magnification fit to SN Refsdal's images S2-SX in the F125W filter (top row) and the F160W filter (middle row) with realizations of the magnification function for each of images S2-SX for 100 draws from the posterior (bottom row). The realizations of the fitted light curves are created according to the procedure described in \S\ref{sec:plotting}. We find that there is considerable uncertainty regarding the effect of microlensing on image SX, as also found in the analysis from \citet{Kelly_2023_ApJ}. \textsc{GausSN} more confidently constraints the relative magnification affecting images S2-S4. There is evidence for microlensing affecting the beginnings of the light curves for images S2 and S4 and the tail of the light curve for image S3.}
    \label{fig:Refsdal-lensing}
\end{figure*}

We show realizations of the sigmoid magnification function using draws from the posteriors, as shown in Figure \ref{fig:Refsdal-lensing}. As in \cite{Kelly_2023_ApJ}, we find considerable uncertainty about the effect of microlensing on image SX. There appears to be evidence for microlensing affecting each of images S2-S4, as well. At early times, there is uncertainty in the relative magnification because of the lack of overlapping data from multiple images. However, even around peak, where there is data from other images, the microlensing effect is uncertain. For image S3, the microlensing effect switches on in the tail, where the light curve appears to begin rising again. As this behavior is not seen in the tails of image S2 or S4, it is attributed to microlensing. The magnification from microlensing does not change the shape of the light curve around peak, so it has a minimal effect on the time delay.

Images S2 and S4 show signs of microlensing affecting the beginnings of the light curves. This behavior suggests that images S2 and S4 may exhibit a faster rise than the other images. Interestingly, the time delays of these two images are more discrepant with the \cite{Kelly_2023_ApJ} results. The rise of the light curves is well-observed for all five images, so we expect that a considerable amount of time-delay constraining power comes from this region of the light curves. Therefore, microlensing affecting images S2 and S4 in this time frame may result in the more significant shift in the time delay from the constant magnification model to the sigmoid magnification model. There is clearly uncertainty in the time-delay fit due to microlensing effects on the light curves, highlighting the importance of accounting for microlensing alongside the time-delay estimate.

%% file: 07_discussion.tex
There now exist several methods for extracting time delays from lensed quasars and glSNe. A range of time-delay estimation methods have been used in analyses of lensed quasars. A number of optimization schemes have been paired with spline and GP models of the underlying light curve shape, due to the stochastic nature of the quasar light curves. Among these methods is the widely popular spline-based method \texttt{PyCS} \citep{Tewes_2013_pycs, Millon_2020}, which was used in the H0LiCOW 2.4\% precision $H_{0}$ estimate \citep{Wong_2019} and adapted for use on SN Refsdal \citep{Kelly_2023_ApJ}. On the other hand, template-based approaches, such as \texttt{sntd}, have been favored in recent analyses of glSNe \citep{Pierel_2022, Kelly_2023_ApJ}. We consider below a few advantages to the \textsc{GausSN} approach, which motivate its use alongside other methods in analyses of glSNe discovered in the future, as well as point out where improvements on \textsc{GausSN} can be made.

\subsection{Template Choice Uncertainty}
SNe are known to have some level of intrinsic variation within each type and a number of methods have been used to extract templates from SN light curves. As a result, there are several templates available for each SN types which one may chose to fit with using a template-based approach for time-delay estimation. However, the effect of fitting for a time delay with an incorrectly chosen template has yet to be explored. This unknown systematic contributes unaccounted for uncertainties to the time-delay estimate, which may be difficult to propagate through to the uncertainty in a motivated way. Even for types of SNe which exhibit a great diversity of light curves, which may be difficult to fit with a template-based approach, \textsc{GausSN} maintains optimal performance.

In addition, template-based approaches depend on reliable redshift estimates to shift the template into the correct frame of reference. In the era of Rubin-LSST and Roman, spectroscopic follow-up will be limited, so spectroscopic redshifts will likely not be available for every object. Therefore, \textsc{GausSN}, a method which is agnostic to source redshift, is an important complement to methods which depend on redshifts.

Furthermore, lensing enables us to probe SNe at higher redshifts than would otherwise be observable by existing telescopes. The redshift evolution of SNe is an open question in the field, with an unaccounted for evolution potentially contributing to systematics in cosmological analyses \citep{Nicolas_2021}. Therefore, the use of templates based on SNe light curves in the nearby universe similarly introduces a potential source of unaccounted for systematic uncertainty in the time-delay estimate. Especially because templates tend to be based on data from very-nearby SNe to ensure it is high quality, the extremity of this difference will be at its peak for glSNe expected to be discovered by Rubin-LSST, up to $z=2$, and by Roman, up to $z=4$.

Being completely independent of templates, \textsc{GausSN} is not subject to this systematic and therefore provides complete Bayesian uncertainties on time-delay estimates without further post-processing to account for such an uncertainty. Given glSNe are attractive for having minimal systematics relative to the distance ladder, keeping sources of systematics to a minimum is important in these analyses.

Finally, if the SN light curve shape is peculiar or without a good template, as was the case with SN Refsdal, template-based approaches must turn to generic templates, such as the Bazin function \citep{Bazin_2009, Karpenka_2013, Villar_2019}. The flexibility of generic SN templates can make fitting the templates to the data tedious and sensitive to fine tuning. Alternatively, a custom template may be created for the system. While this approach was possible for SN Refsdal \citep{Kelly_2023_ApJ}, it will not be feasible in the future as larger numbers of glSNe are discovered. Furthermore, it is difficult to validate an analysis which is based off a custom-made template. \textsc{GausSN} is capable of fitting any object, regardless of its spectroscopic classification, without constructing or fine-tuning a template for the data beforehand.

However, in the very sparse data regime, a template-based approach may be preferable if it is possible to make stronger assumptions about the shape of the underlying light curve to better constrain the time delay. While \textsc{GausSN} performs well on Roman simulations and Rubin-LSST simulations, we cannot predict performance on only a few epochs of data per image, especially if these observations have little overlap in SN light curve phase, because of the limited ability of a GP to extrapolate between sparse observations. As demonstrated in Table \ref{tab:Rubin-missedpeak}, the performance of \textsc{GausSN} is reduced for objects in which pre-peak data is missing for one or both images. In future work, we plan to incorporate SN light curve templates as the mean function of the GP. This functionality would provide a more informative prior over the functions describing the light curve, while still allowing for residual variations which may arise from incorrect template choice or a redshift evolution of SN light curves.

\subsection{Microlensing}

Another advantage to \textsc{GausSN} is the treatment of microlensing. The wide ranging nature of the effect which microlensing from stars and substructure can have on a light curve is extremely difficult to constrain \citep{Pierel_2022}. For this reason, existing time-delay techniques consider only an additional uncertainty due to microlensing rather than treating it alongside the time-delay estimate. By treating macro- and micro-lensing together as one time-varying magnification term, \textsc{GausSN} marginalizes over only the relative magnification realizations which are consistent with the data. Therefore, the uncertainties from microlensing are propagated through the measurement in a fully Bayesian way.

We adopt the sigmoid magnification model because it resembles the microlensing curves shown in \cite{FoxleyMarrable_2018} and \cite{Pierel_2022} and is a convenient fitting function which can be interpreted in a qualitative sense. However, this parametric form is not directly derived from physical models of the microlensing of an expanding photosphere. Furthermore, this treatment does not take advantage of the knowledge of the microlensing magnification distribution. As the photosphere expands, the microlensing effect tends to become less strong because the microcaustics are being averaged over a larger area. Additional parametric models which take advantage of this information can and should be tested within the \textsc{GausSN} framework.

It would also be possible to implement a more flexible and expressive non-parametric microlensing representation within \textsc{GausSN}. Other time-delay estimation methods have used non-parameteric approaches such as splines \citep{Tewes_2013_pycs} or simulations from a Gaussian process \citep{Pierel_2019} to quantify the effect of microlensing. Within \textsc{GausSN} such a representation could be incorporated seamlessly into the Bayesian model. A natural kernel choice in a GP representation of microlensing could be the non-stationary \citet{Gibbs_1997} kernel \citep[see][for an example]{Revsbech_2018}. This would allow for microlensing functions with a short burst of fast magnification change (corresponding to the moment of caustic crossing) and a tendency towards something flatter away from this.

However, any one parameterization of microlensing will not be able to capture the full diversity of microlensing curves possible. In future work, we intend to explore different parametric and non-parametric magnification models, including models which do not assume achromatic microlensing. In addition, we plan to develop methods for Bayesian model averaging \citep[e.g.][for a recent application in astronomy]{Nixon_2023} to marginalize over multiple possible microlensing models when estimating time delays.

\subsection{Fully Bayesian Treatment}
The \textsc{GausSN} framework provides a fully Bayesian treatment of time-delay estimation, but this is just one component of the $H_{0}$ estimate. A model of the gravitational potential of the lens is necessary to estimate a time-delay distance, which can be used to constrain $H_{0}$ \citep[see e.g.][]{Treu_2022, Suyu_2023}. Of course, the time delay and lensing potential are intrinsically linked, meaning there is shared information between these two methods that is not taken advantage of by considering them separately. In the future, a fully Bayesian estimate of the time-delay distance from a combined lens modeling and time-delay estimation method would provide the tightest possible constraints on an $H_{0}$ estimate.

The most outstanding progress in that direction thus far in the field came from \cite{Birrer_2020}, which used a hierarchical Bayesian model to jointly infer $H_{0}$ and galaxy density profiles. \textsc{GausSN} represents a step towards this goal, as well. The \textsc{GausSN} posteriors over the time delay and time-varying magnification parameters could be incorporated into a larger hierarchical Bayesian model for $H_{0}$ to best utilize all the information available.

%% file: 08_conclusions.tex
We have demonstrated that precise and reliable time-delay estimates can be obtained for resolved glSN systems using \textsc{GausSN}, a Bayesian semi-parametric Gaussian Process model. This approach has a number of complementary attributes to existing approaches. In particular, \textsc{GausSN}:
\begin{itemize}
    \item easily fits glSNe of any type, regardless of whether it has a well-understood light curve shape, without fine-tuning or redshift information,

    \item is not sensitive to systematic effects introduced by template selection, such as a potential redshift evolution of SNe light curves or fitting with a template that is not representative of the true underlying light curve shape,

    \item accounts for microlensing alongside time-delay fitting with a modular implementation, which enables any number of microlensing models to be used,

    \item provides fully Bayesian uncertainties for time-delay estimates.
\end{itemize}

\textsc{GausSN} has proven successful on simulations of glSNe as expected from Rubin-LSST and Roman. The Roman simulations from \citetalias{Pierel_2021} are highly realistic, with careful treatment of survey cadence and depth, dust extinction, and microlensing. Without assuming that the SN type is known and a valid template is available, and without knowledge of the redshift, \textsc{GausSN} maintains a similar performance to the template-based approach \texttt{sntd}. While the results from \texttt{sntd} are based on the assumption of the correct SN-type, template and redshift, \textsc{GausSN} is entirely agnostic to this information. Furthermore, these simulations were not generated from the \textsc{GausSN} forward model, so our results demonstrate the ability of \textsc{GausSN} to generalize.

The simplified Rubin-LSST simulations, created for this analysis using templates from \texttt{sncosmo}, emulate the cadence and depth of the survey. While these simulations do not take into account effects such as dust extinction and microlensing, they test \textsc{GausSN}'s performance on sparsely sampled data. \textsc{GausSN} is successful in predicting time delays which are very close to the true time delay, but the uncertainties on the fits are underestimated in the case of some catastrophic failures. Given the current and projected rates of discovery of glSNe, we expect to have the capacity to inspect the fits individually and choose object-specific priors, which we expect will mitigate any highly inaccurate outliers.

We finally demonstrate the performance of \textsc{GausSN} on the five images of SN Refsdal. Using \textsc{GausSN}, we find that the time delay of image SX is highly consistent with the results from \cite{Kelly_2023_ApJ}, regardless of whether a constant or sigmoid magnification model is used. The time delays of images S2-S4 show more dispersion around the values estimated in \cite{Kelly_2023_ApJ}, but not so much to be concerning, especially given the lessened cosmological interest in these images. As in the original analysis, there is considerable uncertainty in the effect of microlensing on image SX. However, even with the uncertainty due to microlensing accounted for, we are still able to estimate the time delay to 1.43\% precision -- a level of precision competitive with the estimates reported in \cite{Kelly_2023_ApJ}. \textsc{GausSN}, by fitting for the time delay alongside a time-varying magnification term using the data from all five images simultaneously, provides competitive time-delay constraints using real glSN data, without compromising on fit quality.

There remain many possibilities for extensions of this work to address i) restrictions on the kernel which reflect known physical characteristics of the system, such as constraining the kernel to positive flux only, ii) a test of additional treatments of microlensing, which ideally would be based on physically-derived parameters, iii) the incorporation of additional spectroscopic or unresolved light curve information, iv) a treatment of the covariance between light curve shapes in different bands, as we know photometric data to be integrals of slices of a continuous SED in time- and wavelength- space, v) incorporation of dust effects from the host and lens into the model, and vi) development of quantitative model checking approaches to assess the reliability of time-delay estimates. We leave addressing these issues to future work.

A percent-level $H_{0}$ measurement will require more than precise and accurate time-delay estimation. Optimistically, even if Rubin-LSST and Roman on their own provide the necessary photometric data for time day estimation measurements, additional data will be needed from other instruments to enable the complex lens modeling required for an $H_{0}$ measurement. Fast and synergistic follow-up of glSNe from a variety of instruments will be needed to complete the full cosmological analysis of these objects.

Despite these challenges, the recent increase in the rate of discovery of glSNe and first measurement of $H_{0}$ from SN Refsdal suggests promise for this young technique. A measurement of $H_{0}$ to percent level precision from glSNe would be an important new piece of evidence for efforts to determine whether the Hubble tension is driven by systematics or new physics. With upcoming instruments, such as Rubin-LSST and Roman, and precise statistical analysis tools for time-delay extraction, such as \textsc{GausSN}, glSNe show promise to reach this benchmark in the coming decade.

%% file: 09_appendixA.tex
\begin{figure}
    \centering
    \includegraphics[width=\linewidth]{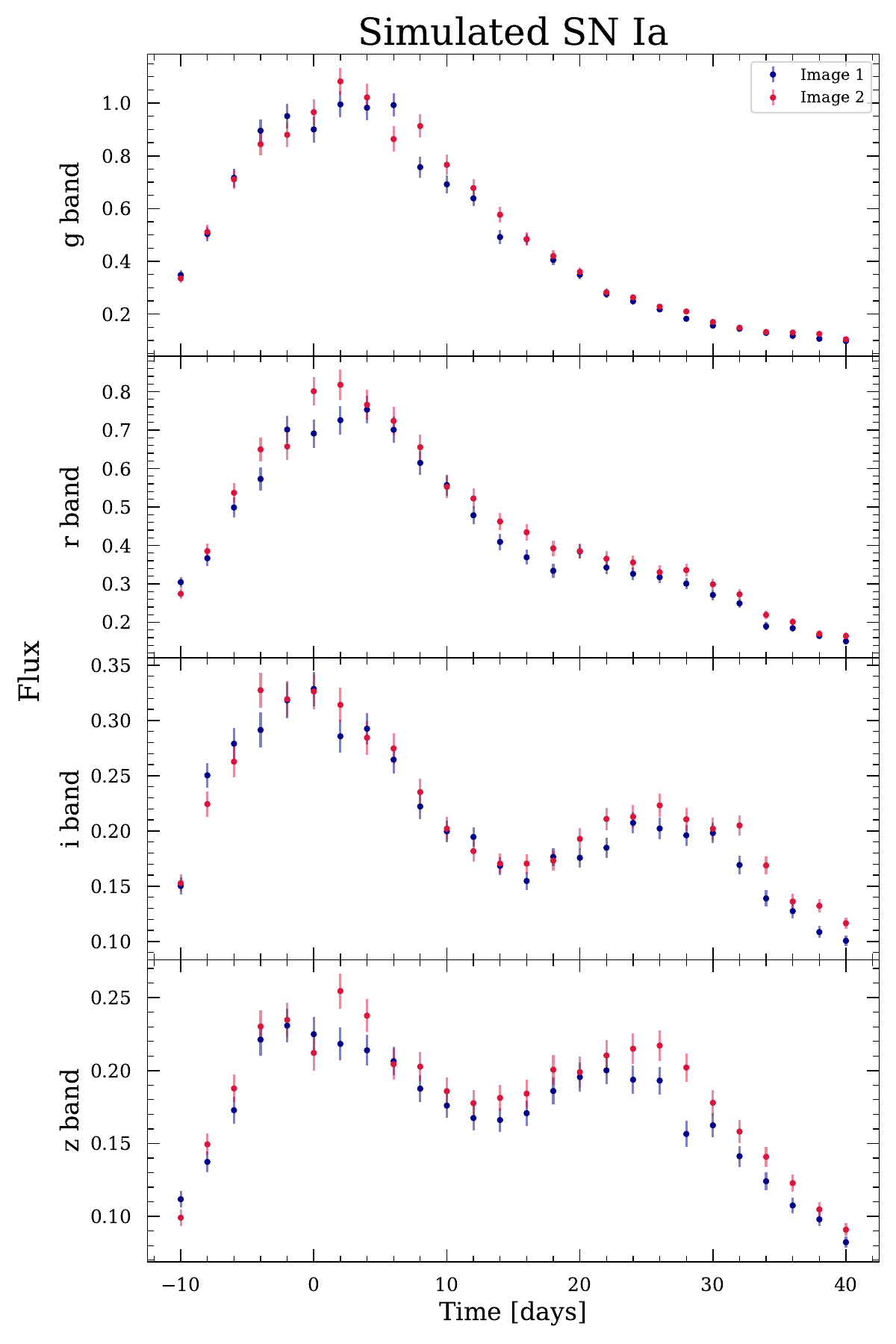}
    \caption{The example data drawn from the underlying light curves shown in the upper panel of Figure \ref{fig:microlensing-example}, where the second image is subject to a time varying magnification function shown in the bottom panel of Figure \ref{fig:microlensing-example}. The light curves are ``observed'' with a 2 day cadence in Rubin-LSST's $griz$ filters. The true time delay is $\Delta = 0$.}
    \label{fig:appendixA-microlensingex}
\end{figure}

To further motivate the inclusion of a time-varying magnification model at the level of time-delay fitting, we demonstrate the error that microlensing can introduce into the time-delay estimate when fitting with a constant magnification model in an example with a known, time-varying magnification term acting on the second image. We sample the light curves of a doubly-imaged SN~Ia at $z=0$, shown in Figure \ref{fig:microlensing-example}, with a two day cadence in Rubin-LSST's $griz$ filters. These images are coincident, with $\Delta_{\text{true}} = 0$, $\beta_{0,\text{true}} = 1$, $\beta_{1,\text{true}} = 0.1$, $r_{\text{true}} = 0.9$, and $t_{0, \text{true}} = 2$. We assume a standard deviation on the error of 5\% of the measured flux. The ``observed'' data from the doubly-imaged system is shown in Figure \ref{fig:appendixA-microlensingex}.

\begin{figure}
    \centering
    \includegraphics[width=\linewidth]{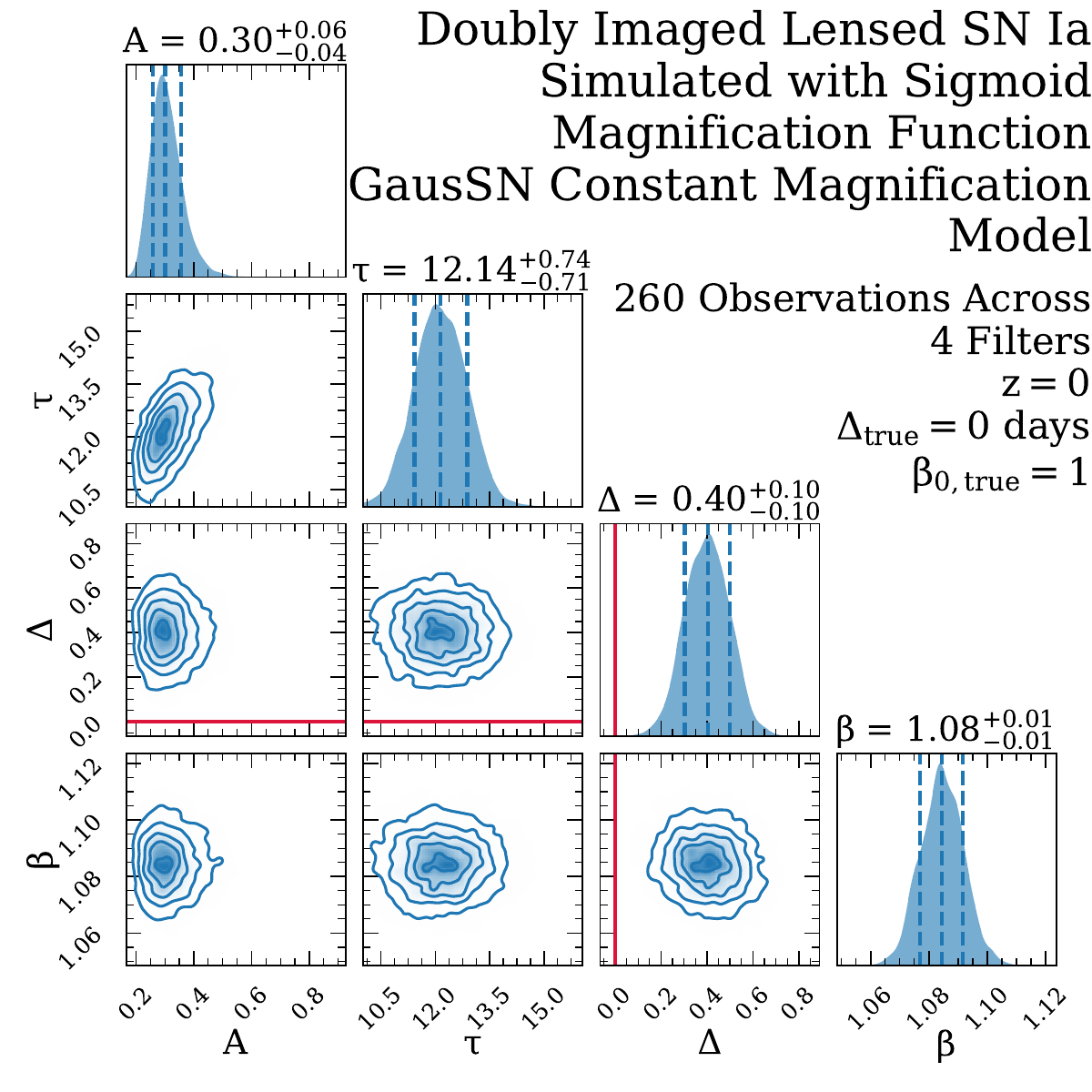}
    \caption{The joint posteriors for the four parameters in the constant magnification model, with the true time delay shown in red. The recovered time delay of $\Delta = 0.40 \pm 0.10$ is 4$\sigma$ from the truth. This bias results from the presence of a time-varying magnification function, which the constant magnification model cannot accommodate.}
    \label{fig:appendixA-corner-constant}
\end{figure}

We first fit the object with the constant magnification model with priors on the parameters of:
\begin{gather}
    A \sim \mathcal{U}(0, 1), \\
    \tau \sim \mathcal{U}(10, 40)\text{ days}, \\
    \Delta \sim \mathcal{U}(-20, 20)\text{ days}, \\
    \beta \sim \mathcal{U}(0.1, 2)
\end{gather}
Figure \ref{fig:appendixA-corner-constant} shows the joint posteriors for the fit. The time delay, $\Delta = 0.42 \pm 0.18$, is overestimated at 4$\sigma$ significance. The magnification is estimated to be $\beta = 1.06 \pm 0.01$, which is approximately the mean of the magnification term. This result shows how the model must compromise the time-delay fit because it lacks flexibility. Therefore, a time-varying magnification source can be a significance source of bias if not accounted for in one's time-delay estimation method.

\begin{figure*}
    \centering
    \includegraphics[width=\linewidth]{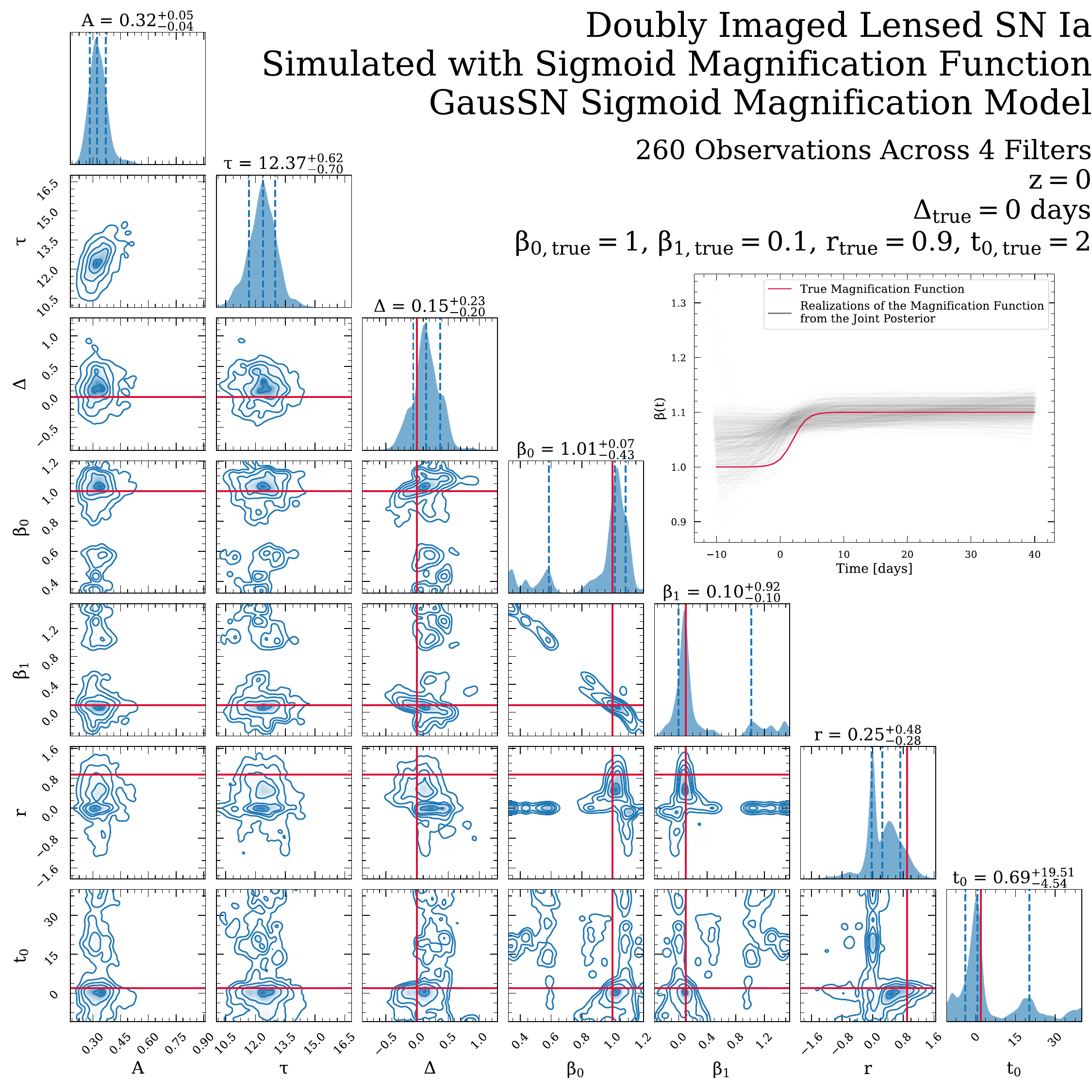}
    \caption{The joint posterior for the mock data shown in Figure \ref{fig:appendixA-microlensingex} for the seven parameters from the sigmoid magnification model. The true values of the time delay and magnification parameters are shown in red. When allowing for a time-varying magnification effect, the time-delay recovery is less biased than when using a constant magnification model (see Figure \ref{fig:appendixA-corner-constant}). The additional flexibility of a time-varying magnification treatment leads to a more accurate posterior distribution for $\Delta$, which better reflects the uncertainty in the estimate due to microlensing. \textbf{Inset:} Realizations of the magnification function based on 1,000 draws from the joint posterior for the time delay and magnification parameters, with the true underlying magnification function over-plotted in red. The fitted magnification function is consistent with the true function, demonstrating \textsc{GausSN}'s ability to constrain the shape of the time-varying magnification acting on the second image.}
    \label{fig:appendixA-corner-sigmoid}
\end{figure*}

We then fit this example system with the sigmoid magnification model to see if there is an improvement in the quality of fit when allowing a more flexible magnification treatment. We use the same priors as above, as well as the following priors for the additional magnification parameters:
\begin{gather}
    \beta_{1} \sim \mathcal{N}(0, 0.5^{2}) \\
    r \sim \mathcal{N}(0, 0.5^{2}), \\
    t_{0} | \, \bm{\hat{T}}, \Delta \sim \mathcal{U}(\min(\bm{\hat{T}}_{\Delta}), \max(\bm{\hat{T}}_{\Delta}))
\end{gather}
The joint posteriors for this fit are shown in Figure \ref{fig:appendixA-corner-sigmoid}. The time delay is estimated to be $\Delta = -0.21^{+0.40}_{-0.51}$, which is both closer to the truth than the constant magnification fit and captures the truth within the $1\sigma$ confidence interval. The estimated uncertainties are larger, but this more appropriately reflects our level of uncertainty due to the difficulty of constraining time-varying magnification from the light curves of the images alone.

Furthermore, we can use this example to check that the magnification curve is well-recovered by \textsc{GausSN}. The inset plot in Figure \ref{fig:appendixA-corner-sigmoid} shows 1,000 realizations of the magnification function based on draws from the posterior distribution of the sigmoid magnification fit to the data. The true magnification function, over-plotted in the figure, is consistent with the fitted magnification curves, demonstrating qualitatively the ability of \textsc{GausSN} to recover the true underlying relative magnification.

%% file: 10_appendixB.tex
A single parameterization of microlensing will not be able to capture the full diversity of effects microlensing can have on glSNe. Therefore, it will be important to test multiple parameterizations of microlensing and marginalize over this choice with Bayesian model averaging to minimize systematics. In this appendix, we provide a brief exploration of a flexible sinusoidal microlensing parameterization. We define the time-varying magnification function as:
\begin{equation}
    \beta(t) = \beta_{0} + \beta_{1}\text{sin}(2\pi(t-t_{0})/\mathbb{T})
\end{equation}
where $\beta_{0}$ controls the macro-scale magnification, $\beta_{1}$ controls the amplitude of the time-varying microlensing magnification effect, $\mathbb{T}$ controls the timescale of the microlensing effect, and $t_{0}$ controls the centering of the microlensing effect. We adopt the following priors on the 7 fitted parameters of the system:
\begin{align}
    A &\sim \mathcal{U}(0, 5) \\
    \tau &\sim \mathcal{U}(10, 40) \\
    \Delta | \, \bm{\hat{T}} &\sim \mathcal{U}(\min(\bm{\hat{t}}_{1}) - \max(\bm{\hat{t}}_{2}), \max(\bm{\hat{t}}_{1}) - \min(\bm{\hat{t}}_{2})) \\
    \beta_{0} &\sim \mathcal{TN}(\mu_{\beta}, 10^{2}, 0, \infty) \\
    \beta_{1} &\sim \mathcal{N}(0, 0.5^{2}) \\
    \mathbb{T}/2\pi &\sim \mathcal{U}(20, 600) \\
    t_{0}|\, \bm{\hat{T}}, \Delta &\sim \mathcal{U}(\min(\bm{\hat{T}}_{\Delta}), \max(\bm{\hat{T}}_{\Delta}))
\end{align}
where $\bm{\hat{T}}$ is given by Equation \ref{eq:time-vector}, $\bm{\hat{T}}_{\Delta}$ is given by Equation \ref{eq:deshifted-time-vector}, and $\mu_{\beta}$, $\bm{\hat{t}}_{1}$, and $\bm{\hat{t}}_{2}$ are defined as in Equations \ref{eq:roman-priors-1}. While the microlensing signal is not expected to be sinusoidal, the smooth and well-behaved curves produced by this function in the regime of timescales of greater than 20 days make them suitable for this application.

\begin{table*}
    \centering
    \caption{Results for the flexible sinusoidal \textsc{GausSN} model on a subset of Rubin-LSST simulations.}
    \begin{threeparttable}
        \begin{tabular}{c|c|c|c|c|c|c|c|c|c}
                & & Sim. & $|\Delta_{\text{fit}} - \Delta_{\text{true}}|$ & $|\Delta_{\text{fit}} - \Delta_{\text{true}}|$ & $|\Delta_{\text{fit}} - \Delta_{\text{true}}|$ & Median & $|\Delta_{\text{fit}} - \Delta_{\text{true}}|/\Delta_{\text{true}}$ & $\Delta_{\text{true}}$ & $\Delta_{\text{true}}$ \\
                SN Type & N & Model\tnote{a} & < 1 day & < 3 days & < 5 days & $\Delta_{\text{fit}} - \Delta_{\text{true}}$ & < 5\% & in 68\% CI\tnote{b} & in 95\% CI\tnote{b} \\
            \hline
            \hline
            Ia & 100 & C & 0.44 & 0.65 & 0.73 & -0.002 & 0.42 & 0.37 & 0.64 \\
             & 103 & A & 0.262 & 0.563 & 0.738 & -0.484 & 0.194 & 0.612 & 0.884 \\
            Ib & 100 & C & 0.35 & 0.62 & 0.76 & -0.213 & 0.4 & 0.53 & 0.82 \\
            Ic & 100 & C & 0.24 & 0.51 & 0.67 & -0.441 & 0.22 & 0.51 & 0.75 \\
            IIb & 100 & C & 0.25 & 0.56 & 0.7 & 0.066 & 0.3 & 0.46 & 0.75 \\
            IIn & 99 & C & 0.303 & 0.687 & 0.788 & 0.572 & 0.343 & 0.535 & 0.768 \\
            IIP & 100 & C & 0.31 & 0.58 & 0.71 & -0.108 & 0.37 & 0.48 & 0.79 \\
            \hline
        \end{tabular}
        \begin{tablenotes}\footnotesize
        \item[a] The model from which the data was simulated, where ``C'' denotes the constant magnification model simulations from this work and ``A'' denotes fully realistic glSN~Ia simulations from \cite{Arendse_2023}.

        \item[b] The fraction of SNe for which $\Delta_{\text{true}}$ falls within the 68\% and 95\% credible intervals (CI) computed from the $16^{\text{th}}$ and $84^{\text{th}}$ percentiles of the posterior samples.
        \end{tablenotes}
    \end{threeparttable}
    \label{tab:appendixB}
\end{table*}

We present the results on a subset of the Rubin-LSST simulations in Table \ref{tab:appendixB}. The results are consistent with those reported for the other magnification models tested, indicating that no strong biases are introduced by this choice of magnification model. With the flexible sinusoidal model, the true time delays fall within the 68\% and 95\% CI of the posteriors for NN\% and NN\% of objects simulated from the constant magnification model, respectively. For the \texttt{lensedSST} simulations, NN\% and NN\% of objects have true time delays that fall within the 68\% and 95\% CI of the posteriors. These values are consistent with what is seen in results from the sigmoid magnification model. Furthermore, the median of the distribution of $\Delta_{\text{fit}} - \Delta_{\text{true}}$ is -0.02 days for the constant magnification simulations overall, although offsets of up to a half a day are seen when broken down by SN type.

There is a slight bias of -0.484 days in the median of the $\Delta_{\text{fit}} - \Delta_{\text{true}}$ distribution for the \texttt{lensedSST} simulations. The fraction of objects with true time delays that fall within the 68\% and 95\% CI -- 61.17\% and 88.35\%, respectively -- remains consistent for the flexible sinusoidal model. Because of the broad consistency of results across multiple magnification models, we conclude that the method is robust for several selection of magnification models.

While the results are consistent with the other choices of magnification models, the uncertainties for all magnification models remain underestimated on both the constant magnification simulations and the \texttt{lensedSST} simulations. The underestimated uncertainties are likely due to the choice of a single microlensing parameterization that cannot capture the full diversity of microlensing effects possible. We reiterate that marginalizing over the choice of microlensing parameterization with Bayesian model averaging, as discussed in \S\ref{sec:discussion}, be necessary to properly account for the uncertainty in modeling the effects of microlensing.

%% file: 00_main.bbl
\begin{thebibliography}{}
\makeatletter
\relax
\def\mn@urlcharsother{\let\do\@makeother \do\$\do\&\do\#\do\^\do\_\do\%\do\~}
\def\mn@doi{\begingroup\mn@urlcharsother \@ifnextchar [ {\mn@doi@}
  {\mn@doi@[]}}
\def\mn@doi@[#1]#2{\def\@tempa{#1}\ifx\@tempa\@empty \href
  {http://dx.doi.org/#2} {doi:#2}\else \href {http://dx.doi.org/#2} {#1}\fi
  \endgroup}
\def\mn@eprint#1#2{\mn@eprint@#1:#2::\@nil}
\def\mn@eprint@arXiv#1{\href {http://arxiv.org/abs/#1} {{\tt arXiv:#1}}}
\def\mn@eprint@dblp#1{\href {http://dblp.uni-trier.de/rec/bibtex/#1.xml}
  {dblp:#1}}
\def\mn@eprint@#1:#2:#3:#4\@nil{\def\@tempa {#1}\def\@tempb {#2}\def\@tempc
  {#3}\ifx \@tempc \@empty \let \@tempc \@tempb \let \@tempb \@tempa \fi \ifx
  \@tempb \@empty \def\@tempb {arXiv}\fi \@ifundefined
  {mn@eprint@\@tempb}{\@tempb:\@tempc}{\expandafter \expandafter \csname
  mn@eprint@\@tempb\endcsname \expandafter{\@tempc}}}

\bibitem[\protect\citeauthoryear{{Arendse} et~al.,}{{Arendse}
  et~al.}{2023}]{Arendse_2023}
{Arendse} N.,  et~al., 2023, preprint, \href
  {https://ui.adsabs.harvard.edu/abs/2023arXiv231204621A} {} (\mn@eprint
  {arXiv} {2312.04621})

\bibitem[\protect\citeauthoryear{{Baklanov}, {Lyskova}, {Blinnikov}  \&
  {Nomoto}}{{Baklanov} et~al.}{2021}]{Baklanov_2021}
{Baklanov} P.,  {Lyskova} N.,  {Blinnikov} S.,   {Nomoto} K.,  2021, \mn@doi
  [\apj] {10.3847/1538-4357/abcd98}, \href
  {https://ui.adsabs.harvard.edu/abs/2021ApJ...907...35B} {907, 35}

\bibitem[\protect\citeauthoryear{{Barbary} et~al.,}{{Barbary}
  et~al.}{2023}]{Barbary_2022}
{Barbary} K.,  et~al., 2023, {SNCosmo}, Zenodo, \mn@doi{10.5281/zenodo.592747}

\bibitem[\protect\citeauthoryear{{Bazin} et~al.,}{{Bazin}
  et~al.}{2009}]{Bazin_2009}
{Bazin} G.,  et~al., 2009, \mn@doi [\aap] {10.1051/0004-6361/200911847}, \href
  {https://ui.adsabs.harvard.edu/abs/2009A&A...499..653B} {499, 653}

\bibitem[\protect\citeauthoryear{{Birrer} \& {Amara}}{{Birrer} \&
  {Amara}}{2018}]{Birrer_2018}
{Birrer} S.,  {Amara} A.,  2018, \mn@doi [Physics of the Dark Universe]
  {10.1016/j.dark.2018.11.002}, \href
  {https://ui.adsabs.harvard.edu/abs/2018PDU....22..189B} {22, 189}

\bibitem[\protect\citeauthoryear{{Birrer} et~al.,}{{Birrer}
  et~al.}{2020}]{Birrer_2020}
{Birrer} S.,  et~al., 2020, \mn@doi [\aap] {10.1051/0004-6361/202038861}, \href
  {https://ui.adsabs.harvard.edu/abs/2020A&A...643A.165B} {643, A165}

\bibitem[\protect\citeauthoryear{{Birrer} et~al.,}{{Birrer}
  et~al.}{2021}]{Birrer_2021}
{Birrer} S.,  et~al., 2021, \mn@doi [J.\ Open Source Software]
  {10.21105/joss.03283}, \href
  {https://ui.adsabs.harvard.edu/abs/2021JOSS....6.3283B} {6, 3283}

\bibitem[\protect\citeauthoryear{{Birrer}, {Dhawan}  \& {Shajib}}{{Birrer}
  et~al.}{2022}]{Birrer_2022}
{Birrer} S.,  {Dhawan} S.,   {Shajib} A.~J.,  2022, \mn@doi [\apj]
  {10.3847/1538-4357/ac323a}, \href
  {https://ui.adsabs.harvard.edu/abs/2022ApJ...924....2B} {924, 2}

\bibitem[\protect\citeauthoryear{{Biswas}, {Setzer}  \& {Azfar}}{{Biswas}
  et~al.}{2019}]{Biswas_2019}
{Biswas} R.,  {Setzer} C.,   {Azfar} F.,  2019, {LSSTDESC/OpSimSummary: 2.0.0},
  Zenodo, \mn@doi{10.5281/zenodo.2671955}

\bibitem[\protect\citeauthoryear{{Biswas}, {Daniel}, {Hlo{\v{z}}ek}, {Kim},
  {Yoachim}  \& {LSST Dark Energy Science Collaboration}}{{Biswas}
  et~al.}{2020}]{Biswas_2020}
{Biswas} R.,  {Daniel} S.~F.,  {Hlo{\v{z}}ek} R.,  {Kim} A.~G.,  {Yoachim} P.,
   {LSST Dark Energy Science Collaboration} 2020, \mn@doi [\apjs]
  {10.3847/1538-4365/ab72f2}, \href
  {https://ui.adsabs.harvard.edu/abs/2020ApJS..247...60B} {247, 60}

\bibitem[\protect\citeauthoryear{{Bonvin} et~al.,}{{Bonvin}
  et~al.}{2017}]{Bonvin_2017}
{Bonvin} V.,  et~al., 2017, \mn@doi [\mnras] {10.1093/mnras/stw3006}, \href
  {https://ui.adsabs.harvard.edu/abs/2017MNRAS.465.4914B} {465, 4914}

\bibitem[\protect\citeauthoryear{{Bonvin} et~al.,}{{Bonvin}
  et~al.}{2018}]{Bonvin_2018}
{Bonvin} V.,  et~al., 2018, \mn@doi [\aap] {10.1051/0004-6361/201833287}, \href
  {https://ui.adsabs.harvard.edu/abs/2018A&A...616A.183B} {616, A183}

\bibitem[\protect\citeauthoryear{{Bonvin}, {Tihhonova}, {Millon}, {Chan},
  {Savary}, {Huber}  \& {Courbin}}{{Bonvin}
  et~al.}{2019a}]{Bonvin_2019_microlens}
{Bonvin} V.,  {Tihhonova} O.,  {Millon} M.,  {Chan} J.~H.~H.,  {Savary} E.,
  {Huber} S.,   {Courbin} F.,  2019a, \mn@doi [\aap]
  {10.1051/0004-6361/201833405}, \href
  {https://ui.adsabs.harvard.edu/abs/2019A&A...621A..55B} {621, A55}

\bibitem[\protect\citeauthoryear{{Bonvin} et~al.,}{{Bonvin}
  et~al.}{2019b}]{Bonvin_2019_lensedqso}
{Bonvin} V.,  et~al., 2019b, \mn@doi [\aap] {10.1051/0004-6361/201935921},
  \href {https://ui.adsabs.harvard.edu/abs/2019A&A...629A..97B} {629, A97}

\bibitem[\protect\citeauthoryear{{Boone}}{{Boone}}{2019}]{Boone_2019}
{Boone} K.,  2019, \mn@doi [\aj] {10.3847/1538-3881/ab5182}, \href
  {https://ui.adsabs.harvard.edu/abs/2019AJ....158..257B} {158, 257}

\bibitem[\protect\citeauthoryear{{Chen} et~al.,}{{Chen}
  et~al.}{2022}]{Chen_2022}
{Chen} W.,  et~al., 2022, \mn@doi [\nat] {10.1038/s41586-022-05252-5}, \href
  {https://ui.adsabs.harvard.edu/abs/2022Natur.611..256C} {611, 256}

\bibitem[\protect\citeauthoryear{{Chen} et~al.,}{{Chen}
  et~al.}{2024}]{Chen_2024}
{Chen} W.,  et~al., 2024, preprint, \href
  {https://ui.adsabs.harvard.edu/abs/2024arXiv240319029C} {} (\mn@eprint
  {arXiv} {2403.19029})

\bibitem[\protect\citeauthoryear{{Chornock} et~al.,}{{Chornock}
  et~al.}{2013}]{Chornock_2013}
{Chornock} R.,  et~al., 2013, \mn@doi [\apj] {10.1088/0004-637X/767/2/162},
  \href {https://ui.adsabs.harvard.edu/abs/2013ApJ...767..162C} {767, 162}

\bibitem[\protect\citeauthoryear{{Craig}, {O'Connor}, {Chakrabarti}, {Rodney},
  {Pierel}, {McCully}  \& {Perez-Fournon}}{{Craig} et~al.}{2021}]{Craig_2021}
{Craig} P.,  {O'Connor} K.,  {Chakrabarti} S.,  {Rodney} S.~A.,  {Pierel}
  J.~R.,  {McCully} C.,   {Perez-Fournon} I.,  2021, preprint, \href
  {https://ui.adsabs.harvard.edu/abs/2021arXiv211101680C} {} (\mn@eprint
  {arXiv} {2111.01680})

\bibitem[\protect\citeauthoryear{Delgado, Saha, Chandrasekharan, Cook, Petry
  \& Ridgway}{Delgado et~al.}{2014}]{Delgado_2014}
Delgado F.,  Saha A.,  Chandrasekharan S.,  Cook K.,  Petry C.,   Ridgway S.,
  2014, in Angeli G.~Z.,  Dierickx P.,  eds,  Proceedings Volume 9150 Vol.
  9150, Modeling, Systems Engineering, and Project Management for Astronomy VI.
  SPIE, p. 915015, \mn@doi{10.1117/12.2056898}

\bibitem[\protect\citeauthoryear{{Di Valentino} et~al.,}{{Di Valentino}
  et~al.}{2021}]{DiValentino_2021}
{Di Valentino} E.,  et~al., 2021, \mn@doi [Classical \& Quantum Gravity]
  {10.1088/1361-6382/ac086d}, \href
  {https://ui.adsabs.harvard.edu/abs/2021CQGra..38o3001D} {38, 153001}

\bibitem[\protect\citeauthoryear{{Diego} et~al.,}{{Diego}
  et~al.}{2016}]{Diego_2016}
{Diego} J.~M.,  et~al., 2016, \mn@doi [\mnras] {10.1093/mnras/stv2638}, \href
  {https://ui.adsabs.harvard.edu/abs/2016MNRAS.456..356D} {456, 356}

\bibitem[\protect\citeauthoryear{{Falco}, {Gorenstein}  \& {Shapiro}}{{Falco}
  et~al.}{1985}]{Falco_1985}
{Falco} E.~E.,  {Gorenstein} M.~V.,   {Shapiro} I.~I.,  1985, \mn@doi [\apjl]
  {10.1086/184422}, \href
  {https://ui.adsabs.harvard.edu/abs/1985ApJ...289L...1F} {289, L1}

\bibitem[\protect\citeauthoryear{{Feroz}, {Hobson}  \& {Bridges}}{{Feroz}
  et~al.}{2009}]{Feroz_2009}
{Feroz} F.,  {Hobson} M.~P.,   {Bridges} M.,  2009, \mn@doi [\mnras]
  {10.1111/j.1365-2966.2009.14548.x}, \href
  {https://ui.adsabs.harvard.edu/abs/2009MNRAS.398.1601F} {398, 1601}

\bibitem[\protect\citeauthoryear{{Filippenko}}{{Filippenko}}{1997}]{Filippenko_1997}
{Filippenko} A.~V.,  1997, \mn@doi [\araa] {10.1146/annurev.astro.35.1.309},
  \href {https://ui.adsabs.harvard.edu/abs/1997ARA&A..35..309F} {35, 309}

\bibitem[\protect\citeauthoryear{{Foreman-Mackey}, {Hogg}, {Lang}  \&
  {Goodman}}{{Foreman-Mackey} et~al.}{2013}]{ForemanMackey_2013}
{Foreman-Mackey} D.,  {Hogg} D.~W.,  {Lang} D.,   {Goodman} J.,  2013, \mn@doi
  [\pasp] {10.1086/670067}, \href
  {https://ui.adsabs.harvard.edu/abs/2013PASP..125..306F} {125, 306}

\bibitem[\protect\citeauthoryear{{Foxley-Marrable}, {Collett}, {Vernardos},
  {Goldstein}  \& {Bacon}}{{Foxley-Marrable}
  et~al.}{2018}]{FoxleyMarrable_2018}
{Foxley-Marrable} M.,  {Collett} T.~E.,  {Vernardos} G.,  {Goldstein} D.~A.,
  {Bacon} D.,  2018, \mn@doi [\mnras] {10.1093/mnras/sty1346}, \href
  {https://ui.adsabs.harvard.edu/abs/2018MNRAS.478.5081F} {478, 5081}

\bibitem[\protect\citeauthoryear{{Frye} et~al.,}{{Frye}
  et~al.}{2023}]{Frye_2023}
{Frye} B.,  et~al., 2023, Transient Name Server AstroNote, \href
  {https://ui.adsabs.harvard.edu/abs/2023TNSAN..96....1F} {96, 1}

\bibitem[\protect\citeauthoryear{Frye et~al.,}{Frye
  et~al.}{2024}]{Frye_2023_ApJ}
Frye B.~L.,  et~al., 2024, \mn@doi [The Astrophysical Journal]
  {10.3847/1538-4357/ad1034}, 961, 171

\bibitem[\protect\citeauthoryear{{Gal-Yam}}{{Gal-Yam}}{2017}]{GalYam_2017}
{Gal-Yam} A.,  2017, in {Alsabti} A.~W.,  {Murdin} P.,  eds, , Handbook of
  Supernovae.
Springer, p.~195, \mn@doi{10.1007/978-3-319-21846-5_35}

\bibitem[\protect\citeauthoryear{{Gibbs}}{{Gibbs}}{1997}]{Gibbs_1997}
{Gibbs} M.~N.,  1997, PhD thesis, Cambridge University

\bibitem[\protect\citeauthoryear{{Goldstein}, {Nugent}, {Kasen}  \&
  {Collett}}{{Goldstein} et~al.}{2018}]{Goldstein_2018}
{Goldstein} D.~A.,  {Nugent} P.~E.,  {Kasen} D.~N.,   {Collett} T.~E.,  2018,
  \mn@doi [\apj] {10.3847/1538-4357/aaa975}, \href
  {https://ui.adsabs.harvard.edu/abs/2018ApJ...855...22G} {855, 22}

\bibitem[\protect\citeauthoryear{{Goldstein}, {Nugent}  \&
  {Goobar}}{{Goldstein} et~al.}{2019}]{Goldstein_2019}
{Goldstein} D.~A.,  {Nugent} P.~E.,   {Goobar} A.,  2019, \mn@doi [\apjs]
  {10.3847/1538-4365/ab1fe0}, \href
  {https://ui.adsabs.harvard.edu/abs/2019ApJS..243....6G} {243, 6}

\bibitem[\protect\citeauthoryear{{Goobar} et~al.,}{{Goobar}
  et~al.}{2017}]{Goobar_2017}
{Goobar} A.,  et~al., 2017, \mn@doi [Science] {10.1126/science.aal2729}, \href
  {https://ui.adsabs.harvard.edu/abs/2017Sci...356..291G} {356, 291}

\bibitem[\protect\citeauthoryear{{Goobar} et~al.,}{{Goobar}
  et~al.}{2023a}]{Goobar_2022}
{Goobar} A.,  et~al., 2023a, \mn@doi [Nature Astronomy]
  {10.1038/s41550-023-01981-3}, \href
  {https://ui.adsabs.harvard.edu/abs/2023NatAs...7.1098G} {7, 1098}

\bibitem[\protect\citeauthoryear{{Goobar} et~al.,}{{Goobar}
  et~al.}{2023b}]{Goobar_2023}
{Goobar} A.,  et~al., 2023b, \mn@doi [Nature Astronomy]
  {10.1038/s41550-023-02034-5}, \href
  {https://ui.adsabs.harvard.edu/abs/2023NatAs...7.1137G} {7, 1137}

\bibitem[\protect\citeauthoryear{{Goodman} \& {Weare}}{{Goodman} \&
  {Weare}}{2010}]{Goodman_2010}
{Goodman} J.,  {Weare} J.,  2010, \mn@doi [Communications in Applied Math.\ \&
  Comput.\ Sci.] {10.2140/camcos.2010.5.65}, \href
  {https://ui.adsabs.harvard.edu/abs/2010CAMCS...5...65G} {5, 65}

\bibitem[\protect\citeauthoryear{{Grayling} et~al.,}{{Grayling}
  et~al.}{2023}]{Grayling_2023}
{Grayling} M.,  et~al., 2023, \mn@doi [\mnras] {10.1093/mnras/stad056}, \href
  {https://ui.adsabs.harvard.edu/abs/2023MNRAS.520..684G} {520, 684}

\bibitem[\protect\citeauthoryear{{Grillo} et~al.,}{{Grillo}
  et~al.}{2016}]{Grillo_2016}
{Grillo} C.,  et~al., 2016, \mn@doi [\apj] {10.3847/0004-637X/822/2/78}, \href
  {https://ui.adsabs.harvard.edu/abs/2016ApJ...822...78G} {822, 78}

\bibitem[\protect\citeauthoryear{{Guy} et~al.,}{{Guy} et~al.}{2007}]{Guy_2007}
{Guy} J.,  et~al., 2007, \mn@doi [\aap] {10.1051/0004-6361:20066930}, \href
  {https://ui.adsabs.harvard.edu/abs/2007A&A...466...11G} {466, 11}

\bibitem[\protect\citeauthoryear{{Handley}, {Hobson}  \& {Lasenby}}{{Handley}
  et~al.}{2015a}]{Handley_2015a}
{Handley} W.~J.,  {Hobson} M.~P.,   {Lasenby} A.~N.,  2015a, \mn@doi [\mnras]
  {10.1093/mnrasl/slv047}, \href
  {https://ui.adsabs.harvard.edu/abs/2015MNRAS.450L..61H} {450, L61}

\bibitem[\protect\citeauthoryear{{Handley}, {Hobson}  \& {Lasenby}}{{Handley}
  et~al.}{2015b}]{Handley_2015b}
{Handley} W.~J.,  {Hobson} M.~P.,   {Lasenby} A.~N.,  2015b, \mn@doi [\mnras]
  {10.1093/mnras/stv1911}, \href
  {https://ui.adsabs.harvard.edu/abs/2015MNRAS.453.4384H} {453, 4384}

\bibitem[\protect\citeauthoryear{{Hojjati} \& {Linder}}{{Hojjati} \&
  {Linder}}{2014}]{Hojjati_2014}
{Hojjati} A.,  {Linder} E.~V.,  2014, \mn@doi [\prd]
  {10.1103/PhysRevD.90.123501}, \href
  {https://ui.adsabs.harvard.edu/abs/2014PhRvD..90l3501H} {90, 123501}

\bibitem[\protect\citeauthoryear{{Hojjati}, {Kim}  \& {Linder}}{{Hojjati}
  et~al.}{2013}]{Hojjati_2013}
{Hojjati} A.,  {Kim} A.~G.,   {Linder} E.~V.,  2013, \mn@doi [\prd]
  {10.1103/PhysRevD.87.123512}, \href
  {https://ui.adsabs.harvard.edu/abs/2013PhRvD..87l3512H} {87, 123512}

\bibitem[\protect\citeauthoryear{{Hounsell} et~al.,}{{Hounsell}
  et~al.}{2018}]{Hounsell_2018}
{Hounsell} R.,  et~al., 2018, \mn@doi [\apj] {10.3847/1538-4357/aac08b}, \href
  {https://ui.adsabs.harvard.edu/abs/2018ApJ...867...23H} {867, 23}

\bibitem[\protect\citeauthoryear{{Hsiao}, {Conley}, {Howell}, {Sullivan},
  {Pritchet}, {Carlberg}, {Nugent}  \& {Phillips}}{{Hsiao}
  et~al.}{2007}]{Hsiao_2007}
{Hsiao} E.~Y.,  {Conley} A.,  {Howell} D.~A.,  {Sullivan} M.,  {Pritchet}
  C.~J.,  {Carlberg} R.~G.,  {Nugent} P.~E.,   {Phillips} M.~M.,  2007, \mn@doi
  [\apj] {10.1086/518232}, \href
  {https://ui.adsabs.harvard.edu/abs/2007ApJ...663.1187H} {663, 1187}

\bibitem[\protect\citeauthoryear{{Hu} \& {Tak}}{{Hu} \& {Tak}}{2020}]{Hu_2020}
{Hu} Z.,  {Tak} H.,  2020, \mn@doi [\aj] {10.3847/1538-3881/abc1e2}, \href
  {https://ui.adsabs.harvard.edu/abs/2020AJ....160..265H} {160, 265}

\bibitem[\protect\citeauthoryear{{Huber} et~al.,}{{Huber}
  et~al.}{2019}]{Huber_2019}
{Huber} S.,  et~al., 2019, \mn@doi [\aap] {10.1051/0004-6361/201935370}, \href
  {https://ui.adsabs.harvard.edu/abs/2019A&A...631A.161H} {631, A161}

\bibitem[\protect\citeauthoryear{{Huber}, {Suyu}, {Noebauer}, {Chan}, {Kromer},
  {Sim}, {Sluse}  \& {Taubenberger}}{{Huber} et~al.}{2021}]{Huber_2021}
{Huber} S.,  {Suyu} S.~H.,  {Noebauer} U.~M.,  {Chan} J.~H.~H.,  {Kromer} M.,
  {Sim} S.~A.,  {Sluse} D.,   {Taubenberger} S.,  2021, \mn@doi [\aap]
  {10.1051/0004-6361/202039218}, \href
  {https://ui.adsabs.harvard.edu/abs/2021A&A...646A.110H} {646, A110}

\bibitem[\protect\citeauthoryear{{Jauzac} et~al.,}{{Jauzac}
  et~al.}{2016}]{Jauzac_2016}
{Jauzac} M.,  et~al., 2016, \mn@doi [\mnras] {10.1093/mnras/stw069}, \href
  {https://ui.adsabs.harvard.edu/abs/2016MNRAS.457.2029J} {457, 2029}

\bibitem[\protect\citeauthoryear{{Karamanis} \& {Beutler}}{{Karamanis} \&
  {Beutler}}{2021}]{karamanis2020ensemble}
{Karamanis} M.,  {Beutler} F.,  2021, \mn@doi [Statistics \& Computing]
  {10.1007/s11222-021-10038-2}, \href
  {https://ui.adsabs.harvard.edu/abs/2020arXiv200206212K} {31, 61}

\bibitem[\protect\citeauthoryear{{Karamanis}, {Beutler}  \&
  {Peacock}}{{Karamanis} et~al.}{2021}]{karamanis2021zeus}
{Karamanis} M.,  {Beutler} F.,   {Peacock} J.~A.,  2021, \mn@doi [\mnras]
  {10.1093/mnras/stab2867}, \href
  {https://ui.adsabs.harvard.edu/abs/2021MNRAS.508.3589K} {508, 3589}

\bibitem[\protect\citeauthoryear{{Karpenka}, {Feroz}  \& {Hobson}}{{Karpenka}
  et~al.}{2013}]{Karpenka_2013}
{Karpenka} N.~V.,  {Feroz} F.,   {Hobson} M.~P.,  2013, \mn@doi [\mnras]
  {10.1093/mnras/sts412}, \href
  {https://ui.adsabs.harvard.edu/abs/2013MNRAS.429.1278K} {429, 1278}

\bibitem[\protect\citeauthoryear{{Kawamata}, {Oguri}, {Ishigaki}, {Shimasaku}
  \& {Ouchi}}{{Kawamata} et~al.}{2016}]{Kawamata_2016}
{Kawamata} R.,  {Oguri} M.,  {Ishigaki} M.,  {Shimasaku} K.,   {Ouchi} M.,
  2016, \mn@doi [\apj] {10.3847/0004-637X/819/2/114}, \href
  {https://ui.adsabs.harvard.edu/abs/2016ApJ...819..114K} {819, 114}

\bibitem[\protect\citeauthoryear{{Keeton} \& {Kochanek}}{{Keeton} \&
  {Kochanek}}{1997}]{Keeton_1997}
{Keeton} C.~R.,  {Kochanek} C.~S.,  1997, \mn@doi [\apj] {10.1086/304583},
  \href {https://ui.adsabs.harvard.edu/abs/1997ApJ...487...42K} {487, 42}

\bibitem[\protect\citeauthoryear{{Kelly} et~al.,}{{Kelly}
  et~al.}{2015}]{Kelly_2015}
{Kelly} P.~L.,  et~al., 2015, \mn@doi [Science] {10.1126/science.aaa3350},
  \href {https://ui.adsabs.harvard.edu/abs/2015Sci...347.1123K} {347, 1123}

\bibitem[\protect\citeauthoryear{{Kelly} et~al.,}{{Kelly}
  et~al.}{2016a}]{Kelly_2016_reappearance}
{Kelly} P.~L.,  et~al., 2016a, \mn@doi [\apjl] {10.3847/2041-8205/819/1/L8},
  \href {https://ui.adsabs.harvard.edu/abs/2016ApJ...819L...8K} {819, L8}

\bibitem[\protect\citeauthoryear{{Kelly} et~al.,}{{Kelly}
  et~al.}{2016b}]{Kelly_2016_spec}
{Kelly} P.~L.,  et~al., 2016b, \mn@doi [\apj] {10.3847/0004-637X/831/2/205},
  \href {https://ui.adsabs.harvard.edu/abs/2016ApJ...831..205K} {831, 205}

\bibitem[\protect\citeauthoryear{{Kelly} et~al.,}{{Kelly}
  et~al.}{2022}]{Kelly_2022_riv}
{Kelly} P.,  et~al., 2022, Transient Name Server Discovery Report, \href
  {https://ui.adsabs.harvard.edu/abs/2022TNSTR2356....1K} {2022-2356, 1}

\bibitem[\protect\citeauthoryear{{Kelly} et~al.,}{{Kelly}
  et~al.}{2023a}]{Kelly_2023_Science}
{Kelly} P.~L.,  et~al., 2023a, \mn@doi [Science] {10.1126/science.abh1322},
  \href {https://ui.adsabs.harvard.edu/abs/2023Sci...380.1322K} {380, abh1322}

\bibitem[\protect\citeauthoryear{{Kelly} et~al.,}{{Kelly}
  et~al.}{2023b}]{Kelly_2023_ApJ}
{Kelly} P.~L.,  et~al., 2023b, \mn@doi [\apj] {10.3847/1538-4357/ac4ccb}, \href
  {https://ui.adsabs.harvard.edu/abs/2023ApJ...948...93K} {948, 93}

\bibitem[\protect\citeauthoryear{Kenworthy et~al.,}{Kenworthy
  et~al.}{2021}]{Kenworthy_2021}
Kenworthy W.~D.,  et~al., 2021, \mn@doi [\apj] {10.3847/1538-4357/ac30d8}, 923,
  265

\bibitem[\protect\citeauthoryear{{Kim} et~al.,}{{Kim} et~al.}{2013}]{Kim_2013}
{Kim} A.~G.,  et~al., 2013, \mn@doi [\apj] {10.1088/0004-637X/766/2/84}, \href
  {https://ui.adsabs.harvard.edu/abs/2013ApJ...766...84K} {766, 84}

\bibitem[\protect\citeauthoryear{{Kromer} \& {Sim}}{{Kromer} \&
  {Sim}}{2009}]{Kromer_2009}
{Kromer} M.,  {Sim} S.~A.,  2009, \mn@doi [\mnras]
  {10.1111/j.1365-2966.2009.15256.x}, \href
  {https://ui.adsabs.harvard.edu/abs/2009MNRAS.398.1809K} {398, 1809}

\bibitem[\protect\citeauthoryear{{Meyer}, {van Dyk}, {Tak}  \&
  {Siemiginowska}}{{Meyer} et~al.}{2023}]{Meyer_2023}
{Meyer} A.~D.,  {van Dyk} D.~A.,  {Tak} H.,   {Siemiginowska} A.,  2023,
  \mn@doi [\apj] {10.3847/1538-4357/acbea1}, \href
  {https://ui.adsabs.harvard.edu/abs/2023ApJ...950...37M} {950, 37}

\bibitem[\protect\citeauthoryear{Millon, Tewes, Bonvin, Lengen  \&
  Courbin}{Millon et~al.}{2020}]{Millon_2020}
Millon M.,  Tewes M.,  Bonvin V.,  Lengen B.,   Courbin F.,  2020, \mn@doi [J.\
  Open Source Software] {10.21105/joss.02654}, 5, 2654

\bibitem[\protect\citeauthoryear{{M{\"o}rtsell} \& {Dhawan}}{{M{\"o}rtsell} \&
  {Dhawan}}{2018}]{Mortsell_2018}
{M{\"o}rtsell} E.,  {Dhawan} S.,  2018, \mn@doi [\jcap]
  {10.1088/1475-7516/2018/09/025}, \href
  {https://ui.adsabs.harvard.edu/abs/2018JCAP...09..025M} {2018, 025}

\bibitem[\protect\citeauthoryear{{Naghib}, {Yoachim}, {Vanderbei}, {Connolly}
  \& {Jones}}{{Naghib} et~al.}{2019}]{Naghib_2019}
{Naghib} E.,  {Yoachim} P.,  {Vanderbei} R.~J.,  {Connolly} A.~J.,   {Jones}
  R.~L.,  2019, \mn@doi [\aj] {10.3847/1538-3881/aafece}, \href
  {https://ui.adsabs.harvard.edu/abs/2019AJ....157..151N} {157, 151}

\bibitem[\protect\citeauthoryear{Neal}{Neal}{2003}]{Neal_2003}
Neal R.~M.,  2003, \mn@doi [Annals of Statistics] {10.1214/aos/1056562461}, 31,
  705

\bibitem[\protect\citeauthoryear{{Nicolas} et~al.,}{{Nicolas}
  et~al.}{2021}]{Nicolas_2021}
{Nicolas} N.,  et~al., 2021, \mn@doi [\aap] {10.1051/0004-6361/202038447},
  \href {https://ui.adsabs.harvard.edu/abs/2021A&A...649A..74N} {649, A74}

\bibitem[\protect\citeauthoryear{{Nixon}, {Welbanks}, {McGill}  \&
  {Kempton}}{{Nixon} et~al.}{2023}]{Nixon_2023}
{Nixon} M.~C.,  {Welbanks} L.,  {McGill} P.,   {Kempton} E. M.~R.,  2023,
  preprint, \href {https://ui.adsabs.harvard.edu/abs/2023arXiv231003713N} {}
  (\mn@eprint {arXiv} {2310.03713})

\bibitem[\protect\citeauthoryear{{Oguri}}{{Oguri}}{2015}]{Oguri_2015}
{Oguri} M.,  2015, \mn@doi [\mnras] {10.1093/mnrasl/slv025}, \href
  {https://ui.adsabs.harvard.edu/abs/2015MNRAS.449L..86O} {449, L86}

\bibitem[\protect\citeauthoryear{{Pascale} et~al.,}{{Pascale}
  et~al.}{2024}]{Pascale_2024}
{Pascale} M.,  et~al., 2024, preprint, \href
  {https://ui.adsabs.harvard.edu/abs/2024arXiv240318902P} {} (\mn@eprint
  {arXiv} {2403.18902})

\bibitem[\protect\citeauthoryear{{Pierel} \& {Rodney}}{{Pierel} \&
  {Rodney}}{2019}]{Pierel_2019}
{Pierel} J.~D.~R.,  {Rodney} S.,  2019, \mn@doi [\apj]
  {10.3847/1538-4357/ab164a}, \href
  {https://ui.adsabs.harvard.edu/abs/2019ApJ...876..107P} {876, 107}

\bibitem[\protect\citeauthoryear{{Pierel} et~al.,}{{Pierel}
  et~al.}{2018}]{Pierel_2018}
{Pierel} J.~D.~R.,  et~al., 2018, \mn@doi [\pasp] {10.1088/1538-3873/aadb7a},
  \href {https://ui.adsabs.harvard.edu/abs/2018PASP..130k4504P} {130, 114504}

\bibitem[\protect\citeauthoryear{{Pierel}, {Rodney}, {Vernardos}, {Oguri},
  {Kessler}  \& {Anguita}}{{Pierel} et~al.}{2021}]{Pierel_2021}
{Pierel} J.~D.~R.,  {Rodney} S.,  {Vernardos} G.,  {Oguri} M.,  {Kessler} R.,
  {Anguita} T.,  2021, \mn@doi [\apj] {10.3847/1538-4357/abd8d3}, \href
  {https://ui.adsabs.harvard.edu/abs/2021ApJ...908..190P} {908, 190}

\bibitem[\protect\citeauthoryear{{Pierel} et~al.}{{Pierel}
  et~al.}{2023a}]{Pierel_2023_Encore}
{Pierel} J.,  et~al., 2023a, {Lensed Supernova Encore at $z=2$! The First
  Galaxy to Host Two Multiply-Imaged Supernovae}, JWST Proposal. Cycle 2, ID.
  \#6549, \url
  {https://www.stsci.edu/jwst/science-execution/program-information?id=6549}

\bibitem[\protect\citeauthoryear{{Pierel} et~al.,}{{Pierel}
  et~al.}{2023b}]{Pierel_2022}
{Pierel} J.~D.~R.,  et~al., 2023b, \mn@doi [\apj] {10.3847/1538-4357/acc7a6},
  \href {https://ui.adsabs.harvard.edu/abs/2023ApJ...948..115P} {948, 115}

\bibitem[\protect\citeauthoryear{{Pierel} et~al.,}{{Pierel}
  et~al.}{2024}]{Pierel_2024}
{Pierel} J.~D.~R.,  et~al., 2024, preprint, \href
  {https://ui.adsabs.harvard.edu/abs/2024arXiv240318954P} {} (\mn@eprint
  {arXiv} {2403.18954})

\bibitem[\protect\citeauthoryear{{Planck Collaboration} et~al.,}{{Planck
  Collaboration} et~al.}{2020}]{Planck_2020}
{Planck Collaboration} et~al., 2020, \mn@doi [\aap]
  {10.1051/0004-6361/201833910}, \href
  {https://ui.adsabs.harvard.edu/abs/2020A&A...641A...6P} {641, A6}

\bibitem[\protect\citeauthoryear{{Polletta} et~al.,}{{Polletta}
  et~al.}{2023}]{Polletta_2023}
{Polletta} M.,  et~al., 2023, \mn@doi [\aap] {10.1051/0004-6361/202346964},
  \href {https://ui.adsabs.harvard.edu/abs/2023A&A...675L...4P} {675, L4}

\bibitem[\protect\citeauthoryear{{Qu}, {Sako}, {M{\"o}ller}  \& {Doux}}{{Qu}
  et~al.}{2021}]{Qu_2021}
{Qu} H.,  {Sako} M.,  {M{\"o}ller} A.,   {Doux} C.,  2021, \mn@doi [\aj]
  {10.3847/1538-3881/ac0824}, \href
  {https://ui.adsabs.harvard.edu/abs/2021AJ....162...67Q} {162, 67}

\bibitem[\protect\citeauthoryear{{Quimby} et~al.,}{{Quimby}
  et~al.}{2013}]{Quimby_2013}
{Quimby} R.~M.,  et~al., 2013, \mn@doi [\apjl] {10.1088/2041-8205/768/1/L20},
  \href {https://ui.adsabs.harvard.edu/abs/2013ApJ...768L..20Q} {768, L20}

\bibitem[\protect\citeauthoryear{{Rasmussen} \& {Williams}}{{Rasmussen} \&
  {Williams}}{2006}]{Rasmussen_2006}
{Rasmussen} C.~E.,  {Williams} C. K.~I.,  2006, {Gaussian Processes for Machine
  Learning}.
MIT Press

\bibitem[\protect\citeauthoryear{{Refsdal}}{{Refsdal}}{1964}]{Refsdal_1964}
{Refsdal} S.,  1964, \mn@doi [\mnras] {10.1093/mnras/128.4.307}, \href
  {https://ui.adsabs.harvard.edu/abs/1964MNRAS.128..307R} {128, 307}

\bibitem[\protect\citeauthoryear{{Revsbech}, {Trotta}  \& {van Dyk}}{{Revsbech}
  et~al.}{2018}]{Revsbech_2018}
{Revsbech} E.~A.,  {Trotta} R.,   {van Dyk} D.~A.,  2018, \mn@doi [\mnras]
  {10.1093/mnras/stx2570}, \href
  {https://ui.adsabs.harvard.edu/abs/2018MNRAS.473.3969R} {473, 3969}

\bibitem[\protect\citeauthoryear{{Riess} et~al.,}{{Riess}
  et~al.}{2022}]{Riess_2022}
{Riess} A.~G.,  et~al., 2022, \mn@doi [\apjl] {10.3847/2041-8213/ac5c5b}, \href
  {https://ui.adsabs.harvard.edu/abs/2022ApJ...934L...7R} {934, L7}

\bibitem[\protect\citeauthoryear{{Rodney} et~al.,}{{Rodney}
  et~al.}{2016}]{Rodney_2016}
{Rodney} S.~A.,  et~al., 2016, \mn@doi [\apj] {10.3847/0004-637X/820/1/50},
  \href {https://ui.adsabs.harvard.edu/abs/2016ApJ...820...50R} {820, 50}

\bibitem[\protect\citeauthoryear{{Rodney}, {Brammer}, {Pierel}, {Richard},
  {Toft}, {O'Connor}, {Akhshik}  \& {Whitaker}}{{Rodney}
  et~al.}{2021}]{Rodney_2021}
{Rodney} S.~A.,  {Brammer} G.~B.,  {Pierel} J. D.~R.,  {Richard} J.,  {Toft}
  S.,  {O'Connor} K.~F.,  {Akhshik} M.,   {Whitaker} K.~E.,  2021, \mn@doi
  [Nature Astronomy] {10.1038/s41550-021-01450-9}, \href
  {https://ui.adsabs.harvard.edu/abs/2021NatAs...5.1118R} {5, 1118}

\bibitem[\protect\citeauthoryear{{Rubin Observatory Survey Cadence Optimization
  Committee}}{{Rubin Observatory Survey Cadence Optimization
  Committee}}{2023}]{PSTN-055}
{Rubin Observatory Survey Cadence Optimization Committee} 2023, {Survey Cadence
  Optimization Committee’s Phase 2 Recommendations}, \url
  {https://pstn-055.lsst.io/}

\bibitem[\protect\citeauthoryear{{Sharon} \& {Johnson}}{{Sharon} \&
  {Johnson}}{2015}]{Sharon_2015}
{Sharon} K.,  {Johnson} T.~L.,  2015, \mn@doi [\apjl]
  {10.1088/2041-8205/800/2/L26}, \href
  {https://ui.adsabs.harvard.edu/abs/2015ApJ...800L..26S} {800, L26}

\bibitem[\protect\citeauthoryear{{Skilling}}{{Skilling}}{2004}]{Skilling_2004}
{Skilling} J.,  2004, in {Fischer} R.,  {Preuss} R.,   {Toussaint} U.~V.,  eds,
   American Institute of Physics Conference Series Vol. 735, Bayesian Inference
  and Maximum Entropy Methods in Science and Engineering: 24th International
  Workshop on Bayesian Inference and Maximum Entropy Methods in Science and
  Engineering. pp 395--405, \mn@doi{10.1063/1.1835238}

\bibitem[\protect\citeauthoryear{Skilling}{Skilling}{2006}]{Skilling_2006}
Skilling J.,  2006, \mn@doi [Bayesian Analysis] {10.1214/06-BA127}, 1, 833

\bibitem[\protect\citeauthoryear{{Speagle}}{{Speagle}}{2020}]{Speagle_2020}
{Speagle} J.~S.,  2020, \mn@doi [\mnras] {10.1093/mnras/staa278}, \href
  {https://ui.adsabs.harvard.edu/abs/2020MNRAS.493.3132S} {493, 3132}

\bibitem[\protect\citeauthoryear{{Suyu}, {Goobar}, {Collett}, {More}  \&
  {Vernardos}}{{Suyu} et~al.}{2023}]{Suyu_2023}
{Suyu} S.~H.,  {Goobar} A.,  {Collett} T.,  {More} A.,   {Vernardos} G.,  2023,
  preprint, \href {https://ui.adsabs.harvard.edu/abs/2023arXiv230107729S} {}
  (\mn@eprint {arXiv} {2301.07729})

\bibitem[\protect\citeauthoryear{{Tak}, {Mandel}, {van Dyk}, {Kashyap}, {Meng}
  \& {Siemiginowska}}{{Tak} et~al.}{2017}]{Tak_2017}
{Tak} H.,  {Mandel} K.,  {van Dyk} D.~A.,  {Kashyap} V.~L.,  {Meng} X.-L.,
  {Siemiginowska} A.,  2017, \mn@doi [Annals of Applied Statistics]
  {doi:10.1214/17-AOAS1027}, \href
  {https://ui.adsabs.harvard.edu/abs/2016arXiv160201462T} {11, 1309}

\bibitem[\protect\citeauthoryear{{Tewes}, {Courbin}  \& {Meylan}}{{Tewes}
  et~al.}{2013}]{Tewes_2013_pycs}
{Tewes} M.,  {Courbin} F.,   {Meylan} G.,  2013, \mn@doi [\aap]
  {10.1051/0004-6361/201220123}, \href
  {https://ui.adsabs.harvard.edu/abs/2013A&A...553A.120T} {553, A120}

\bibitem[\protect\citeauthoryear{{Treu} et~al.,}{{Treu}
  et~al.}{2016}]{Treu_2016}
{Treu} T.,  et~al., 2016, \mn@doi [\apj] {10.3847/0004-637X/817/1/60}, \href
  {https://ui.adsabs.harvard.edu/abs/2016ApJ...817...60T} {817, 60}

\bibitem[\protect\citeauthoryear{{Treu}, {Suyu}  \& {Marshall}}{{Treu}
  et~al.}{2022}]{Treu_2022}
{Treu} T.,  {Suyu} S.~H.,   {Marshall} P.~J.,  2022, \mn@doi [\aapr]
  {10.1007/s00159-022-00145-y}, \href
  {https://ui.adsabs.harvard.edu/abs/2022A&ARv..30....8T} {30, 8}

\bibitem[\protect\citeauthoryear{{Vernardos} \& {Fluke}}{{Vernardos} \&
  {Fluke}}{2014}]{Vernardos_Fluke_2014}
{Vernardos} G.,  {Fluke} C.~J.,  2014, \mn@doi [Astronomy \& Computing]
  {10.1016/j.ascom.2014.05.002}, \href
  {https://ui.adsabs.harvard.edu/abs/2014A&C.....6....1V} {6, 1}

\bibitem[\protect\citeauthoryear{{Vernardos}, {Fluke}, {Bate}  \&
  {Croton}}{{Vernardos} et~al.}{2014}]{Vernardos_2014}
{Vernardos} G.,  {Fluke} C.~J.,  {Bate} N.~F.,   {Croton} D.,  2014, \mn@doi
  [\apjs] {10.1088/0067-0049/211/1/16}, \href
  {https://ui.adsabs.harvard.edu/abs/2014ApJS..211...16V} {211, 16}

\bibitem[\protect\citeauthoryear{{Vernardos}, {Fluke}, {Bate}, {Croton}  \&
  {Vohl}}{{Vernardos} et~al.}{2015}]{Vernardos_2015}
{Vernardos} G.,  {Fluke} C.~J.,  {Bate} N.~F.,  {Croton} D.,   {Vohl} D.,
  2015, \mn@doi [\apjs] {10.1088/0067-0049/217/2/23}, \href
  {https://ui.adsabs.harvard.edu/abs/2015ApJS..217...23V} {217, 23}

\bibitem[\protect\citeauthoryear{{Villar} et~al.,}{{Villar}
  et~al.}{2019}]{Villar_2019}
{Villar} V.~A.,  et~al., 2019, \mn@doi [\apj] {10.3847/1538-4357/ab418c}, \href
  {https://ui.adsabs.harvard.edu/abs/2019ApJ...884...83V} {884, 83}

\bibitem[\protect\citeauthoryear{{Vincenzi}, {Sullivan}, {Firth},
  {Guti{\'e}rrez}, {Frohmaier}, {Smith}, {Angus}  \& {Nichol}}{{Vincenzi}
  et~al.}{2019}]{Vincenzi_2019}
{Vincenzi} M.,  {Sullivan} M.,  {Firth} R.~E.,  {Guti{\'e}rrez} C.~P.,
  {Frohmaier} C.,  {Smith} M.,  {Angus} C.,   {Nichol} R.~C.,  2019, \mn@doi
  [\mnras] {10.1093/mnras/stz2448}, \href
  {https://ui.adsabs.harvard.edu/abs/2019MNRAS.489.5802V} {489, 5802}

\bibitem[\protect\citeauthoryear{{Weisenbach}, {Collett}, {Sainz de Murieta},
  {Krawczyk}, {Vernardos}, {Enzi}  \& {Lundgren}}{{Weisenbach}
  et~al.}{2024}]{Weisenbach_2024}
{Weisenbach} L.,  {Collett} T.,  {Sainz de Murieta} A.,  {Krawczyk} C.,
  {Vernardos} G.,  {Enzi} W.,   {Lundgren} A.,  2024, preprint, \href
  {https://ui.adsabs.harvard.edu/abs/2024arXiv240303264W} {} (\mn@eprint
  {arXiv} {2403.03264})

\bibitem[\protect\citeauthoryear{{Wojtak}, {Hjorth}  \& {Gall}}{{Wojtak}
  et~al.}{2019}]{Wojtak_2019}
{Wojtak} R.,  {Hjorth} J.,   {Gall} C.,  2019, \mn@doi [\mnras]
  {10.1093/mnras/stz1516}, \href
  {https://ui.adsabs.harvard.edu/abs/2019MNRAS.487.3342W} {487, 3342}

\bibitem[\protect\citeauthoryear{{Wong} et~al.,}{{Wong}
  et~al.}{2020}]{Wong_2019}
{Wong} K.~C.,  et~al., 2020, \mn@doi [\mnras] {10.1093/mnras/stz3094}, \href
  {https://ui.adsabs.harvard.edu/abs/2020MNRAS.498.1420W} {498, 1420}

\bibitem[\protect\citeauthoryear{{Woosley}, {Pinto}  \& {Ensman}}{{Woosley}
  et~al.}{1988}]{Woosley_1988}
{Woosley} S.~E.,  {Pinto} P.~A.,   {Ensman} L.,  1988, \mn@doi [\apj]
  {10.1086/165908}, \href
  {https://ui.adsabs.harvard.edu/abs/1988ApJ...324..466W} {324, 466}

\makeatother
\end{thebibliography}
